%% file: MAIN.tex
\documentclass{article} 
\usepackage{iclr2027_conference,times}

\input{math_commands.tex}

\usepackage{url}
\usepackage{enumitem}
\usepackage{graphicx}
\usepackage[dvipsnames]{xcolor}
\usepackage[colorlinks,citecolor=gray]{hyperref}
\usepackage{booktabs,tabularx}
\usepackage{wrapfig}

\title{Receptive-field-constrained stimulus optimization for human early and intermediate visual cortex}

\author{%
  \normalfont
  \begin{tabular}[t]{@{}l@{\hspace{3em}}l@{}}
    \textbf{Junru Zhao\textsuperscript{\textdagger}} 
      & \textbf{Hanfei Guo\textsuperscript{\textdagger}} \\
    Carnegie Mellon University 
      & Carnegie Mellon University \\
    \texttt{junruz@andrew.cmu.edu}
      & \texttt{hanfeig@andrew.cmu.edu} \\[1.5em]
    \textbf{Andrew Luo} 
      & \textbf{Margaret M. Henderson} \\
    University of Hong Kong 
      & Carnegie Mellon University \\
    \texttt{aluo@hku.hk}
      & \texttt{mmhender@cmu.edu}
  \end{tabular}%
}

\iclrfinalcopy 
\begin{document}

\maketitle
\begingroup
\renewcommand{\thefootnote}{\textdagger}
\footnotetext{These authors contributed equally to this work.}
\endgroup

\input{sections/00_abstract}
\input{sections/01_intro}
\input{sections/02_related_work}

\input{sections/03_method}

\input{sections/04_experiments}
\input{sections/05_results}

\input{sections/06_discussion}

\clearpage
\bibliography{iclr2027_conference}
\bibliographystyle{iclr2027_conference}

\appendix
\clearpage
\input{sections/99_appendix}


\end{document}

%% file: math_commands.tex
\usepackage{amsmath,amsfonts,bm}

\def\eqref#1{equation~\ref{#1}}

\def\1{\bm{1}}

\DeclareMathAlphabet{\mathsfit}{\encodingdefault}{\sfdefault}{m}{sl}
\SetMathAlphabet{\mathsfit}{bold}{\encodingdefault}{\sfdefault}{bx}{n}



%% file: sections/00_abstract.tex
\begin{abstract}

 An ongoing challenge in sensory neuroscience is to characterize the feature dimensions encoded by cortical populations. Recent approaches probe feature selectivity in a data-driven way, by synthesizing a most-exciting-input (MEI) for a target neural population. While this approach has been successfully applied to human higher visual cortex using fMRI data, generating MEIs for early- and mid-level retinotopic visual areas requires additional modeling constraints due to small receptive field sizes. To address this challenge, we introduce two novel MEI generation frameworks, Receptive Field Diffusion for Visual Exploration (\textbf{RF-DiVE}) and Receptive Field Gradient Optimization (\textbf{RF-GO}). Both methods use a population receptive field (pRF)-constrained voxelwise encoding model; RF-DiVE combines this with a pretrained latent diffusion model, while RF-GO uses regularized gradient ascent. When applied to single voxels in retinotopically defined areas V1-hV4, using data from the Natural Scenes Dataset, we obtain MEIs that exhibit consistent structure within the pRF, suggesting selectivity for local features like contour, color, and texture. We systematically compare MEIs generated by RF-DiVE and RF-GO using two encoding backbones, performing \textit{in-silico} validation of predicted responses to MEIs using independent encoding models. Across all methods and all visual areas, MEIs elicit higher model-predicted responses than the most activating natural images. We further find that the choice of generation framework and encoding backbone differentially affects MEI properties, including their visual appearance, structural interpretability, and cross-model generalizability. These results offer a new approach for performing data-driven characterization of spatial and feature selectivity across human visual cortex.
\end{abstract}

%% file: sections/01_intro.tex
\section{Introduction}
\input{fig_text/teaser}

Characterizing feature selectivity in the visual cortex has long been a central challenge in sensory neuroscience.
Traditional approaches measure neural selectivity using small sets of hand-selected stimuli (e.g., oriented gratings, isolated images of faces or objects), which can bias results toward established hypotheses and fail to capture more novel aspects of selectivity. This issue may be especially pronounced when investigating intermediate stages of the visual system (e.g., V4), where the space of mid-level features encoded is high-dimensional, nonlinear, and unlikely to be fully described by interpretable dimensions.
To overcome these challenges, recent approaches use stimulus optimization to probe neural feature selectivity in a more data-driven way. Stimulus optimization frameworks use computational encoding models to generate a most-exciting-input (MEI) for a target neural population measured with either invasive single-unit or fMRI recordings, using a forward encoding model combined with either gradient ascent or a generative model \citep{bashivan2019neural, ponce2019evolving, walker2019inception, cowley2026compact, gu2022neurogen, ratan2021computational}. A recent method (BrainDiVE) uses diffusion models \citep{rombach2022high} to generate fMRI-based MEIs with more realistic semantic information compared to previous approaches \citep{henderson2026diffusion, hwang2026silico, luo2023brain}. This diffusion-based approach has thus far been applied to regions of higher visual cortex, but has not yet been applied to early and intermediate visual areas.

Extending this approach to early and intermediate visual cortex requires new modeling strategies. In particular, early and intermediate visual responses are strongly shaped by retinotopic organization and small population receptive fields (pRFs; \cite{dumoulin2008population}), constraints which are not modeled in existing fMRI stimulus optimization methods. Further, neural selectivity varies widely across sub-populations in early and intermediate cortical regions, necessitating fine-grained analysis of response properties at the single voxel level, in contrast to previous approaches which have averaged across voxels in a region. Moreover, because early visual cortex regions are not strongly selective for semantic information, it is not yet clear whether the natural image prior imposed by a diffusion model is required in order to generate interpretable MEIs, or whether a gradient-based method can provide complementary insight.

Here, we introduce and compare two MEI-generation frameworks for single voxels in early and intermediate human visual cortex: receptive field diffusion for visual exploration (\textbf{RF-DiVE}), and receptive field gradient optimization (\textbf{RF-GO}). Both frameworks use a subject-specific, voxelwise feature-weighted receptive field (fwRF) encoding model \citep{st2018feature} to guide stimulus optimization. By spatially pooling features of a deep convolutional neural network according to an estimated voxel-specific pRF, this model incorporates both feature and spatial selectivity to predict fMRI responses. RF-GO uses these predictions to directly optimize images through regularized gradient ascent, whereas RF-DiVE uses response gradients to guide a pretrained latent diffusion model (LDM) with a natural-image prior. Targeting individual voxels allows both approaches to probe variation in selectivity within cortical areas while accounting for differences in receptive-field location and size. Together, these frameworks extend model-guided stimulus optimization to the spatially specific and heterogeneous representations found across V1, V2, V3, and hV4. 

Our contributions are as follows:

\begin{itemize}
\item We introduce two new frameworks, RF-DiVE and RF-GO, enabling spatially-constrained MEI synthesis for individual voxels in early and intermediate retinotopic cortex.
\item Applying our method to data from the Natural Scenes Dataset (NSD; \cite{allen2022massive}), we generate single-voxel MEIs that exhibit coherent, spatially localized structure, appearing to depict mid-level properties such as orientation, curvature, color, and form.
\item We provide \textit{in-silico} validation of these MEIs using an independent test brain encoder and show MEIs consistently elicit higher predicted responses than natural-image controls. 
\item We provide a direct comparison of gradient-based (RF-GO) and diffusion-based (RF-DiVE) optimization methods for the same brain regions, which reveal differences in model generalization performance, visual interpretability, and low-level visual statistics, which also depend on the DNN backbone used.
\item We use a human behavioral study to provide initial characterization of the visual properties captured by these MEIs, showing that RF-DiVE MEIs for hV4 voxels depict more three-dimensional form than those for V1 voxels, indicating our method can capture expected differences between early and intermediate visual cortex.

\end{itemize}

%% file: fig_text/teaser.tex
\begin{figure}[t]
    \centering
    \includegraphics[width=\linewidth]{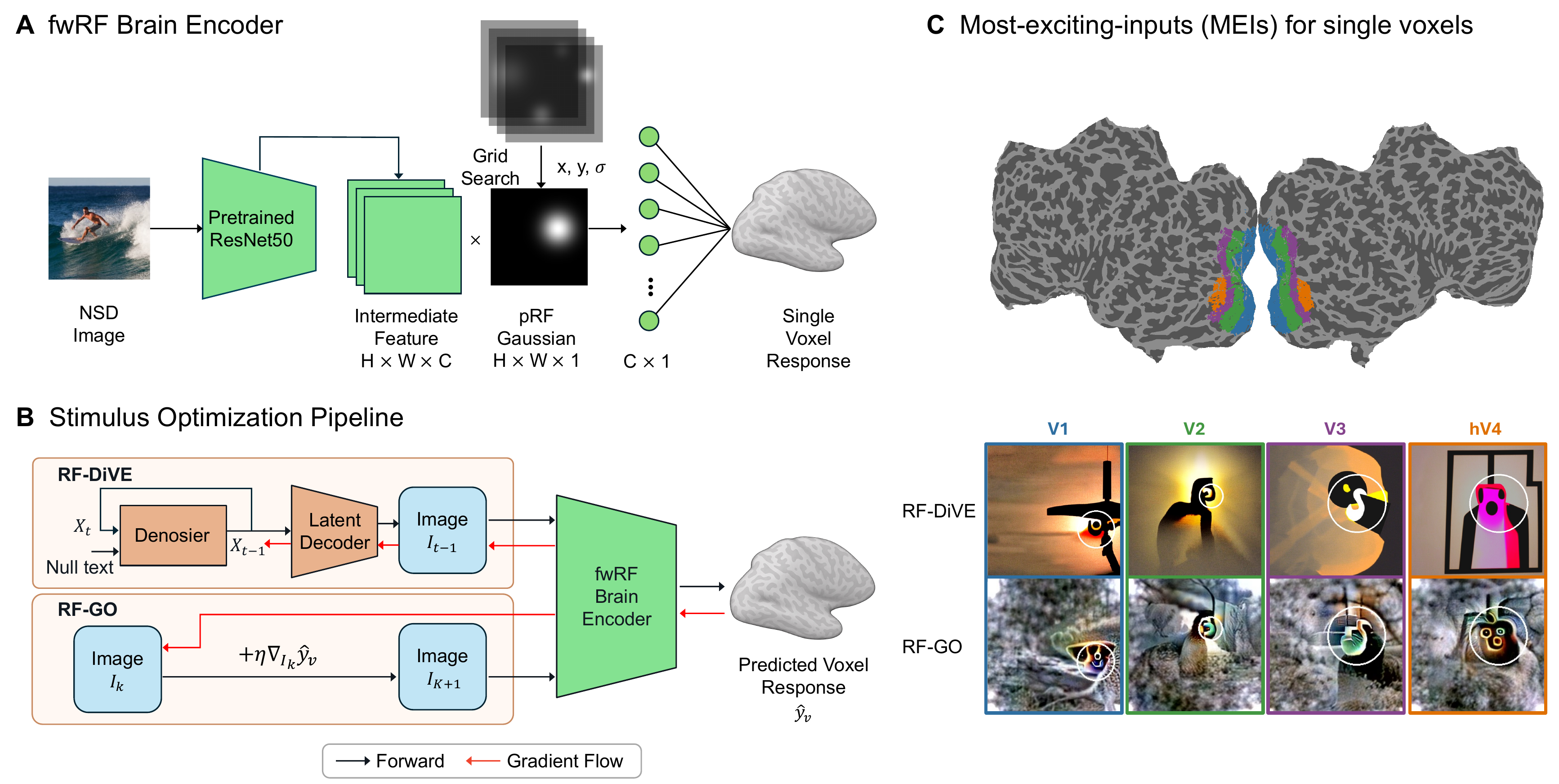}
    \caption{\textbf{Our stimulus optimization pipeline.} \textbf{(A)} Our encoding model predicts activity in early visual cortex using a deep CNN backbone, constrained by a Gaussian that models voxel population receptive field (pRF). \textbf{(B)} Visualization of our \textbf{RF-DiVE} and \textbf{RF-GO} frameworks. In RF-DiVE, the optimization process is constrained by a diffusion model, while RF-GO optimizes the image using MACO. \textbf{(C)} Examples of most-exciting-images (MEIs)
    for individual voxels in V1, V2, V3, and hV4. Each column corresponds to one visual area; white circle labels the voxel's pRF}
    \label{fig:teaser}
\end{figure}

%% file: sections/02_related_work.tex
\section{Related work}

\textbf{Feature selectivity in the visual hierarchy.} In classic models, representational complexity is thought to increase along the primate visual cortex hierarchy, where V1 encodes basic features such as contrast, orientation, and retinotopic position \citep{Hubel1962, Carandini2005}, V4 encodes contour, shape, color, and texture \citep{pasupathy2020visual, Conway2007SpecializedCortex}; and higher visual areas represent objects and semantic categories \citep{GrillSpector2014, Desimone1984}. However, the feature space represented in intermediate areas like V4 remains poorly understood, and recent work also shows that V1 can encode more complex properties such as curvature \citep{tang2018complex, ding2026functional}. Our work addresses these questions with a data-driven approach for characterizing feature selectivity in early and intermediate visual cortex.

\textbf{Voxelwise encoding models for fMRI data.}  Forward encoding models (i.e., brain encoders) provide a powerful way to model neural activity, providing an image-computable mapping from stimulus to brain response by extracting features from deep neural network (DNN) models. Task-optimized DNNs have proven effective as feature extractors for brain encoders because of their hierarchical representations similar to the primate ventral visual stream \citep{naselaris2011encoding, khaligh2014deep, yamins2014performance, yamins2016using, eickenberg2017seeing, schrimpf2018brain, zhuang2021unsupervised, conwell2024large}. Standard encoding models linearly map DNN features to voxel responses, often collapsing spatial information by averaging. In contrast, feature-weighted receptive field models explicitly capture voxelwise spatial and feature selectivity through a pRF model \citep{st2018feature}. We adopt this framework with multiple pretrained visual DNN backbones.

\textbf{Stimulus optimization frameworks.} Stimulus optimization synthesizes MEIs by maximizing brain-encoder responses, providing a data-driven probe of neural selectivity. Prior work has used regularized gradient ascent \citep{bashivan2019neural, walker2019inception, willeke2026deep, cowley2026compact, khosla2026higher} and generative models such as GANs and diffusion models \citep{ponce2019evolving, gu2022neurogen, ratan2021computational, luo2023brain, henderson2026diffusion, hwang2026silico}. Strong natural-image priors imposed by generative models can improve interpretability, whereas weaker gradient-based priors may reveal selectivity outside the natural-image distribution. Related work has also perturbed existing images to modulate regional responses \citep{Prince2026, garcia2025brainactiv}. For gradient-based methods, recent closed-loop results further suggest strong backbone dependence, with adversarially-trained networks performing well for neural control \citep{Prince2026}. Here, we compare gradient- and diffusion-based MEIs in visual properties and \textit{in silico} generalization using ADV-RN50 \citep{robustness} and DINO-RN50 \citep{caron2021emerging} brain encoders.

%% file: sections/03_method.tex
\section{Method}

In this section, we first describe our approach for constructing an image-computable pRF- constrained brain encoder, which can predict the response in early and intermediate visual fMRI voxels for image stimuli. 

Then, we introduce two frameworks that use the pRF-constrained brain encoder to generate MEIs for early and intermediate visual cortex voxels: diffusion-based RF-DiVE framework and gradient-based RF-GO framework.

\label{sec:fwrf}
\textbf{pRF-constrained brain encoder.} The brain encoder was implemented using a CNN-based fwRF model ~\citep{st2018feature}, as shown in Figure~\ref{fig:teaser}A. Specifically, the input image was first processed by a CNN
 
to obtain intermediate feature maps with height, width, and channel dimensions ($h, w, c$). The pRF for each voxel was modeled as a 2D isotropic Gaussian, with center ($x, y$) and size ($\sigma$) parameters. To extract pRF-specific features, we took a dot product between the pRF (scaled to size $h$ x  $w$) and the spatial dimensions of the CNN feature maps, achieving spatial pooling that reduced the features to a channel dimension ($c$). Features were concatenated across multiple layers in the channel dimension. The resulting feature representation was then linearly combined with channel weights and biases (estimated using ridge regression) to predict the response of each target voxel. The pRF parameters for each voxel were determined by grid search over a set of candidate pRFs in a log-polar grid \citep{henderson2023low, henderson2023texture}; see Section \ref{sec: expts}.

\textbf{Receptive Field Diffusion for Visual Exploration (RF-DiVE).} RF-DiVE uses a pretrained LDM ~\citep{rombach2022high} as in BrainDiVE \citep{luo2023brain}. At each step of generation, the LDM latent representation is passed through the latent decoder to produce an image, which is then evaluated by a fwRF brain encoder. The predicted response of a target voxel was used as the optimization objective, and gradients were propagated through the brain encoder and latent decoder, perturbing the denoising procedure toward images that maximize the target voxel’s predicted activity (Figure~\ref{fig:teaser}B).

\label{sec:GO_gen}
\textbf{Receptive Field Gradient Optimization (RF-GO).} RF-GO uses encoder gradients to optimize image Fourier phase to maximize predicted voxel responses (Figure~\ref{fig:teaser}B). To avoid high-frequency artifacts common with gradient-based synthesis, MEIs were generated using Magnitude-Constrained Optimization (MACO, \cite{fel2023unlocking}), which regularizes the image by fixing the Fourier magnitude spectrum to an ImageNet-derived natural-image template while optimizing only the Fourier phase.

%% file: sections/04_experiments.tex
\section{Experiments}
\label{sec: expts}

\textbf{Dataset and partitions.} All modeling was performed using the NSD dataset, a large-scale 7T fMRI dataset in which 8 participants each viewed up to 10,000 natural images (shown at 8.4$^\circ$ visual angle). We used 4 subjects (S1, S2, S5, and S7) who each viewed all 10k images each repeated 3 times. Out of the 10k images for each subject, 1000 images were shared across the 4 subjects and 9000 were unique. For each subject, we split the 10k images randomly in half resulting in 2 partitions (P1 and P2), where each set contains 4500 unique images and 500 shared images. Within each partition, the 4500 unique images were further split into training and nested held-out sets with ratio 9:1, and the 500 shared images were used as test images. 

We used single-image $\beta$ weights for each voxel computed using the GLMSingle pipeline \citep{Prince2022ImprovingGLMsingle}, which we \textit{z}-scored for each voxel within each session, then averaged over repeats of each image.  
To define ROIs, we used the NSD’s subject-specific V1-hV4 definitions (combining dorsal and ventral subdivisions and two hemispheres), which were manually delineated from independent population receptive field (pRF) mapping data as described by \cite{allen2022massive}.

\textbf{Brain encoders, backbones, and training.} 
We constructed four brain encoder backbone types for each subject. The first two brain encoders (``generation models'') were used for MEI generation, and were based on ADV-RN50 \citep{robustness} and DINO-RN50 \citep{caron2021emerging} backbones, each trained using the data from P1 only. A separate encoder (``ranking model'') was constructed based on OpenCLIP-RN50 \citep{ilharco_gabriel_2021_5143773}, also trained on P1, and was used to select the top MEIs across random seeds. Finally, an independent test encoder (``test model'') was based on a different convolutional architecture, OpenCLIP ConvNeXt-Base \citep{ilharco_gabriel_2021_5143773}, trained on P2, and was used to independently evaluate the model-predicted responses elicited by the MEIs. See Appendix~\ref{app:encoding_performance} for the overall $R^2$ of each model backbone.

Each encoding model was constructed using the fwRF framework (Section \ref{sec:fwrf}).  Input images were transformed according to the preprocessing associated with the pretrained backbone before extracting activations of first ReLU and layer 1-4 for RN-50, stem and stage 1-4 for ConvNext-Base. For each voxel, feature maps were spatially pooled using a Gaussian pRF shared across layers, concatenated, standardized channel-wise, and mapped to predicted fMRI responses through a linear readout with an intercept. Readout weights were fitted using ridge regression. The optimal pRF position and size ($x, y$, $\sigma$), as well as regularization strength ($\lambda$) were selected based on minimum prediction error on the nested held-out set within the partition.  This procedure for fitting the best pRF was performed separately for each of the generation and test models, so pRF could vary depending on backbone. The ranking encoders were fitted using the pRFs of their corresponding generation encoder.  
For each subject and each area, we selected the top 20 voxels based on the generation models' test-set \(R^2\) within P1, which were then used for MEI generation. Generation models with different backbones (ADV/DINO) can select voxels differently, and their estimated pRFs for the same voxel can also vary (see Appendix~\ref{app:prf_coverage} ).

\textbf{Most-Exciting-Image (MEI) optimization.} For RF-DiVE, we used Stable Diffusion v2.1 with a DPM-Solver multistep scheduler, 100 denoising steps, an empty text prompt, and classifier-free guidance disabled. At each step, gradients from a subject-specific encoder (``generation model'') guided generation to maximize the predicted response of a single target voxel. Brain-guidance and self-attention-guidance scales were set to 300 and 0.75, respectively. 
For RF-GO, the Fourier phases were optimized for 100 steps using NAdam with a learning rate of 1.0. The objective was to maximize the predicted response of the target voxel. For both RF-DiVE and RF-GO, for each voxel, we generated 1,000 images with different random seeds.

To control for low-level brightness and contrast, during optimization, every image presented to the encoding model was projected to a full-image mean luma of \(116/255\) and RMS luma contrast of \(57.48/255\), corresponding to the mean luma and mean RMS luma contrast computed from natural images in the training sets. This full-image constraint was applied again before ranking and evaluation (Appendix~\ref{supp:luma_constraint}). 
Additional controls for luma and contrast within the pRF region itself are provided in the Appendix~\ref{app:constraint_effects}.
Unless otherwise specified, quantitative analyses were performed on the top 10 MEIs selected according to the predicted response based on the ``ranking model'' (see above section). Additional ablations using MEIs selected by the generation model are provided in the Appendix~\ref{app:responses} and~\ref{app:laion_advantages}.

To provide a baseline comparison for the predicted response to MEIs, we used the same ranking model to identify the top-10 predicted natural images from a large image set. Natural images were either sampled from the NSD stimulus set \citep{allen2022massive}, using P2 test set, or from 500 randomly sampled images from the LAION-fMRI stimulus set (\cite{zerbe2026laionfmri, schuhmann2022laion5bopenlargescaledataset}), with LAION-fMRI providing a complementary natural-image control.

%% file: sections/05_results.tex
\section{Results}

\subsection{Generated MEIs show both spatial and feature specificity}
\begin{figure}[ht]
    \centering
    \includegraphics[width=\linewidth]{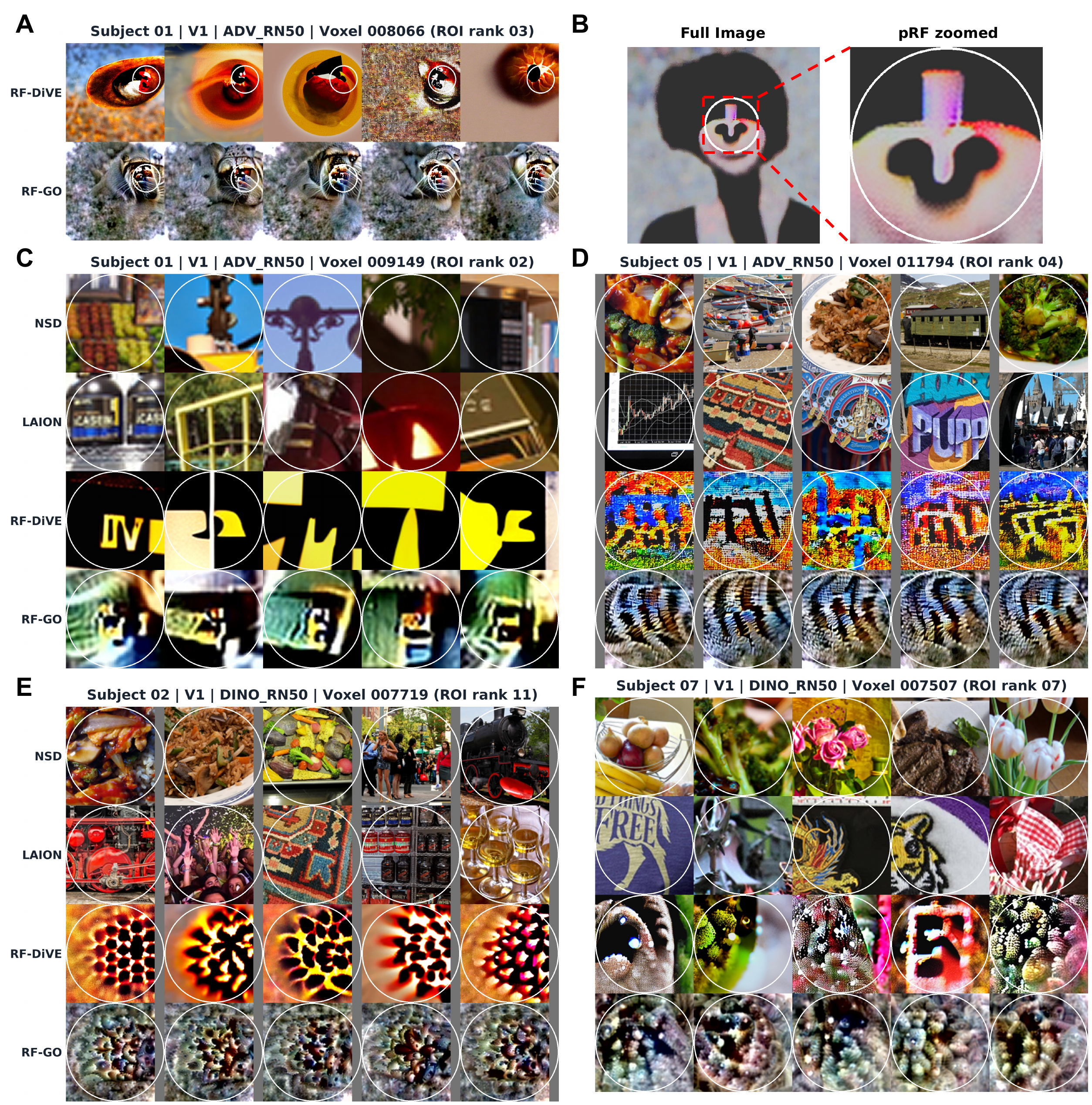}
    \caption{\textbf{Spatial and feature specificity of V1 MEIs.}
\textbf{(A)} Top 5 MEIs for an example voxel using both frameworks.
The MEIs show consistent patterns within the pRF (white circle, $3\sigma$ range of the Gaussian)
and greater variation outside it.
\textbf{(B)} Full-image and pRF-zoomed views of an example MEI.
\textbf{(C--F)} pRF-zoomed top 5 natural images and MEIs for four example V1 voxels, shown for both ADV-RN50 and DINO-RN50 backbones. See Appendix~\ref{app:additional_meis} for more examples.} 
    \label{fig:mei_1}
\end{figure}

Figure~\ref{fig:mei_1}A shows five MEIs generated for an example V1 voxel using both RF-DiVE and RF-GO. The generated MEIs exhibit consistent feature patterns within the voxel's pRF (white circle), while remaining largely random outside it, demonstrating that the pRF constraint imposed by the encoding model led to spatially specific MEIs. Figure~\ref{fig:mei_1}B illustrates a zoomed view of the pRF region within a full MEI; this zoomed view is used to display images in later figures.

Beyond spatial specificity, the MEIs also exhibit coherent structure that captures distinct visual features, including color, form, and texture. Comparing across methods, the MEIs resulting from RF-DiVE often appear more coherent and more colorful than the RF-GO MEIs (we discuss these differences in more detail in the next section), but we also observe shared structure in the MEIs for a given voxel that is common across both methods. Figure~\ref{fig:mei_1}C shows MEIs that depict high-contrast yellow--black rectilinear and angular contours, Figure~\ref{fig:mei_1}D shows multicolored bent lines and curves that create abstract patterns, Figure~\ref{fig:mei_1}E shows clusters of dark dots on warm-colored backgrounds; and Figure~\ref{fig:mei_1}F shows dense, pinkish or greenish speckled patterns. These patterns are evident in both the RF-DiVE and RF-GO MEIs, and similar features are also visible in the top natural images for each voxel. For example, the top images for the voxel in Figure~\ref{fig:mei_1}C include a bright yellow railing, and the voxel in Figure~\ref{fig:mei_1}F includes pink flowers with green stems. Notably, inspecting the top natural images alone does not always reveal a coherent set of shared features, which highlights how MEI generation may be needed to uncover the underlying features driving a neural population.

\begin{figure}[h]
    \centering
    \includegraphics[width=\linewidth]{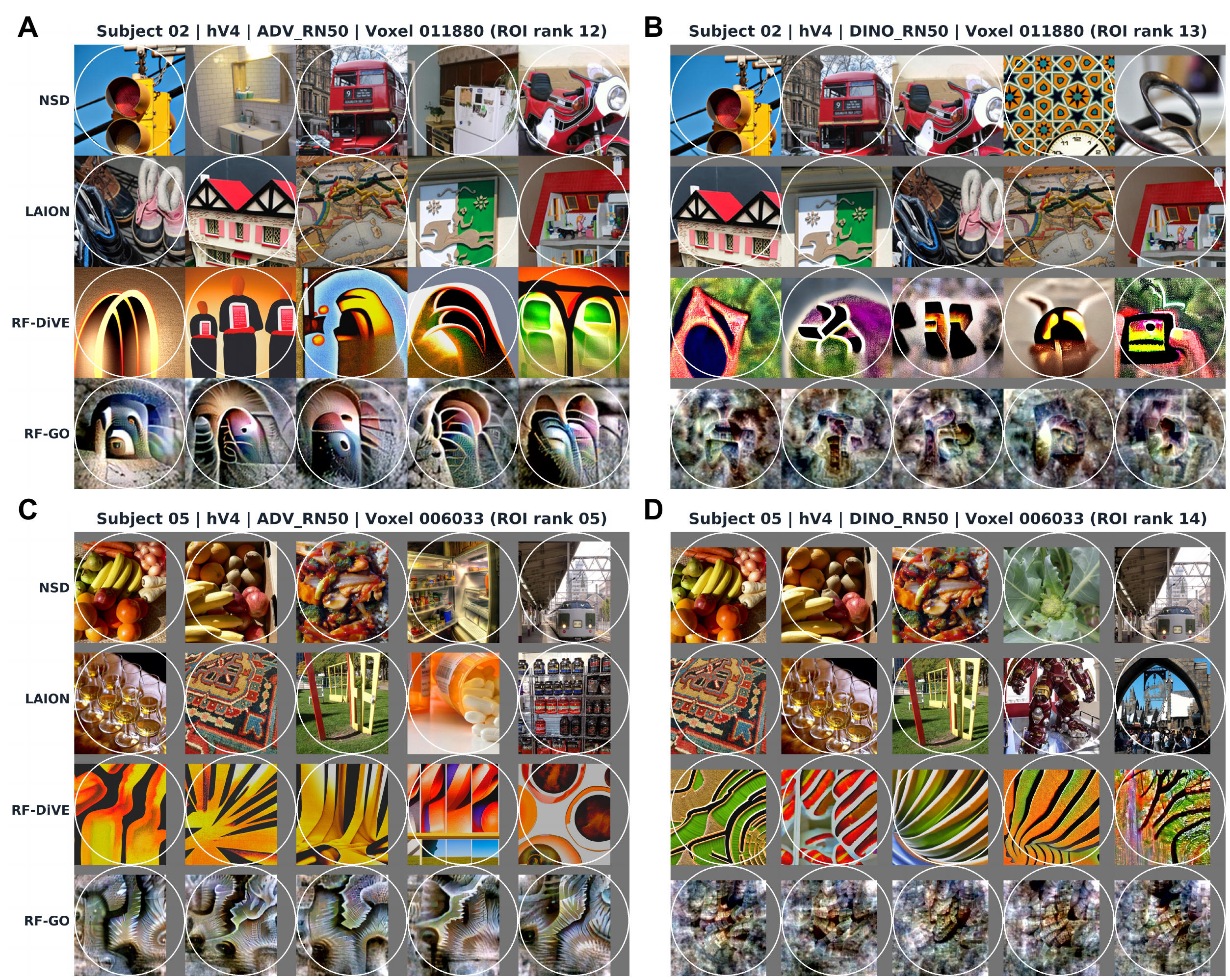}
    \caption{\textbf{Feature specificity of hV4 MEIs across generation
encoders.}
Top 5 natural images and MEIs are shown for two example hV4 voxels (\textbf{A-B} show the same voxel, \textbf{C-D} show the same voxel). Note the pRF is fit separately for each backbone and the rank number of voxels is assigned separately for each backbone. See Appendix~\ref{app:additional_meis} for more examples, including from V2 and V3.} 
    \label{fig:mei_2}
\end{figure}

The MEIs generated for hV4 voxels capture additional intermediate-level visual structure. 
In Figure~\ref{fig:mei_2}, we show MEIs generated for two example hV4 voxels using both DINO and ADV generation backbones, revealing feature content that is consistent across backbone choices. The voxel depicted in Figure~\ref{fig:mei_2}A-B exhibits smooth, arch-like forms and saturated colors, while the voxel in  Figure~\ref{fig:mei_2}C-D exhibits orange--yellow radiating stripes and curved bands. Again, the top natural images show related features (top of a dollhouse for A-B, round fruits for C-D). 

\subsection{Generated MEIs produce higher predicted voxel response}
\begin{figure}[h]
    \centering
    \includegraphics[width=0.95\linewidth]{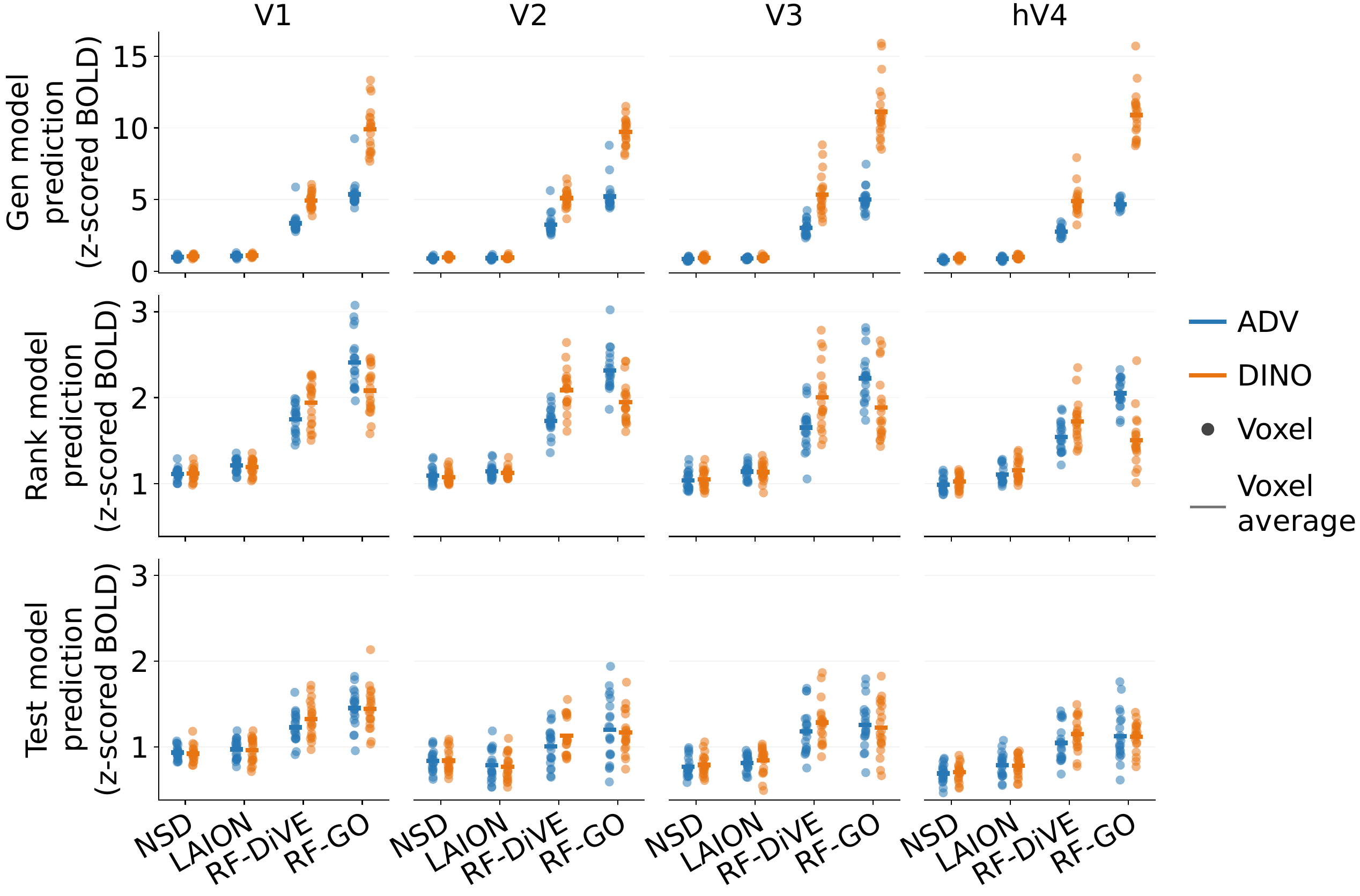}
    \caption{\textbf{Predicted responses to top 10 images for S1.} Top-10 ranking model-selected images for each image group were evaluated by the generation, ranking, and independent test encoders. Each point represents mean prediction across images for one voxel; horizontal bar = mean across voxels in S1. See Appendix~\ref{app:responses} for the other NSD subjects.}
    \label{fig:s1_test}
\end{figure}
To rigorously assess whether the generated MEIs generalize beyond the models used for optimization, we perform \textit{in-silico} evaluation using an independent test encoder with a different architecture, pretraining dataset, and NSD training partition. Figure~\ref{fig:s1_test} compares predicted voxel responses for the top-20 voxels (ranked by $R^2$) in each visual area in S1, elicited by RF-DiVE and RF-GO MEIs versus natural images from NSD and LAION-fMRI, using the generation, ranking, and independent test encoders. Figure~\ref{fig:test_advantage} depicts the advantage for MEIs over natural images (predicted difference score), across all 4 NSD participants. Across these results, both RF-DiVE and RF-GO exhibit a predicted-response advantage over natural images from the stricter control dataset, LAION-fMRI. This difference is especially prominent when evaluated using the generation model, and is attenuated on the ranking and testing model. However, the predicted difference score remains positive for all participants, even in the most challenging case of generalization to the test model.
Although there is variability across participants, this advantage is significantly greater than zero for the majority of subjects in all visual areas and for all generation methods (one-sample \textit{t}-test; p$<$0.01; FDR corrected; see Appendix~\ref{app:paired_tests}, Table~\ref{tab:rf_paired_voxel_tests} for test statistics). These results indicate that the MEIs consistently elicit higher model-predicted responses than the evaluated natural image controls.

\subsection{Effects of different encoder backbones and generation frameworks}
A key strength of our study is the breadth of conditions under which MEIs were generated, spanning different generation methods, brain-encoder backbones, subjects, visual areas, and target voxels. We systematically analyze the effects of these factors on MEI properties and their predicted responses.

\begin{figure}[h]
    \centering
    \includegraphics[width=0.95\linewidth]{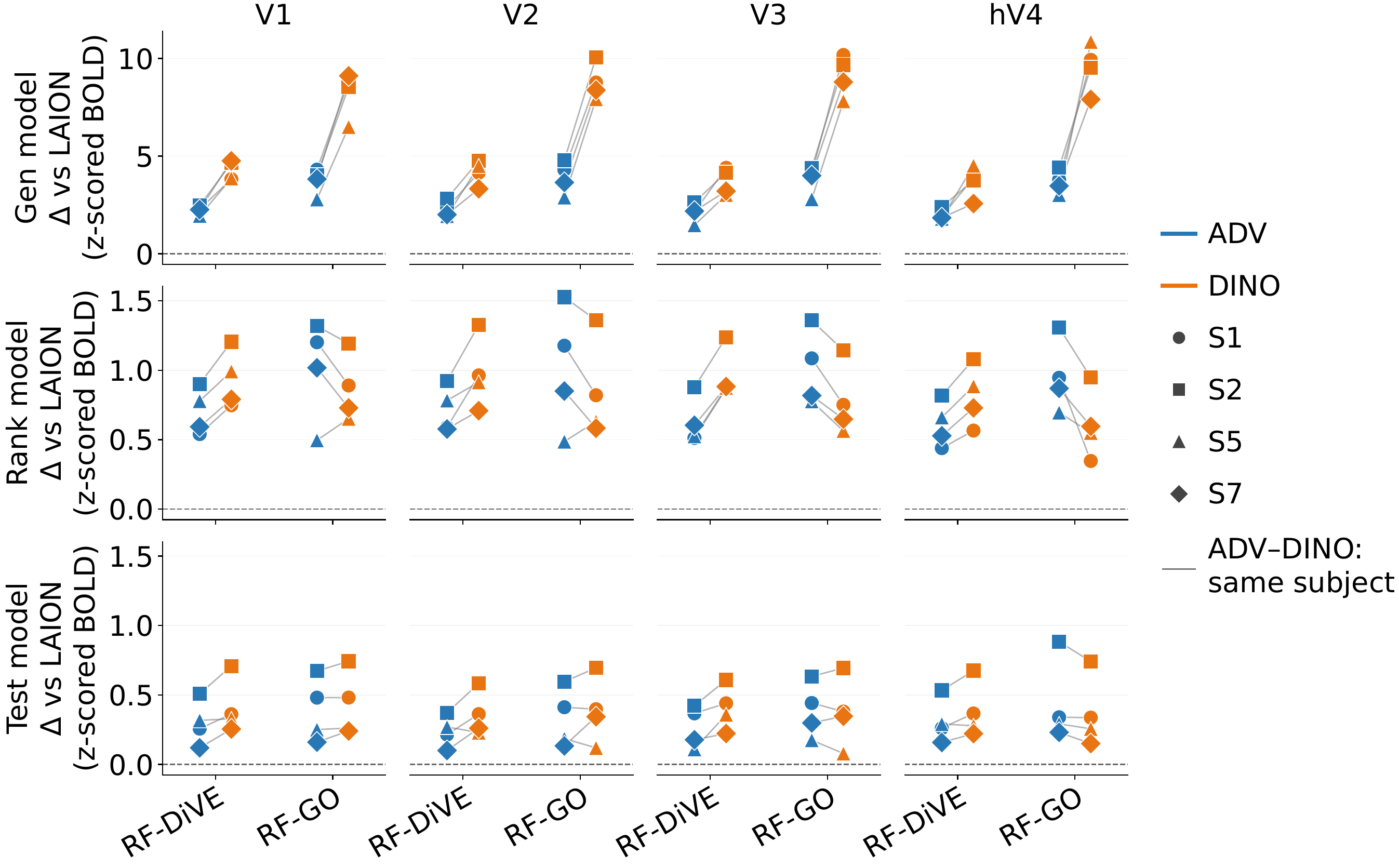}
    \caption{\textbf{Predicted difference score (advantage of MEIs over LAION-fMRI) across subjects.} For each voxel, we computed the predicted-response difference between MEIs and natural images. Points reflect the mean of these differences across voxels within each subject and ROI.}
    \label{fig:test_advantage}
\end{figure}

Figure~\ref{fig:test_advantage} shows a comparison of the predicted difference score of MEIs over LAION-fMRI images, across different encoder backbones and generation frameworks. When evaluated using the generation encoder, RF-GO achieved larger difference scores than RF-DiVE with both backbones. However, this advantage was not consistently retained under the ranking or independent test encoder, suggesting that the learned naturalistic prior enforced by the RF-DiVE method may result in weaker activation of the generation model but improved generalization across backbones, compared to RF-GO.

We also observed effects of model backbone that depended on the generation framework. For RF-DiVE, DINO yielded larger scores than ADV under the generation and ranking encoders, and this tendency was retained in most independent-test comparisons, with DINO having consistently higher difference score than ADV for the majority of participants in all areas (see Appendix~\ref{app:laion_advantages}). However, this pattern differed for RF-GO. For RF-GO, DINO also showed larger difference score than ADV on the generation encoder, but this pattern partially inverted on the ranking and testing model, with DINO having similar or worse difference score relative to ADV in all regions. This may indicate that the DINO backbone, when paired with the RF-GO method, is prone to model-specific biases that impair cross-model generalization, whereas the ADV model results in more robustly generalizable RF-GO MEIs; this is broadly consistent with other recent work \citep{Prince2026}. 
These patterns were similar across all regions tested.

\begin{figure}[h]
    \vspace{-0.1in}
    \centering
    \includegraphics[width=0.85\linewidth]{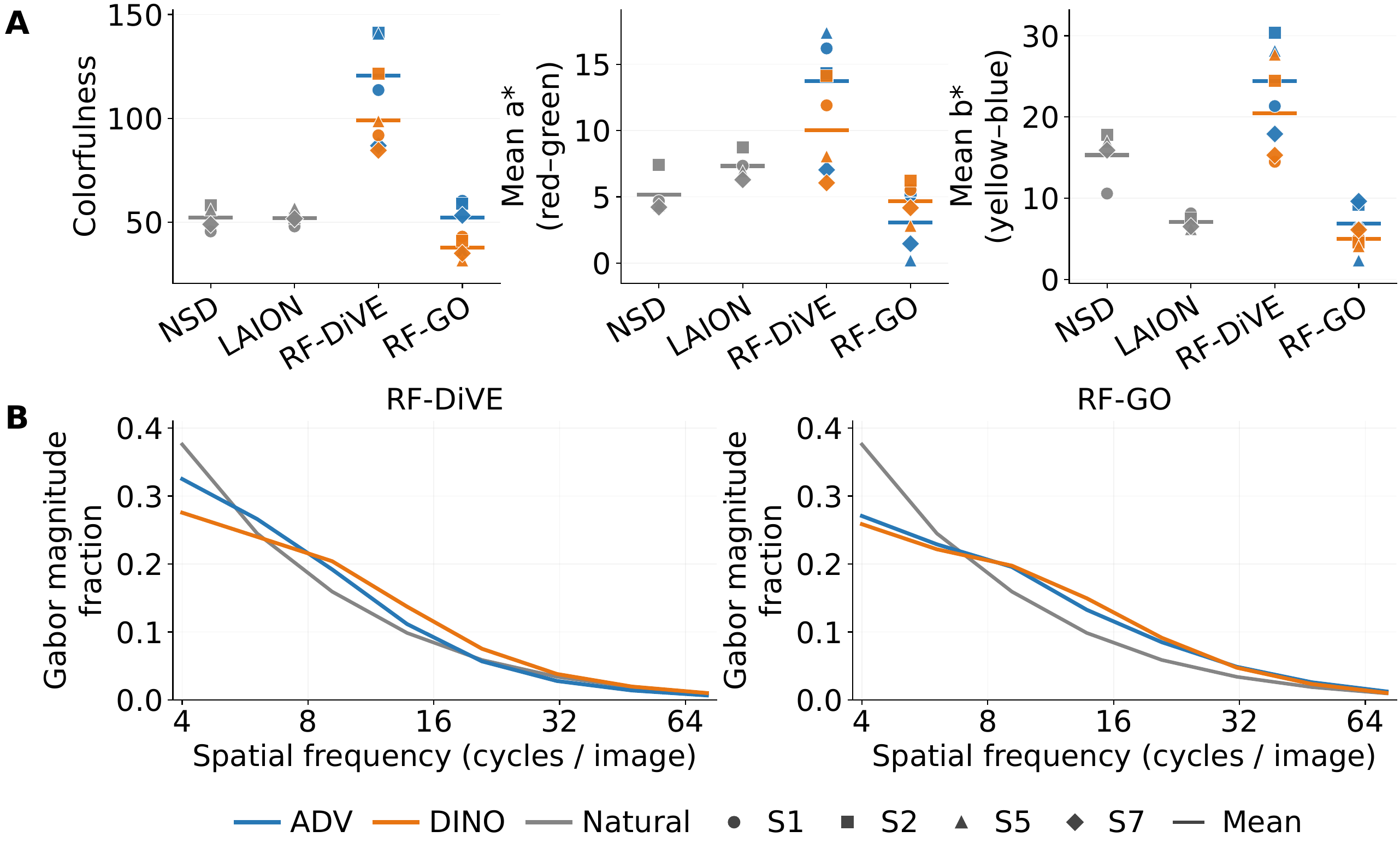}
    \caption{\textbf{Chromatic and spatial-frequency statistics for V1 MEIs.} \textbf{(A)} Colorfulness \citep{hasler2003measuring} and CIELAB color statistics. \textbf{(B)} Normalized Gabor magnitude profiles across spatial frequencies, shown for RF-DiVE and RF-GO. Similar patterns were observed in V2, V3, and hV4.}
\label{fig:image_statistics}
\end{figure}

Given these differences between methods and backbones in model generalization, we next examined differences in the basic visual properties of generated MEIs (see Appendices~\ref{app:color_orientation} and~\ref{supp:structure_comparisons} for image quantification and additional comparisons). Comparing chromatic statistics across generation frameworks and natural-image controls (Figure~\ref{fig:image_statistics}A) shows that RF-DiVE MEIs exhibit higher colorfulness than RF-GO MEIs and natural images. RF-DiVE also shows higher subject-averaged $a^*$ and $b^*$ values, indicating a stronger tendency toward warm colors, particularly red and yellow. These differences are observed with both generation backbones. Figure~\ref{fig:image_statistics}B compares the spatial-frequency profiles of the generated and natural images using normalized Gabor response magnitudes. Relative to natural images, both RF-DiVE and RF-GO MEIs show a smaller relative contribution at the lowest measured spatial frequency and a larger contribution at intermediate frequencies. Within each generation framework, MEIs generated using DINO exhibit a further shift toward intermediate and higher spatial frequencies compared with ADV generated MEIs.

\subsection{Feature differences across regions}

The MEIs from each region depict varied mid-level features that could provide new insight into how feature coding evolves along the visual hierarchy. As a first step toward interpreting the MEI features, we conducted a human behavioral study comparing the perceived three-dimensional form of MEIs generated for V1 and hV4 voxels (see Appendix~\ref{app:behavioral_experiment}). 
\begin{wrapfigure}{r}{0.7\textwidth}
    \centering
    \vspace{-8pt}
    \includegraphics[width=\linewidth]{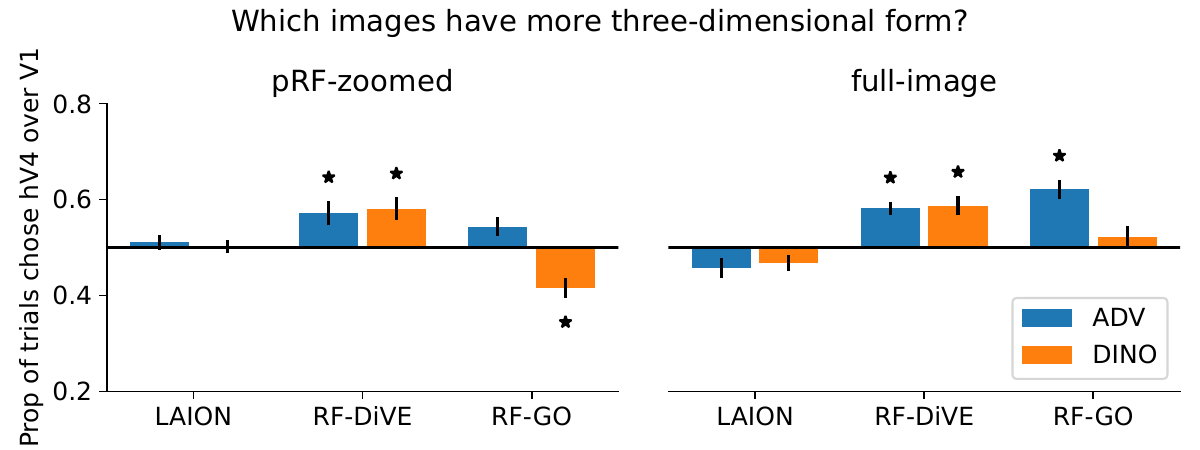}
    \caption{\textbf{Depiction of form in MEIs.}
    Proportion of trials in which participants selected the
    hV4 images as having more three-dimensional form than the V1 images,
    using pRF-zoomed images (left) or full images (right). Bar height and error bars indicate mean$\pm$SEM across 32 participants; asterisk indicates a significant difference from 0.50 (one-sample \textit{t}-test; \textit{p}$<$0.01).}
    \label{fig:behav}
    \vspace{-8pt}
\end{wrapfigure}This design was motivated by past evidence that sensitivity to shape and form increases from V1 to hV4 \citep{Connor2007, pasupathy2020visual}.
In a forced-choice task, participants (n=32 in each of two experiments) more frequently judged hV4 RF-DiVE MEIs as exhibiting greater three-dimensional form than V1 MEIs, across both encoder backbones and viewing conditions. In contrast, no reliable hV4-over-V1 preference was observed for natural images from LAION. Results for RF-GO were less consistent across encoder backbones and viewing conditions, but showed a preference for hV4 in the ADV backbone in the full image condition. These findings suggest that RF-DiVE MEIs may more effectively capture interpretable differences between regions compared to natural images, again illustrating the benefit of MEI synthesis for capturing visual properties that are not readily detectable from natural images alone. Within the RF-GO method, the ADV backbone may also be more effective at capturing these interpretable properties than DINO. Further work is needed to confirm and expand these findings.

%% file: sections/06_discussion.tex
\section{Discussion}
We introduced RF-DiVE and RF-GO, two frameworks for generating most-exciting-images for individual fMRI voxels in early and intermediate human visual cortex. Across V1-hV4, the resulting MEIs contained consistent structure within the estimated pRF and captured diverse features including contour, color, texture, and form. Moreover, MEIs generated by both methods produced higher predicted responses than the strongest natural-image controls across subjects, visual areas, and encoder backbones. These findings extend model-guided stimulus optimization from region-level analyses of higher visual cortex to spatially localized representations at the single-voxel level.

Our analyses also show that an MEI is shaped jointly by the target voxel, the brain-encoder backbone, and the optimization framework. RF-GO had a higher difference score under the generation encoder, but showed relatively weaker generalization to the independent test encoder; this effect was more pronounced for DINO than for ADV. The two frameworks also produced systematically different image statistics: RF-DiVE generated more colorful and warmer images, whereas DINO MEIs shifted toward higher spatial frequencies relative to ADV. At the same time, visual comparison of the MEIs across methods (Figures \ref{fig:mei_1}-\ref{fig:mei_2}) does suggest structural elements such as curves and texture that are common across methods. These patterns suggest that combining multiple methods for MEI generation may be an effective strategy to separate the most robust, generalizable properties of neural tuning from those that are idiosyncratic to one method. 

Several additional limitations suggest directions for future work. First, further validation of the generated MEIs will require closed-loop experiments with human participants. Second, our current quantitative analyses focus on relatively low-level visual properties, including Gabor features, color, edges, and curvature, and therefore provide limited insight into intermediate-level visual features such as texture, contour, and surface structure. Finally, our perceptual experiment was preliminary. Larger-scale behavioral studies, together with more systematic measurements of visual properties such as contour, texture, and shape, will be needed to determine which visual dimensions reliably differentiate cortical areas and generation methods.

%% file: sections/99_appendix.tex
\section{Appendix}

\label{app:appendix}
\noindent\textbf{Contents of this appendix}\par
\begin{list}{}{
    \setlength{\leftmargin}{3em}
    \setlength{\labelwidth}{2.5em}
    \setlength{\labelsep}{0.5em}
    \setlength{\itemsep}{3pt}
    \setlength{\parsep}{0pt}
    \setlength{\topsep}{6pt}
}
    \item[\ref{app:encoding_performance}] Encoding Model Performance \dotfill \pageref{app:encoding_performance}

    \item[\ref{supp:luma_constraint}] Luma mean and RMS-contrast constraints \dotfill \pageref{supp:luma_constraint}

    \item[\ref{app:paired_tests}] Paired-voxel comparisons with natural images \dotfill \pageref{app:paired_tests}

    \item[\ref{app:constraint_effects}] Effects of Final Luma Constraints \dotfill \pageref{app:constraint_effects}

    \item[\ref{app:prf_coverage}] Fitted pRF coverage \dotfill \pageref{app:prf_coverage}

    \item[\ref{app:responses}] Predicted responses across subjects and selection conditions \dotfill \pageref{app:responses}

    \item[\ref{app:laion_advantages}] Predicted-response advantages over LAION \dotfill \pageref{app:laion_advantages}

    \item[\ref{app:color_orientation}] Additional chromatic and orientation statistics \dotfill \pageref{app:color_orientation}

    \item[\ref{supp:structure_comparisons}] Additional structural image comparisons \dotfill \pageref{supp:structure_comparisons}

    \item[\ref{app:behavioral_experiment}] Behavioral experiment to test differences between region MEIs \dotfill \pageref{app:behavioral_experiment}

    \item[\ref{app:additional_meis}] Additional MEIs \dotfill \pageref{app:additional_meis}
\end{list}

\subsection{Encoding Model Performance}
\label{app:encoding_performance}

We report voxelwise encoding performance on 500 held-out test images for each subject and training partition. The reported $R^2$ values are not normalized by noise ceiling. Each figure contains a $4 \times 4$ grid: rows correspond to subjects S1, S2, S5, and S7, and columns to ADV-RN50, DINO-RN50, OpenCLIP-RN50, and OpenCLIP-ConvNeXt-Base. Points represent voxels within the retinotopically defined area with positive noise ceiling and finite evaluation values; these are not restricted to the voxels selected for MEI generation. Red dashed lines show ordinary least-squares fits, and black dashed lines indicate $R^2$ equal to noise ceiling. Panel annotations report the mean, median, and number of voxels. Finite observations outside the displayed axis limits remain included in the summary statistics and regression fits.

\newcommand{\EncodingEvalFigure}[3]{%
    \begin{figure}[!htbp]
        \centering
        \includegraphics[ width=0.8\linewidth, height=0.78\textheight, keepaspectratio ]{appendix_figures/#1}
        \caption{Held-out encoding performance in #2 for training partition #3. Voxelwise $R^2$ is plotted against noise ceiling.}
        \label{fig:encoding-#2-partition-#3}
    \end{figure}
}

\EncodingEvalFigure{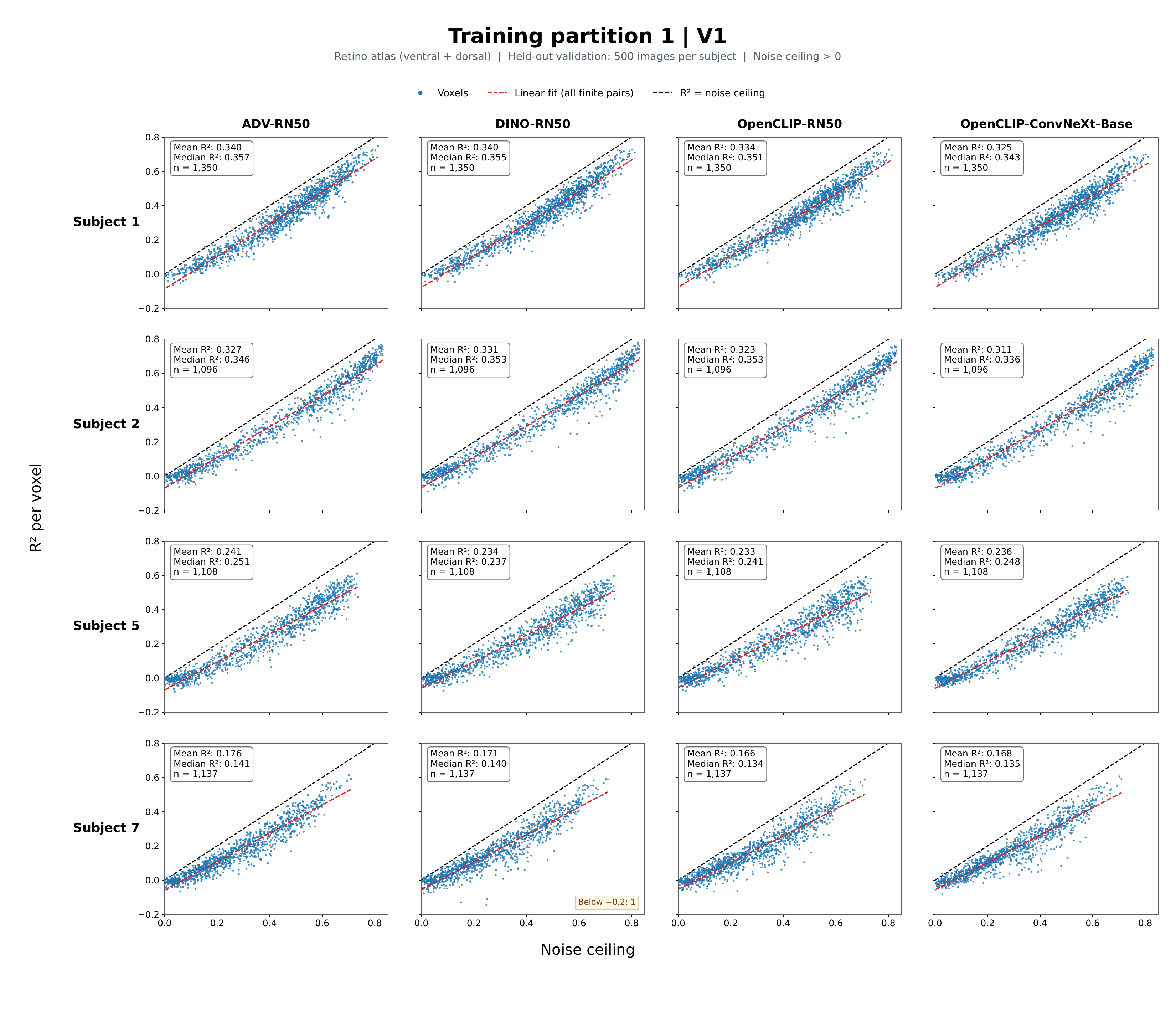}{V1}{1}
\EncodingEvalFigure{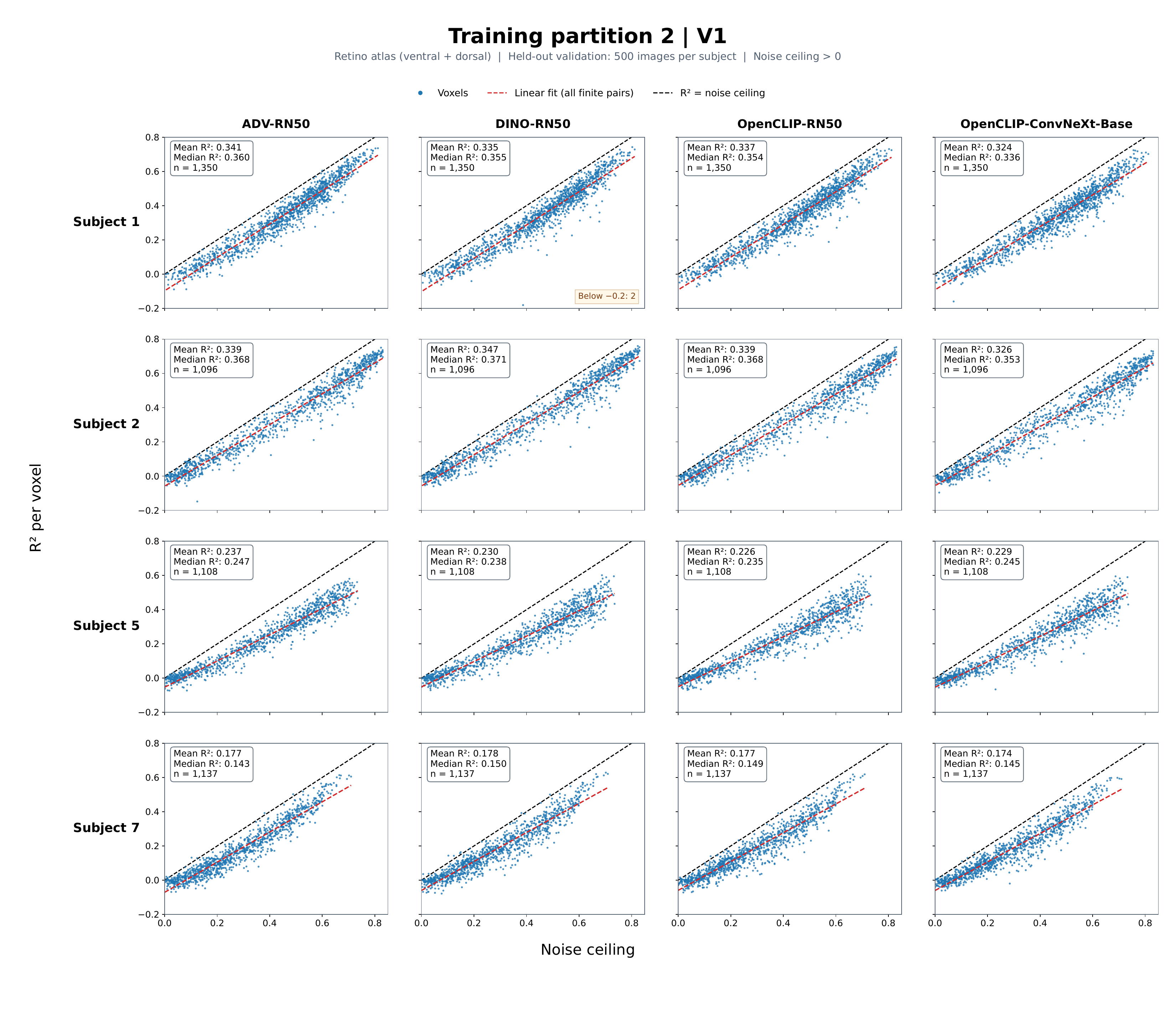}{V1}{2}

\EncodingEvalFigure{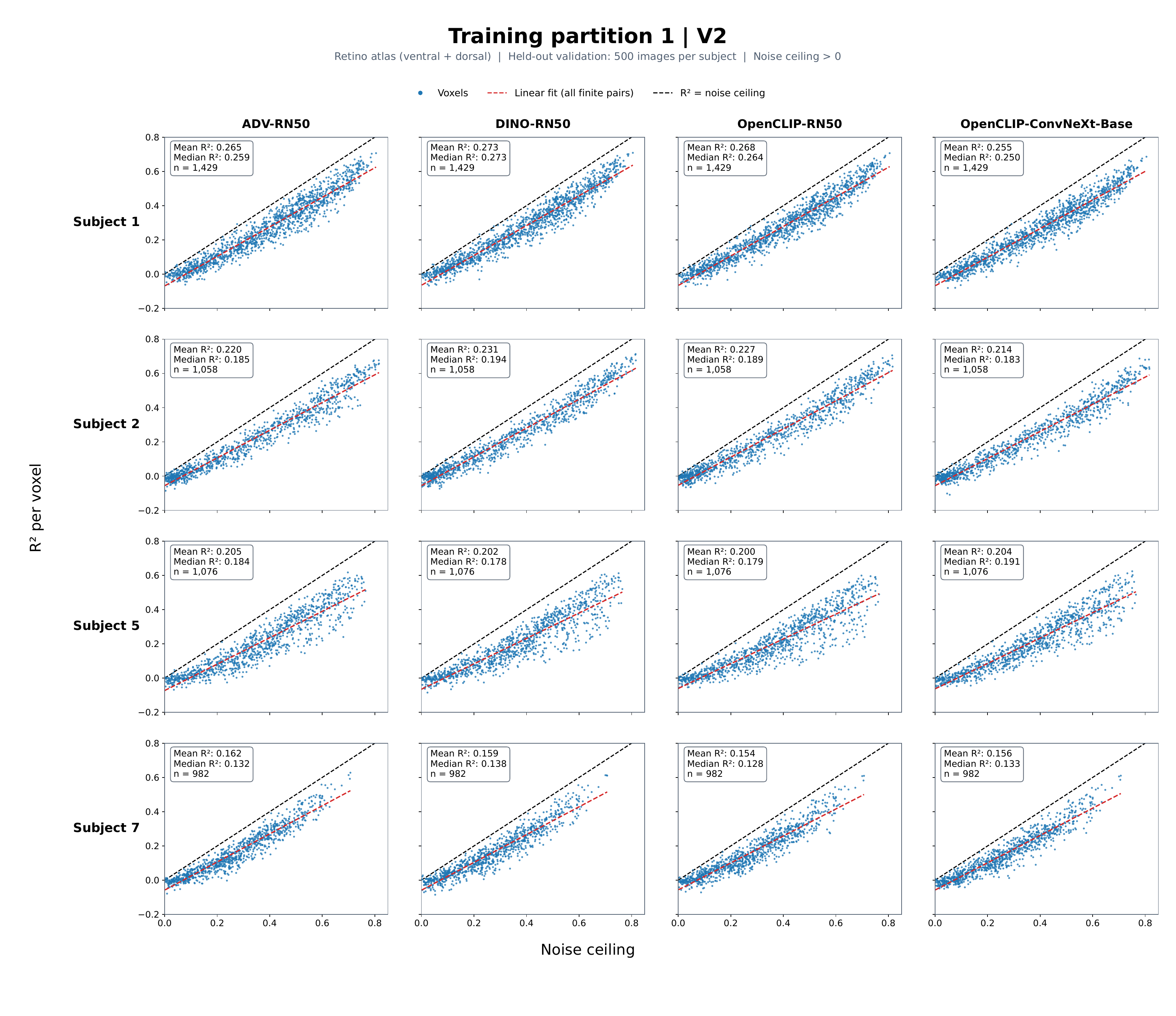}{V2}{1}
\EncodingEvalFigure{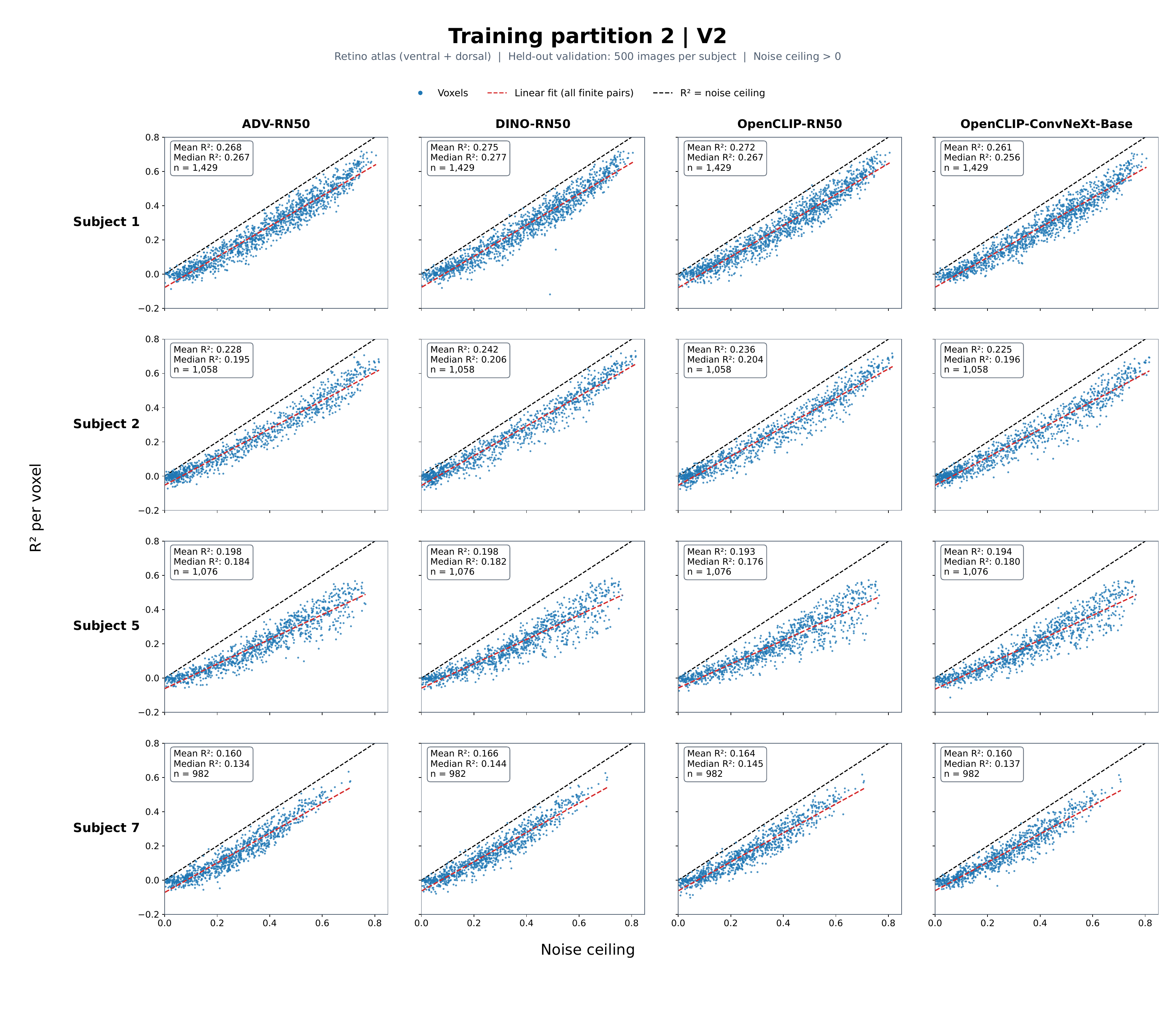}{V2}{2}

\EncodingEvalFigure{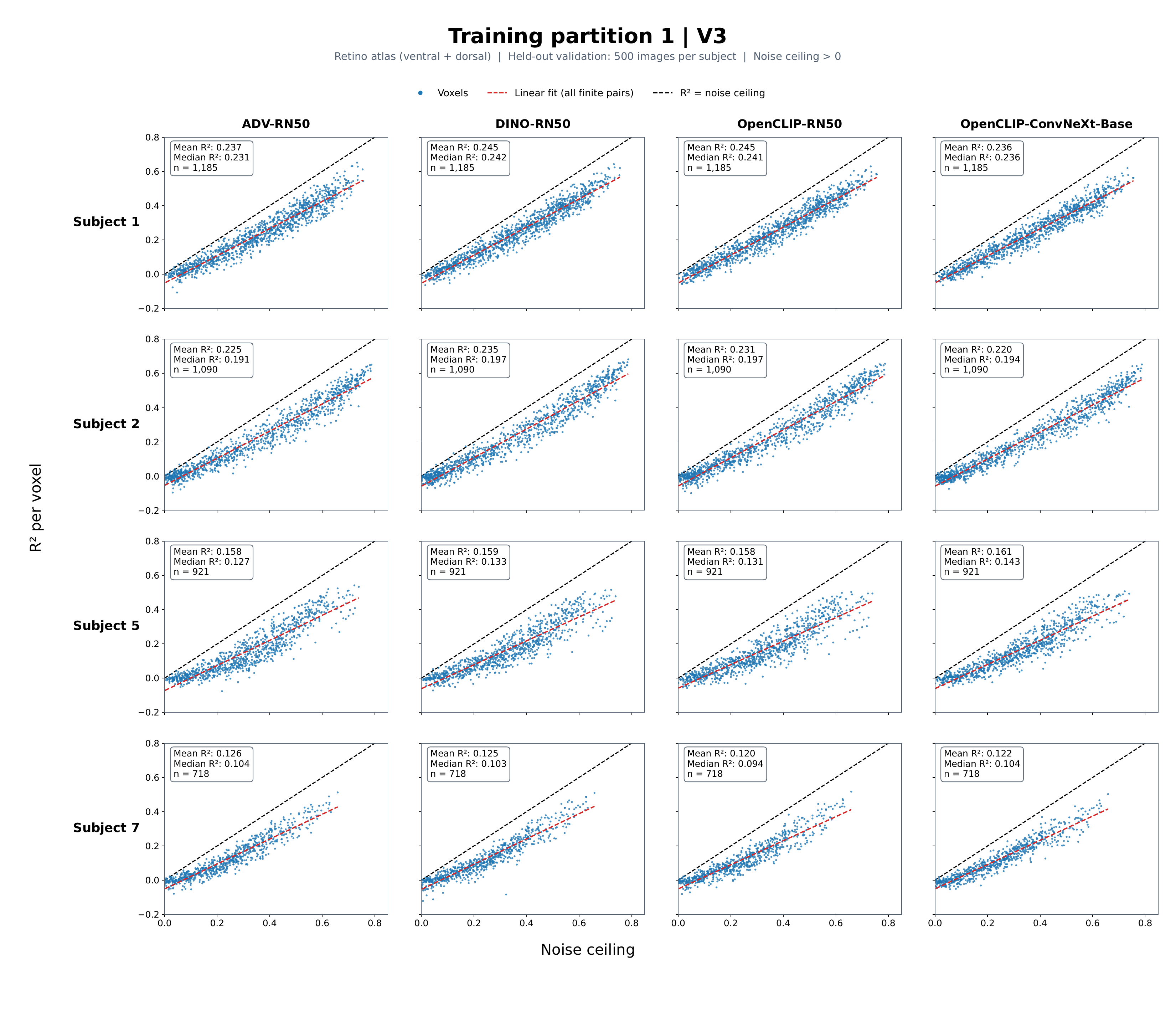}{V3}{1}
\EncodingEvalFigure{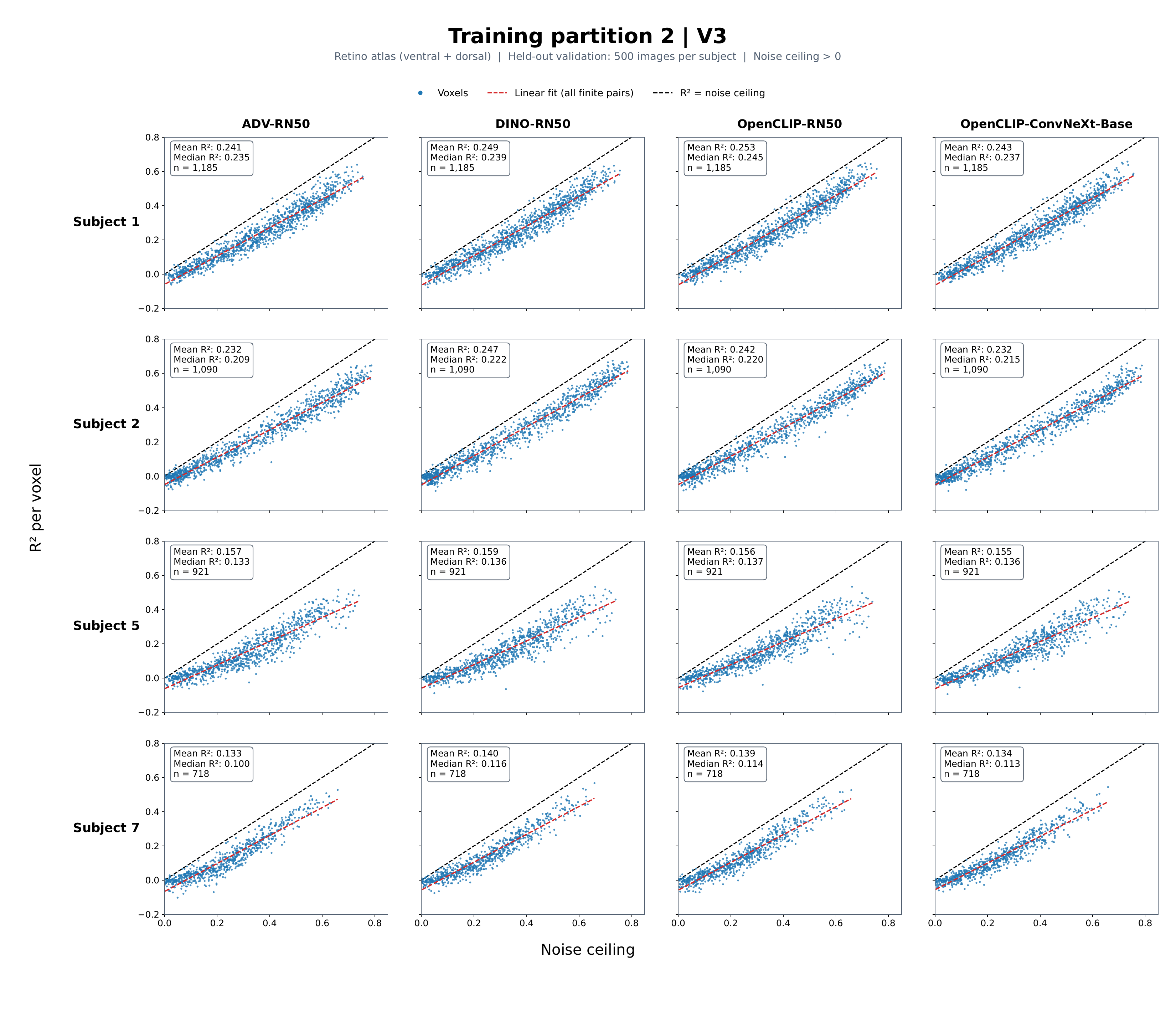}{V3}{2}

\EncodingEvalFigure{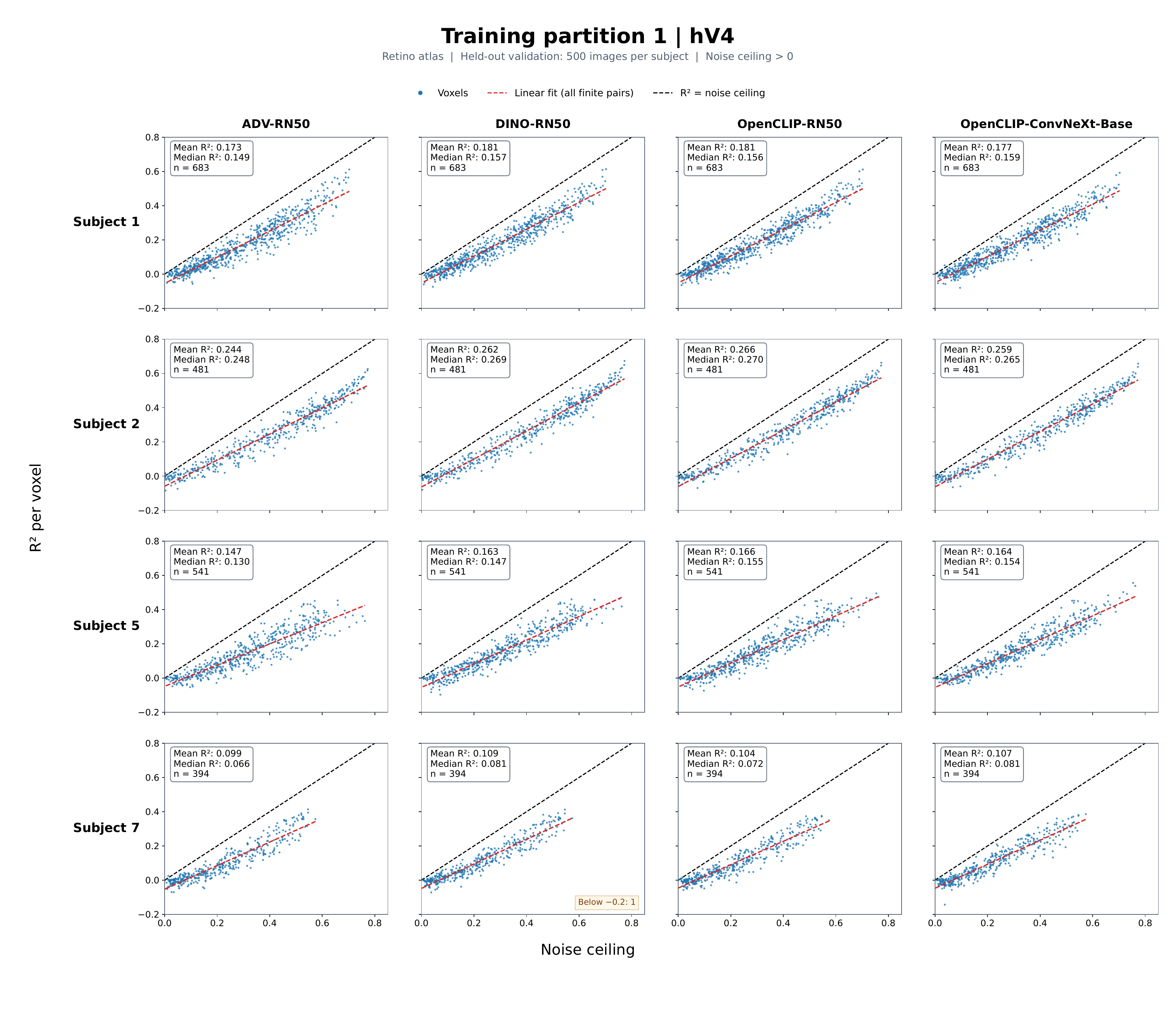}{hV4}{1}
\EncodingEvalFigure{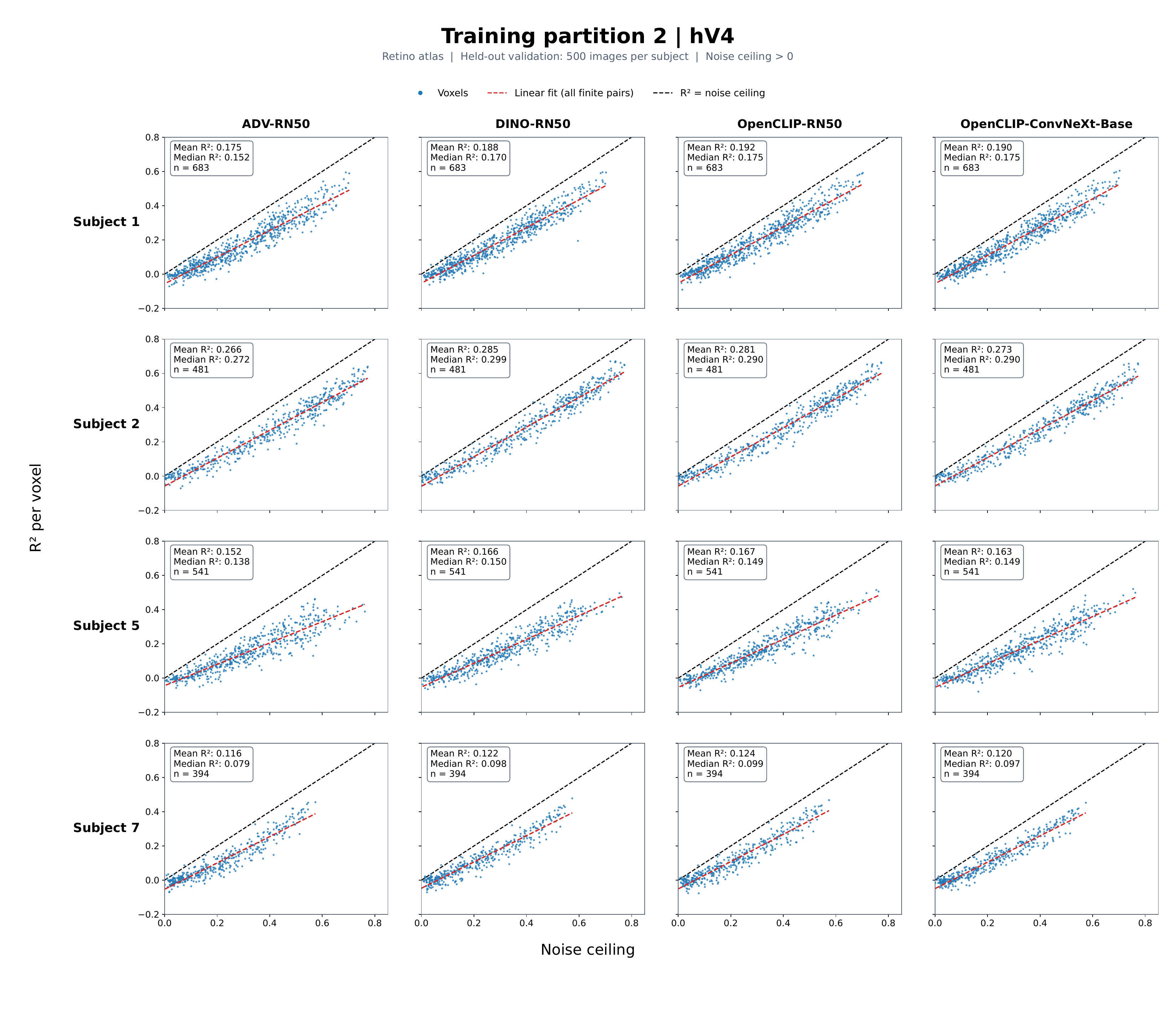}{hV4}{2}

\clearpage
\subsection{Luma mean and RMS-contrast constraints}
\label{supp:luma_constraint}

\paragraph{Luma definition and constraint statistics.}
For an RGB image \(I\in[0,1]^{3\times H\times W}\), let \(\Omega\) denote its \(N=HW\) spatial locations. We computed luma directly from gamma-coded RGB values using Rec.~709 coefficients:
\begin{equation}
Y_p=\mathbf a^{\mathsf T}I_p, \qquad \mathbf a=(0.2126,\,0.7152,\,0.0722)^{\mathsf T}.
\label{eq:rec709_luma}
\end{equation}
This quantity is distinct from linear-light luminance. We defined the chromatic residual as \(D_p=I_p-Y_p\mathbf 1_3\), where \(\mathbf 1_3=(1,1,1)^{\mathsf T}\), so that \(\mathbf a^{\mathsf T}D_p=0\).

For a scalar field \(L\) and nonnegative spatial weights \(q\) satisfying \(\sum_p q_p=1\), we defined
\begin{equation}
\mu_q(L)=\sum_{p\in\Omega}q_pL_p, \qquad C_q(L)= \left[ \sum_{p\in\Omega}q_p \left(L_p-\mu_q(L)\right)^2 \right]^{1/2}.
\label{eq:weighted_luma_statistics}
\end{equation}
Full-image statistics used uniform weights \(g_p=1/N\). For pRF-weighted statistics, the generator model's fitted pRF was mapped to image space and normalized to obtain weights \(w_p\). These soft weights were used continuously, without conversion to a binary aperture. The target mean and RMS contrast were \(\mu^\star=116/255\) and \(C^\star=57.48/255\), respectively.

\paragraph{Projection during MACO optimization.}
At each MACO optimization step, the current raw image was projected before evaluation by the generator encoding model. Starting from \(Z^{(0)}=Y\), the scalar-luma adjustment was
\begin{equation}
Z^{(k+1)}_p= \operatorname{clip}_{[0,1]} \left[ \mu^\star+ \frac{C^\star} {\max\!\left(C_g(Z^{(k)}),\epsilon\right)} \left(Z^{(k)}_p-\mu_g(Z^{(k)})\right) \right],
\label{eq:iterative_scalar_projection}
\end{equation}
where \(\epsilon=10^{-8}\). Because scalar-luma clipping can perturb the statistics, the initial adjustment was followed by up to 50 compensation iterations. The loop terminated early when both scalar-luma statistics were within \(3/255\) of their targets for all images in the batch.

The RGB image presented to the model was
\begin{equation}
P_g(I)_p= \operatorname{clip}_{[0,1]} \left[ Z^{(\mathrm{final})}_p\mathbf 1_3+D_p \right].
\label{eq:rgb_reconstruction}
\end{equation}
Thus, the original chromatic residual was restored with a fixed coefficient of one, rather than explicitly scaled with contrast. RGB clipping can nevertheless alter both chromatic residuals and luma statistics. At optimization step \(t\), the predicted voxel response \(\widehat r_v(P_g(I(\phi_t)))\) was evaluated, and gradients were propagated through the executed projection operations to the MACO Fourier-phase parameters \(\phi_t\).

After the 100th phase update, the image was rendered again without an additional projection, quantized to uint8, and saved losslessly as a reusable parent image. Both final branches below were derived independently from this same parent. The parent therefore omitted only the final projection; its synthesis still used the full-image projection described above.

\paragraph{Final constraint branches.}
Let \(Y\) and \(D\) now denote the luma and chromatic residual recomputed from the saved parent image. The \emph{full-image} branch applied the iterative scalar-luma projection in~\eqref{eq:iterative_scalar_projection}, followed by RGB reconstruction and clipping as in~\eqref{eq:rgb_reconstruction}.

The \emph{full-image+pRF} branch instead fitted four parameters jointly. We defined the local adjustment envelope \(E_p=(w_p/\max_{r\in\Omega}w_r)^{1/2}\) and used
\begin{align}
Y^{\mathrm{local}}_p &=Y_p+(e^\beta-1)E_p \left(Y_p-\mu_w(Y)\right)+\delta E_p, \label{eq:local_luma_transform}\\
Z_p(\boldsymbol\theta) &=\operatorname{clip}_{[0,1]} \left[ \mu_g(Y^{\mathrm{local}})+\eta +e^\gamma \left(Y^{\mathrm{local}}_p-\mu_g(Y^{\mathrm{local}})\right) \right], \label{eq:global_luma_transform}\\
I'_p(\boldsymbol\theta) &=\operatorname{clip}_{[0,1]} \left[ Z_p(\boldsymbol\theta)\mathbf 1_3+D_p \right], \qquad \boldsymbol\theta=(\beta,\gamma,\delta,\eta). \label{eq:final_rgb_projection}
\end{align}
Here, \(\beta\) and \(\delta\) parameterized local contrast and mean adjustments, whereas \(\gamma\) and \(\eta\) parameterized global contrast and mean adjustments. All four parameters were fitted jointly because their effects on the target statistics are coupled. Each solver evaluation used the same parent \(Y\) and \(D\), rather than recursively transforming the previous RGB output. Both spatial weightings used the same targets \(\mu^\star\) and \(C^\star\).

\paragraph{Numerical solution and acceptance.}
For the full-image+pRF branch, fitting statistics were evaluated on the clipped scalar-luma field \(Z(\boldsymbol\theta)\), before chromatic-residual restoration and RGB clipping. We minimized \(\|\mathbf r(\boldsymbol\theta)\|_2^2\), where
\begin{equation}
\mathbf r(\boldsymbol\theta)=
\begin{bmatrix}
\left(\mu_g(Z(\boldsymbol\theta))-\mu^\star\right)/C^\star\\[4pt]
\left(\mu_w(Z(\boldsymbol\theta))-\mu^\star\right)/C^\star\\[4pt]
\log\!\left(C_g(Z(\boldsymbol\theta))/C^\star\right)\\[4pt]
\log\!\left(C_w(Z(\boldsymbol\theta))/C^\star\right)
\end{bmatrix}.
\label{eq:four_statistic_residual}
\end{equation}
The four-parameter fit used a GPU-batched Levenberg--Marquardt solver with an analytic Jacobian \(J=\partial\mathbf r/\partial\boldsymbol\theta \in\mathbb R^{4\times4}\) and double-precision arithmetic. Parameter bounds were \(\beta,\gamma\in[-8,8]\) and \(\delta,\eta\in[-1,1]\), with at most 100 solver iterations per image. The full-image branch used the scalar-luma compensation procedure rather than this four-parameter solver.

Solver convergence and scalar-luma target checks were required before final RGB acceptance checks. Because chromatic-residual restoration and RGB clipping can alter luma statistics, satisfying the scalar-luma targets alone did not guarantee acceptance.

Final RGB outputs were quantized to uint8, and luma \(\widehat Y\) was recomputed from these exact quantized RGB values normalized to \([0,1]\). A candidate was accepted only if it passed the preceding numerical checks and satisfied
\begin{equation}
\max_{q\in\mathcal Q} \max\left\{ \left|\mu_q(\widehat Y)-\mu^\star\right|, \left|C_q(\widehat Y)-C^\star\right| \right\} \leq \frac{3}{255},
\label{eq:constraint_acceptance}
\end{equation}
where \(\mathcal Q=\{g\}\) for the full-image branch and \(\mathcal Q=\{g,w\}\) for the full-image+pRF branch. Ranking and subsequent evaluation used only accepted, quantized images, and retained images were saved as lossless uint8 RGB PNGs.

\subsection{Paired-voxel comparisons with natural images}
\label{app:paired_tests}

We compare the independent test encoder's predictions for MEIs and natural images separately for each subject, ROI, generation backbone, generation method, and final-constraint condition. Both image sets are selected by the ranking encoder. For each source, predictions form an $n\times10$ array: we average the ten selected images within each voxel and pair the resulting means by global voxel ID. Defining their difference as $d_v$, we use the two-sided paired $t$ statistic
\[
 t=\frac{\bar d}{s_d/\sqrt{n}},\qquad
 \bar d=\frac{1}{n}\sum_{v=1}^{n}d_v,\qquad
 s_d^2=\frac{1}{n-1}\sum_{v=1}^{n}(d_v-\bar d)^2,
\]
with $n-1$ degrees of freedom. Most conditions contain 20 paired voxels; incomplete diffusion results yield 18 or 19 in the remaining conditions. Natural images are unmodified. Benjamini--Hochberg (BH) correction is applied jointly to the 256 comparisons in Table~\ref{tab:rf_paired_voxel_tests}. These are exploratory within-subject tests over selected voxels, not population-level tests over subjects. They assume independent voxel differences; spatial dependence is not accounted for.

\IfFileExists{tables/paired_voxel_tests.tex}
  {\input{tables/paired_voxel_tests.tex}}
  {\input{table/paired_voxel_tests.tex}}

\subsection{Effects of Final Luma Constraints}
\label{app:constraint_effects}

We compared images before final adjustment with two final-constraint conditions: Full, which controls full-image mean luma and RMS contrast, and Full+pRF, which additionally controls these statistics within the generator's soft pRF. The targets were 116 for mean luma and 57.48 for RMS contrast, with a tolerance of $\pm 3$ in 0--255 units. For generated images, Parent denotes the saved endpoint before final adjustment, not synthesis without constraints.

Selected image identities were held fixed across conditions, retaining the intersection of images that passed both final projections. Natural images underwent the corresponding adjustments. Responses were evaluated using the independent test encoder, without re-ranking after adjustment. The predicted-response advantage of MEIs over natural-image controls generally persisted under both constraint conditions.

\begin{figure}[htbp]
    \centering
    \includegraphics[width=\linewidth] {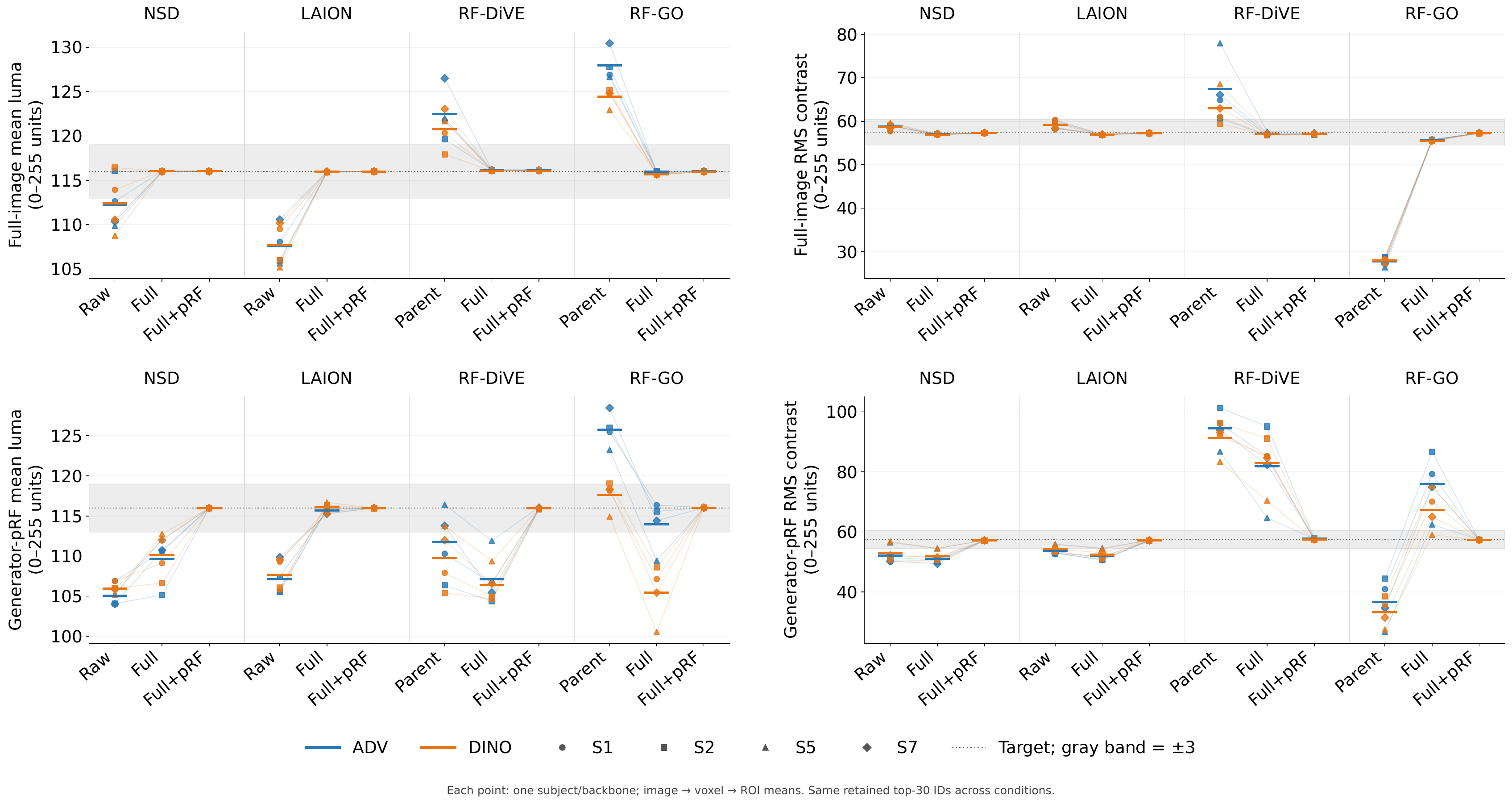}
    \caption{ Verification of full-image and pRF-weighted luma constraints. Each point summarizes one subject and generation backbone; horizontal bars indicate means across subjects. Statistics were averaged over images, voxels, and ROIs in sequence. Dotted lines and shaded bands indicate targets and tolerances. Measurements use saved-image resolution; subsequent encoder-input resizing can change RMS contrast. }
    \label{fig:constraint_verification}
\end{figure}

For the following response comparisons, each point represents one voxel's mean prediction over the retained images from its top-10 selection. Lines connect the same voxel across final-constraint conditions, and horizontal bars indicate means across voxels. Blue and orange denote ADV and DINO generation backbones, respectively. The retained image count can be below ten when a projection fails.

\begin{figure}[htbp]
    \centering
    \includegraphics[width=0.8\linewidth] {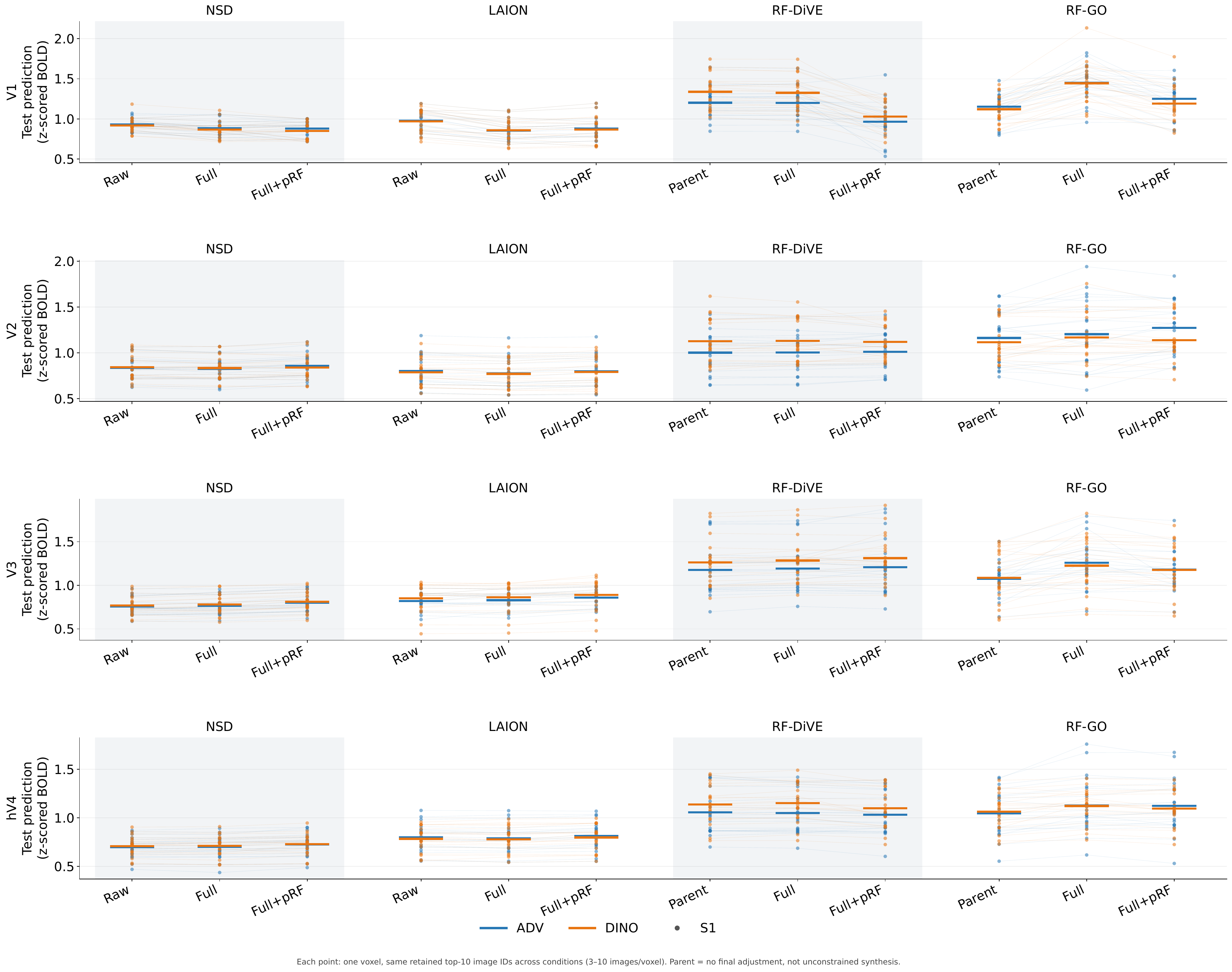}
    \caption{ Independent test-encoder predictions across final-constraint conditions for S1. Rows show V1, V2, V3, and hV4; image sources comprise NSD, LAION-fMRI, RF-DiVE, and RF-GO. }
    \label{fig:constraint_response_s1}
\end{figure}

\begin{figure}[htbp]
    \centering
    \includegraphics[width=0.8\linewidth] {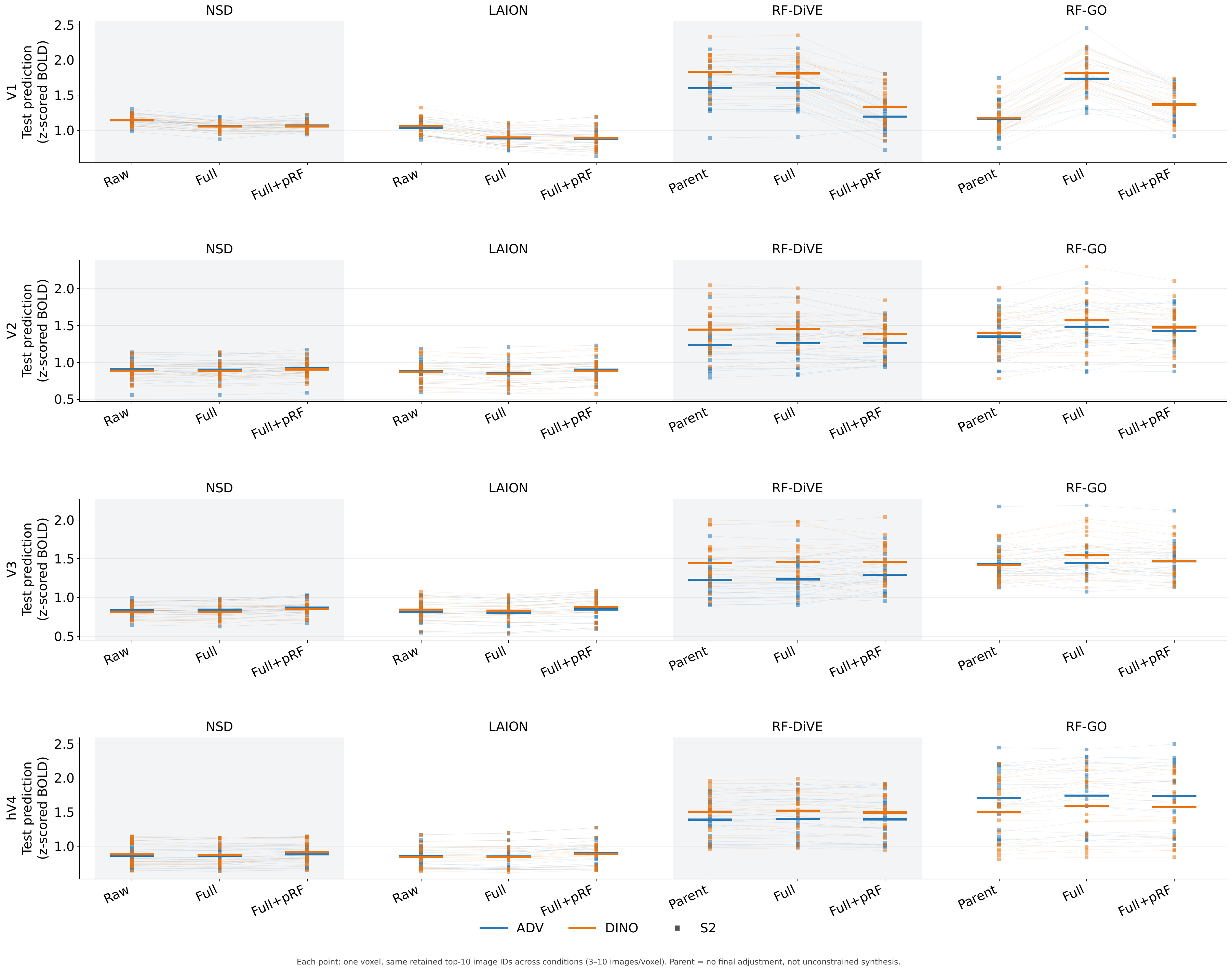}
    \caption{ Independent test-encoder predictions across final-constraint conditions for S2, using the same layout and conventions as Figure~\ref{fig:constraint_response_s1}. }
    \label{fig:constraint_response_s2}
\end{figure}

\begin{figure}[htbp]
    \centering
    \includegraphics[width=0.8\linewidth] {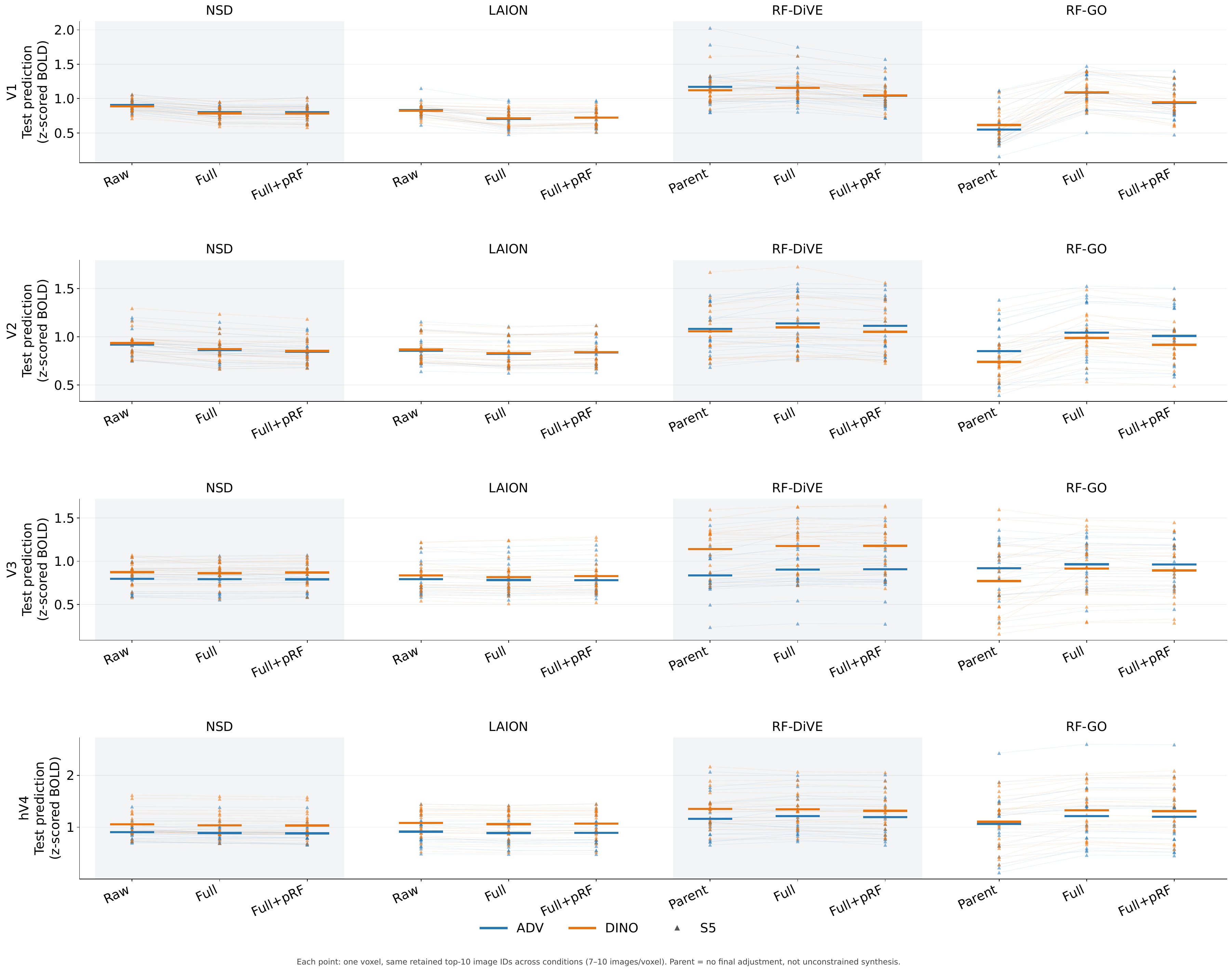}
    \caption{ Independent test-encoder predictions across final-constraint conditions for S5, using the same layout and conventions as Figure~\ref{fig:constraint_response_s1}. }
    \label{fig:constraint_response_s5}
\end{figure}

\begin{figure}[htbp]
    \centering
    \includegraphics[width=0.8\linewidth] {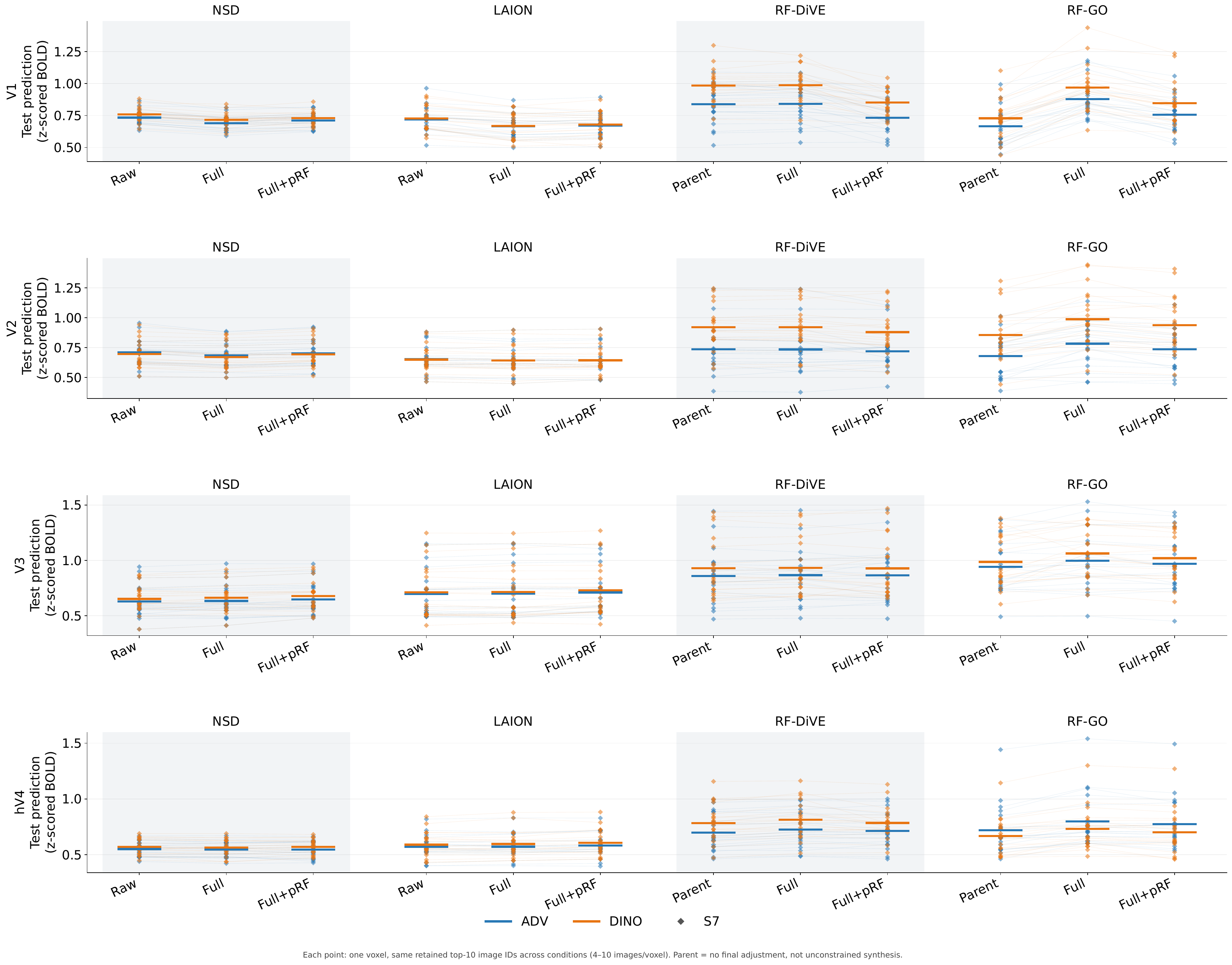}
    \caption{ Independent test-encoder predictions across final-constraint conditions for S7, using the same layout and conventions as Figure~\ref{fig:constraint_response_s1}. }
    \label{fig:constraint_response_s7}
\end{figure}

\subsection{Fitted pRF coverage}
\label{app:prf_coverage}

Each subject is shown with ADV above DINO. Within each backbone block, columns show V1, V2, V3, and hV4; the upper row shows generator pRFs, also used by the refitted ranker, and the lower row shows the test encoder's own pRFs for the same 20 voxel IDs. Dots mark centers and circles have radius $2\sigma$, where $\sigma$ is the fitted Gaussian width. Colors identify voxels within each ROI. Dashed squares mark the $8.4^\circ$-wide input field; the displayed field is $12.6^\circ$ wide (1.5 times the input width). These are fitted feature-pooling pRFs, not full CNN effective receptive fields.

\begin{figure}[!htbp]
  \centering
  \small
  \setlength{\abovecaptionskip}{3pt}
  \setlength{\belowcaptionskip}{2pt}
\includegraphics[width=0.90\linewidth,height=0.30\textheight,keepaspectratio]{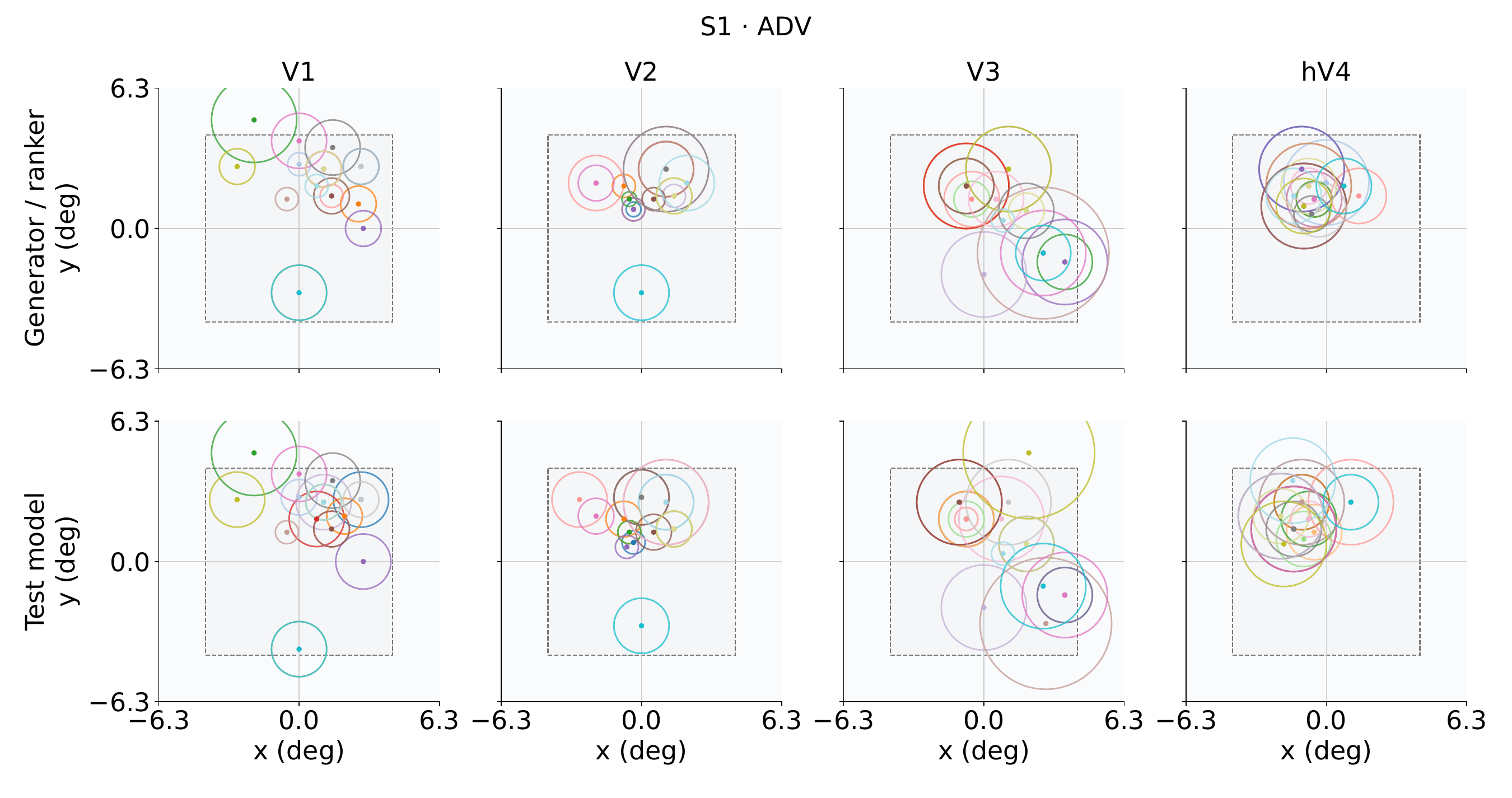}\par
  \includegraphics[width=0.90\linewidth,height=0.30\textheight,keepaspectratio]{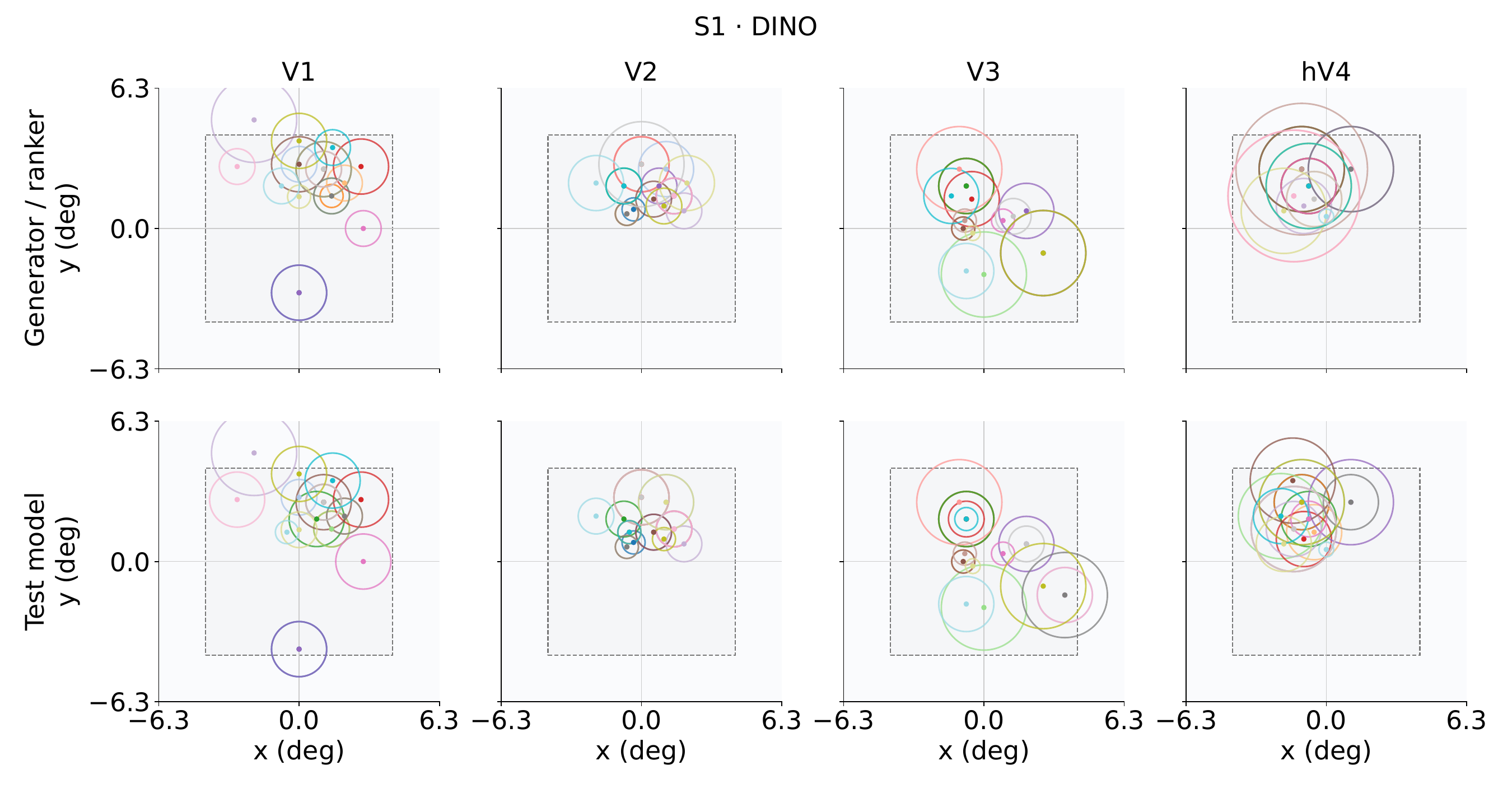}
  \caption{S1 pRF coverage: ADV above DINO.}
  \label{fig:app_prf_S1}
\end{figure}

\begin{figure}[!htbp]
  \centering
  \small
  \setlength{\abovecaptionskip}{3pt}
  \setlength{\belowcaptionskip}{2pt}
\includegraphics[width=0.90\linewidth,height=0.30\textheight,keepaspectratio]{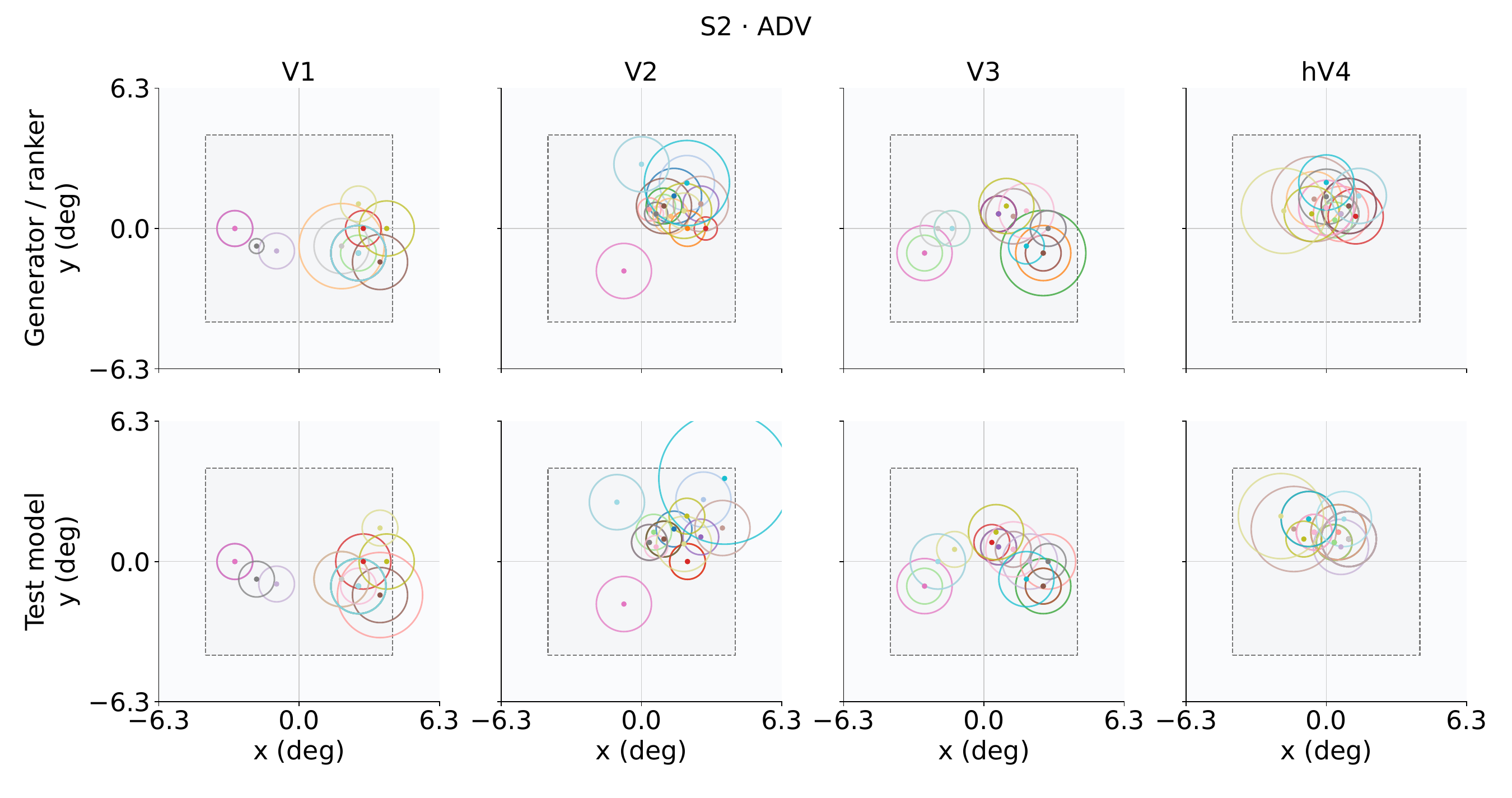}\par
  \includegraphics[width=0.90\linewidth,height=0.30\textheight,keepaspectratio]{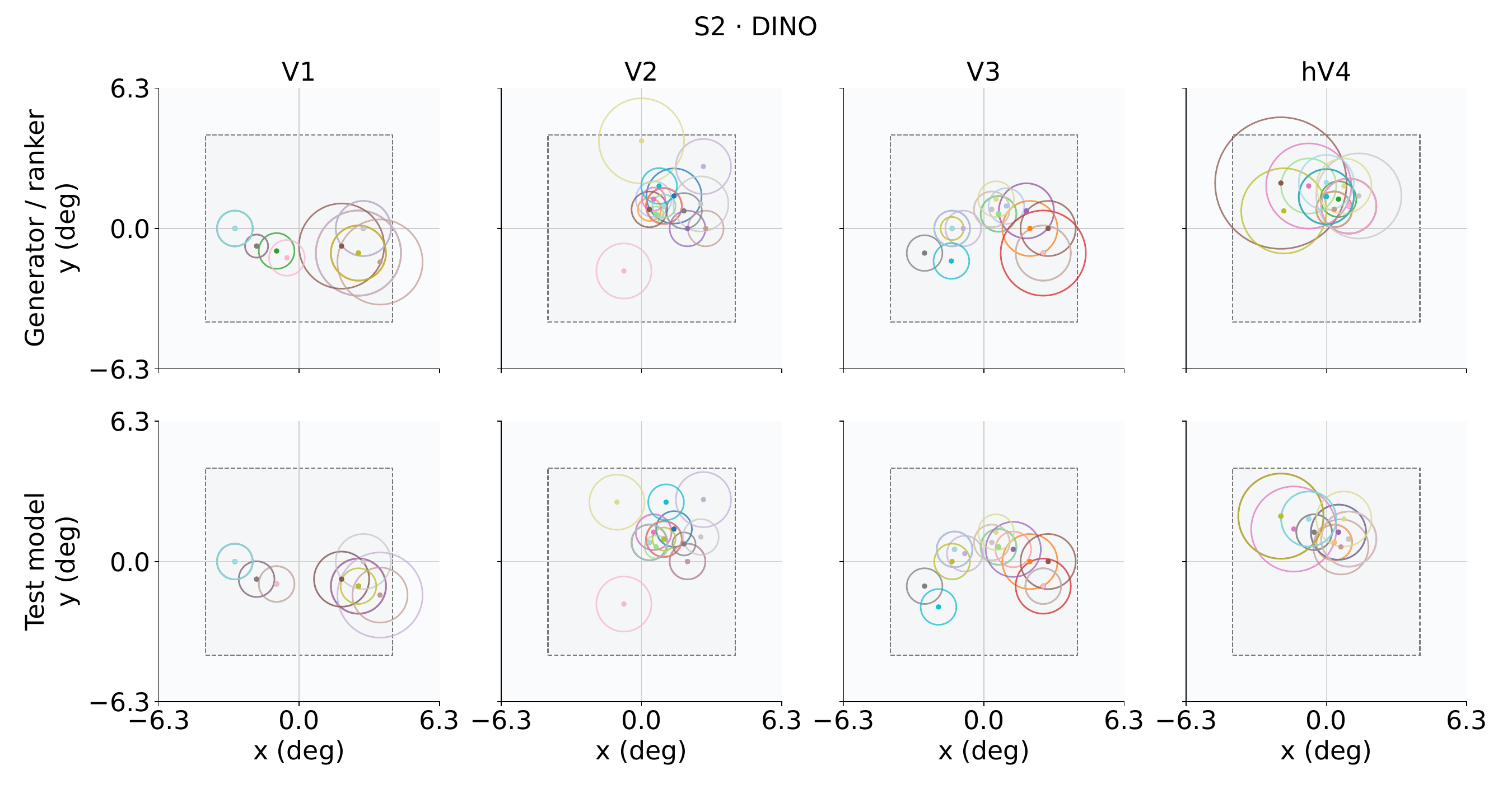}
  \caption{S2 pRF coverage: ADV above DINO.}
  \label{fig:app_prf_S2}
\end{figure}

\begin{figure}[!htbp]
  \centering
  \small
  \setlength{\abovecaptionskip}{3pt}
  \setlength{\belowcaptionskip}{2pt}
\includegraphics[width=0.90\linewidth,height=0.30\textheight,keepaspectratio]{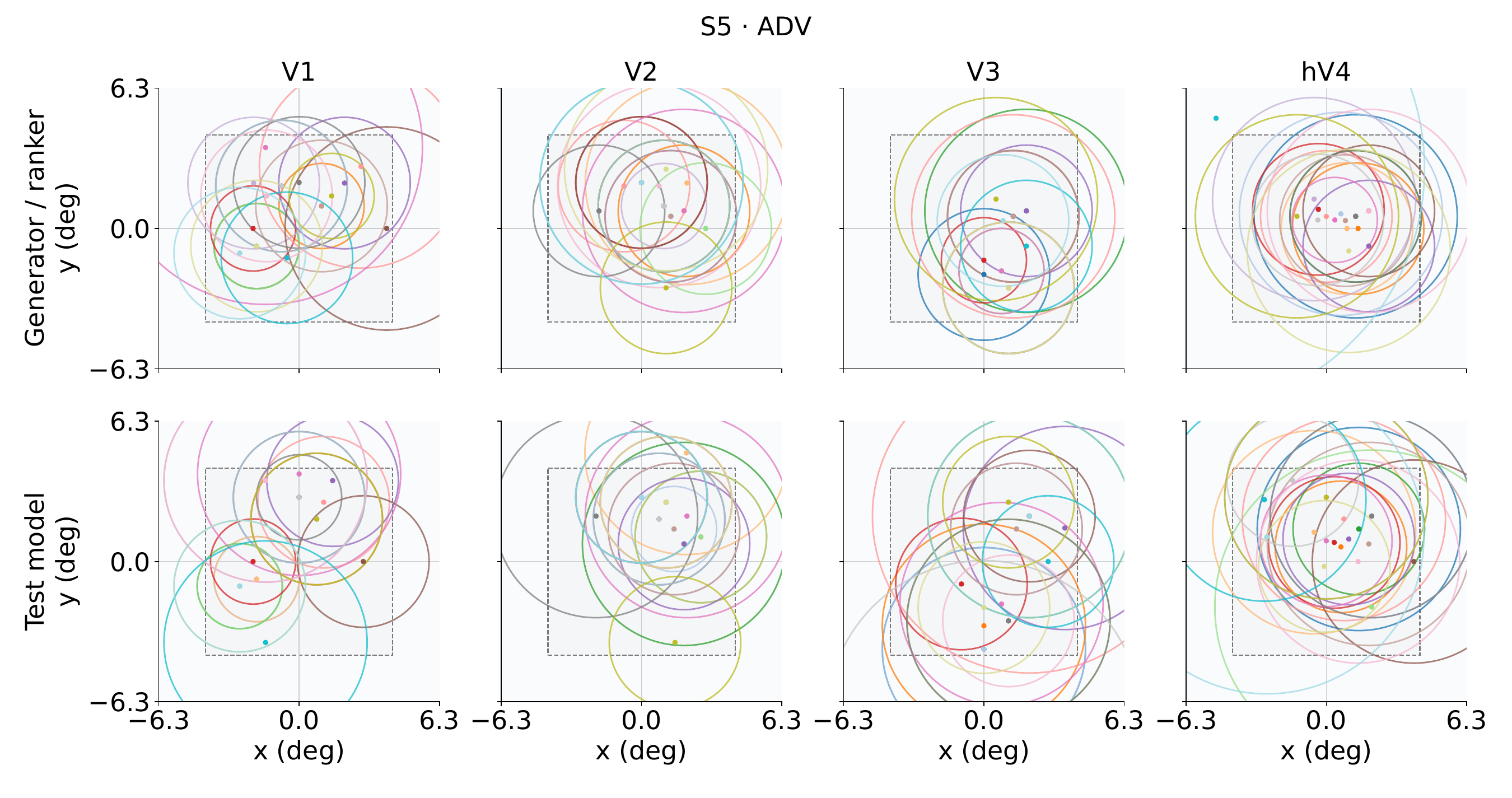}\par
  \includegraphics[width=0.90\linewidth,height=0.30\textheight,keepaspectratio]{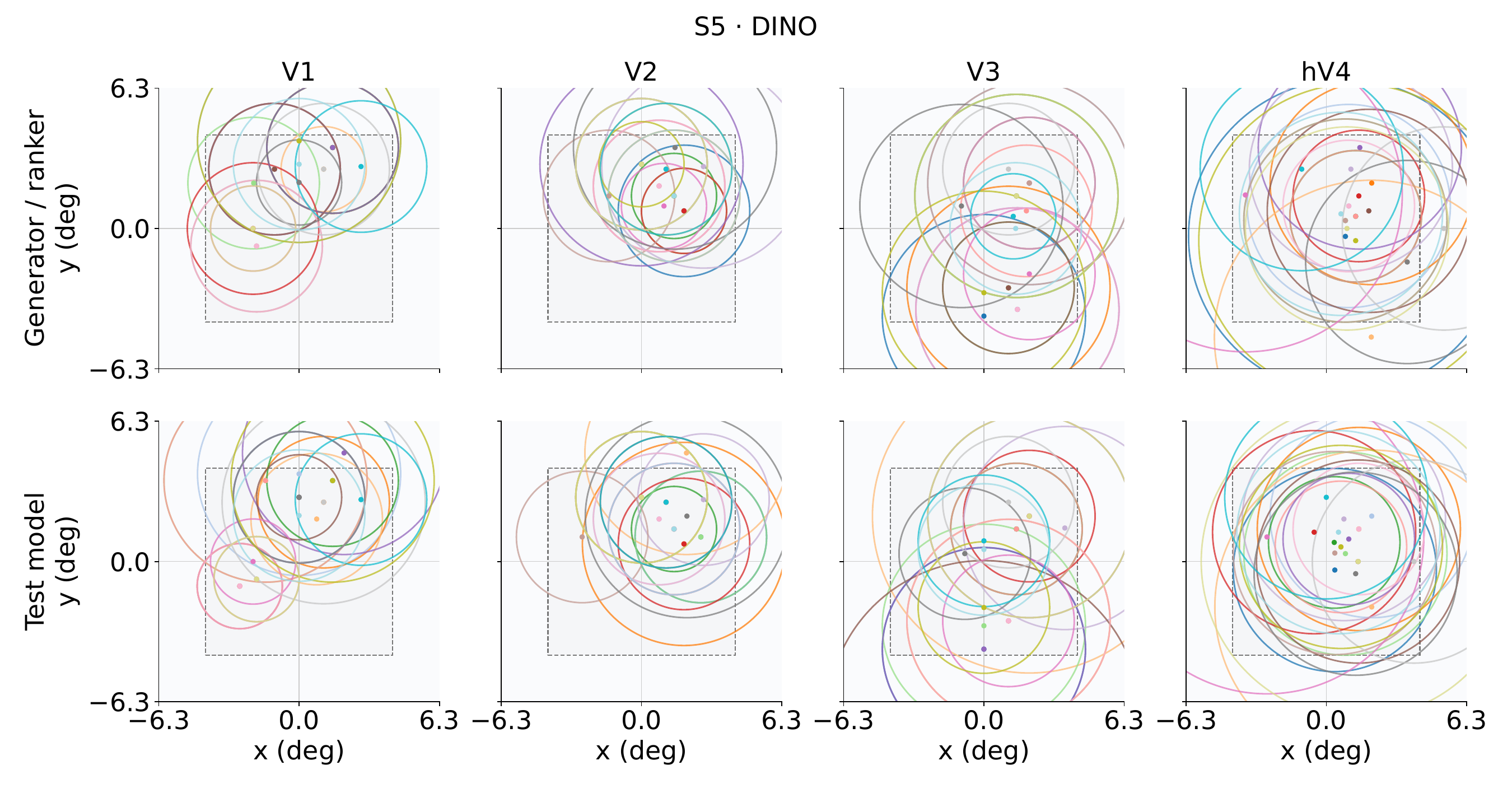}
  \caption{S5 pRF coverage: ADV above DINO.}
  \label{fig:app_prf_S5}
\end{figure}

\begin{figure}[!htbp]
  \centering
  \small
  \setlength{\abovecaptionskip}{3pt}
  \setlength{\belowcaptionskip}{2pt}
\includegraphics[width=0.90\linewidth,height=0.30\textheight,keepaspectratio]{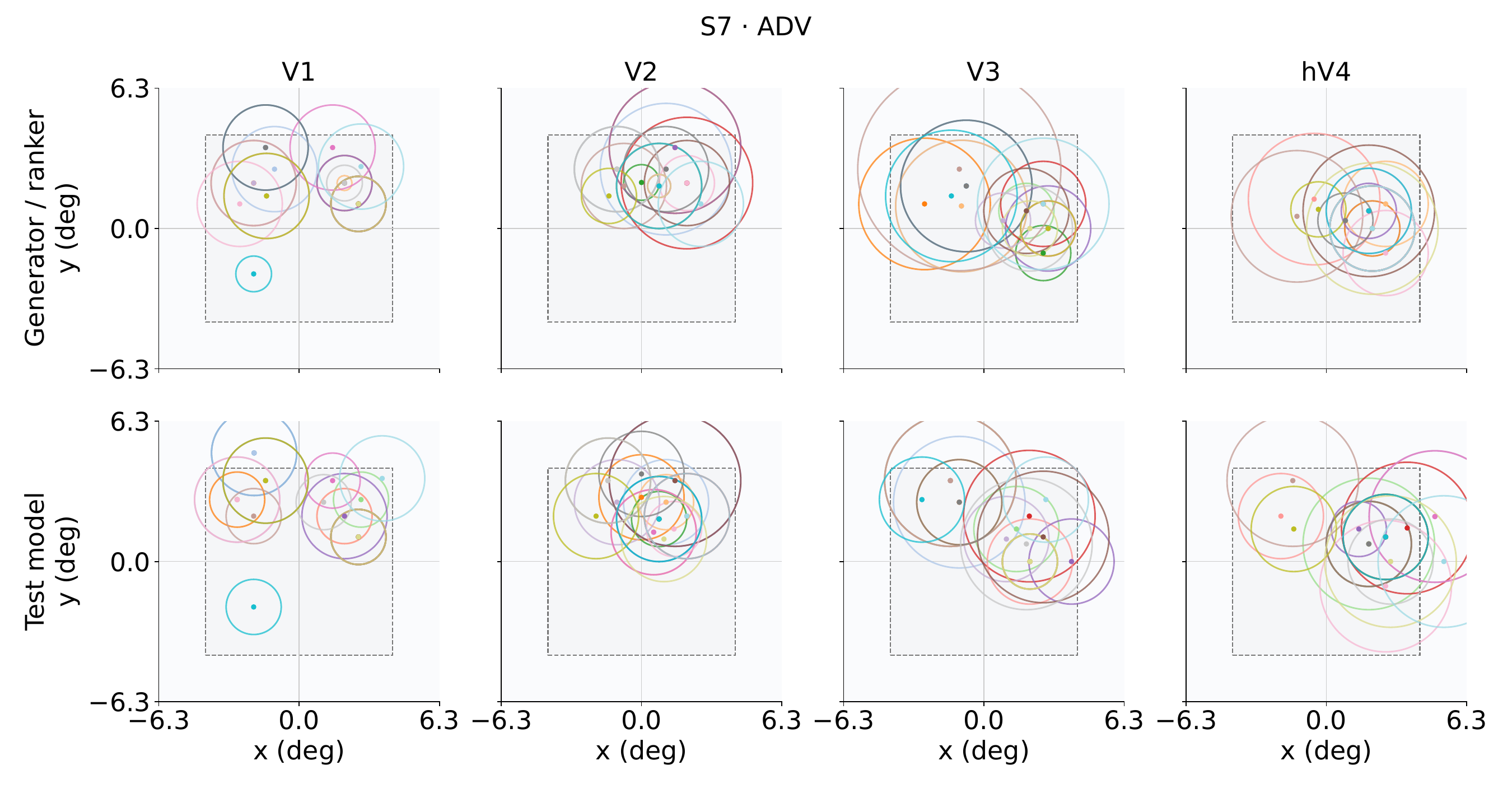}\par
  \includegraphics[width=0.90\linewidth,height=0.30\textheight,keepaspectratio]{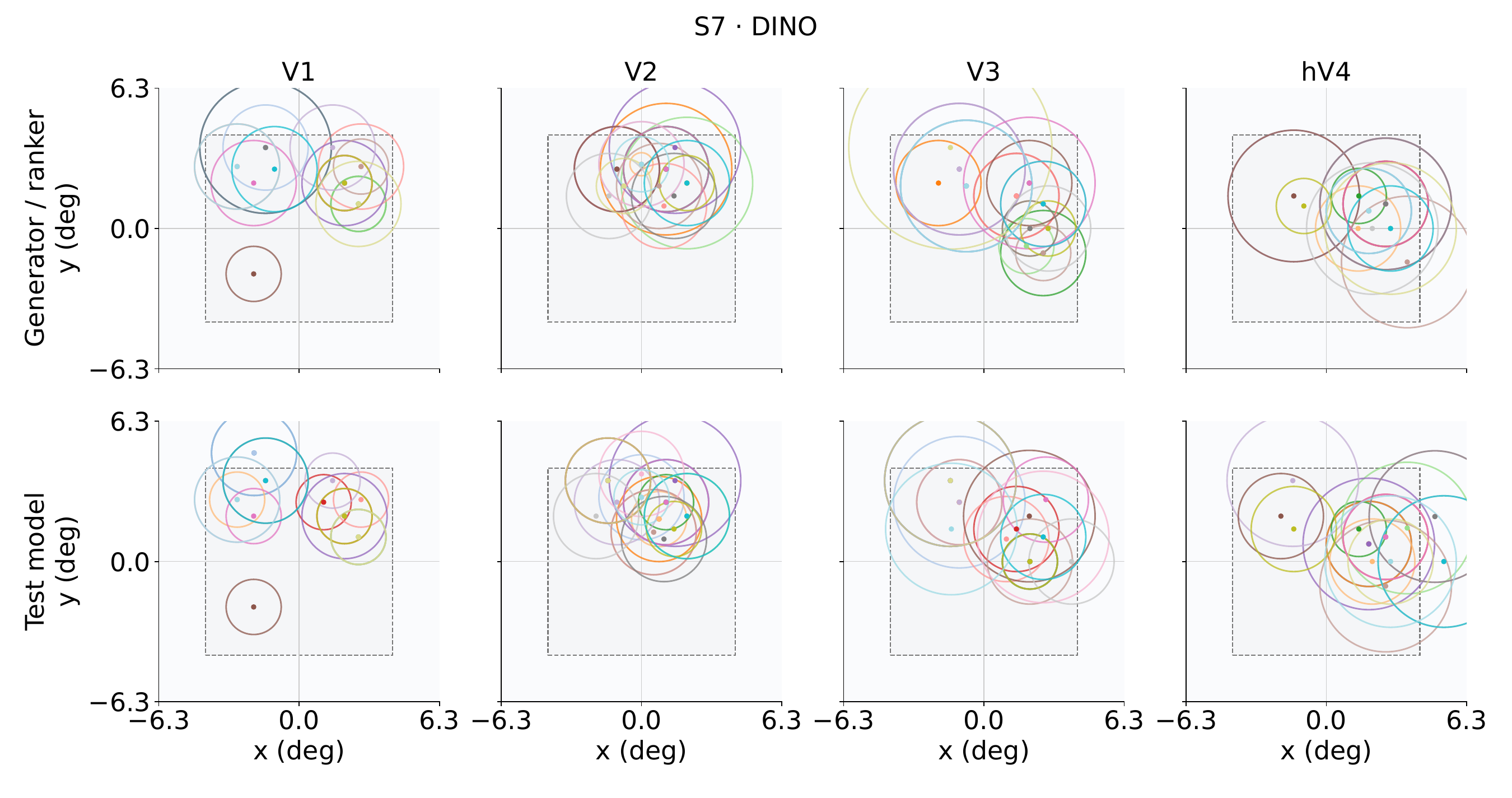}
  \caption{S7 pRF coverage: ADV above DINO.}
  \label{fig:app_prf_S7}
\end{figure}

\clearpage
\subsection{Predicted responses across subjects and selection conditions}
\label{app:responses}

Rows show generation, ranking, and independent test encoder predictions; columns show V1, V2, V3, and hV4. Each point is one voxel's mean prediction across ten selected images, and short horizontal bars indicate means across available voxels. Blue denotes ADV and orange denotes DINO, including their corresponding natural-image readouts. Images from every source are selected by either the ranking or generation encoder, as specified in each caption; the same selected images are then evaluated by all three encoders. Full-image constraints control global mean luma and RMS contrast; full-image + pRF constraints additionally control the corresponding pRF-weighted statistics. Natural controls remain unmodified in both cases. The S1 ranker-selected, full-image result is shown in the main text and is not repeated here.

\begin{figure}[!htbp]
  \centering
  \small
  \setlength{\abovecaptionskip}{3pt}
  \setlength{\belowcaptionskip}{2pt}
\includegraphics[width=0.94\linewidth,height=0.425\textheight,keepaspectratio]{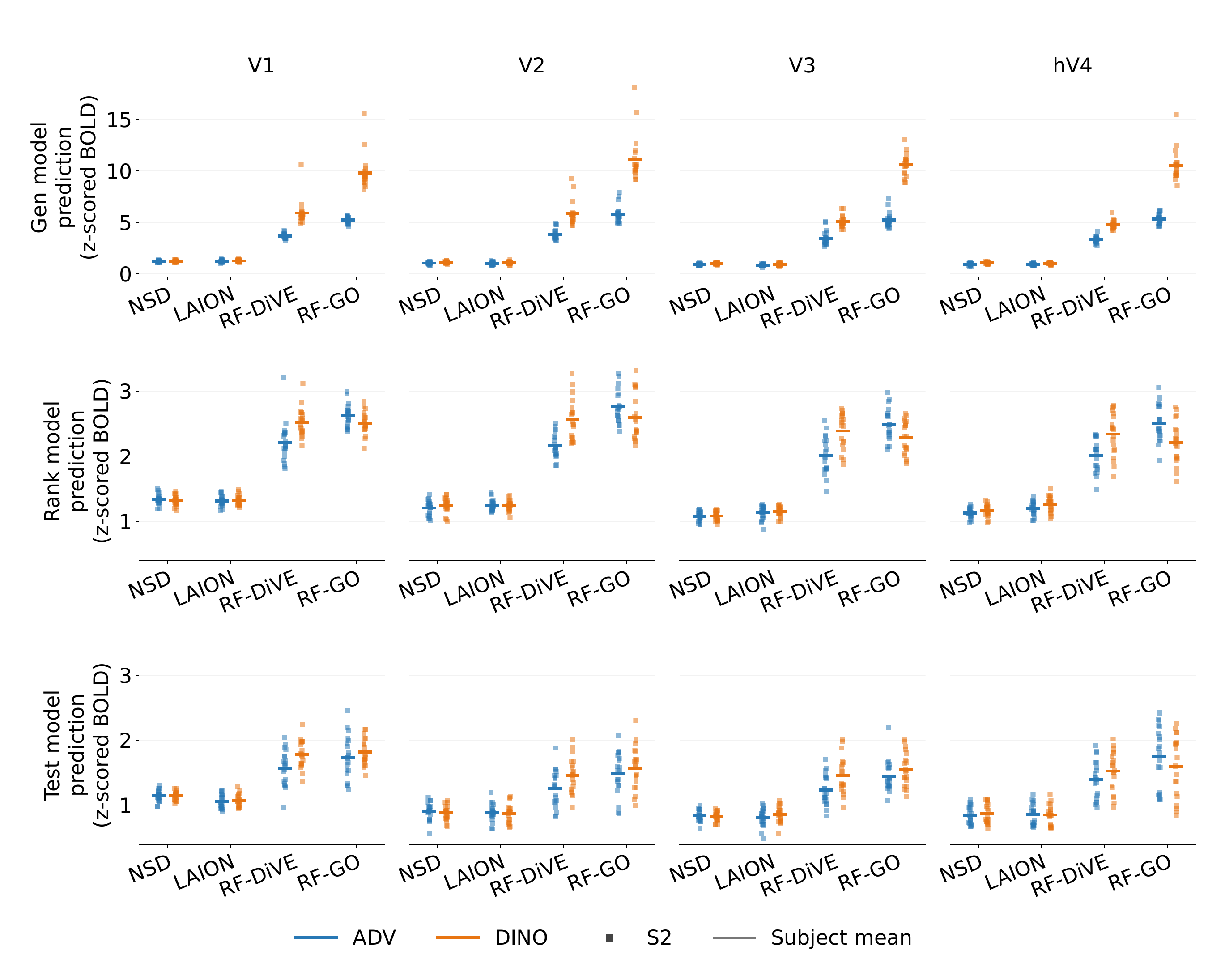}
  \caption{S2 responses: ranker-selected top-10 images; full-image constraints.}
  \label{fig:app_response_S2_rank_full}
\end{figure}

\begin{figure}[!htbp]
  \centering
  \small
  \setlength{\abovecaptionskip}{3pt}
  \setlength{\belowcaptionskip}{2pt}
\includegraphics[width=0.94\linewidth,height=0.425\textheight,keepaspectratio]{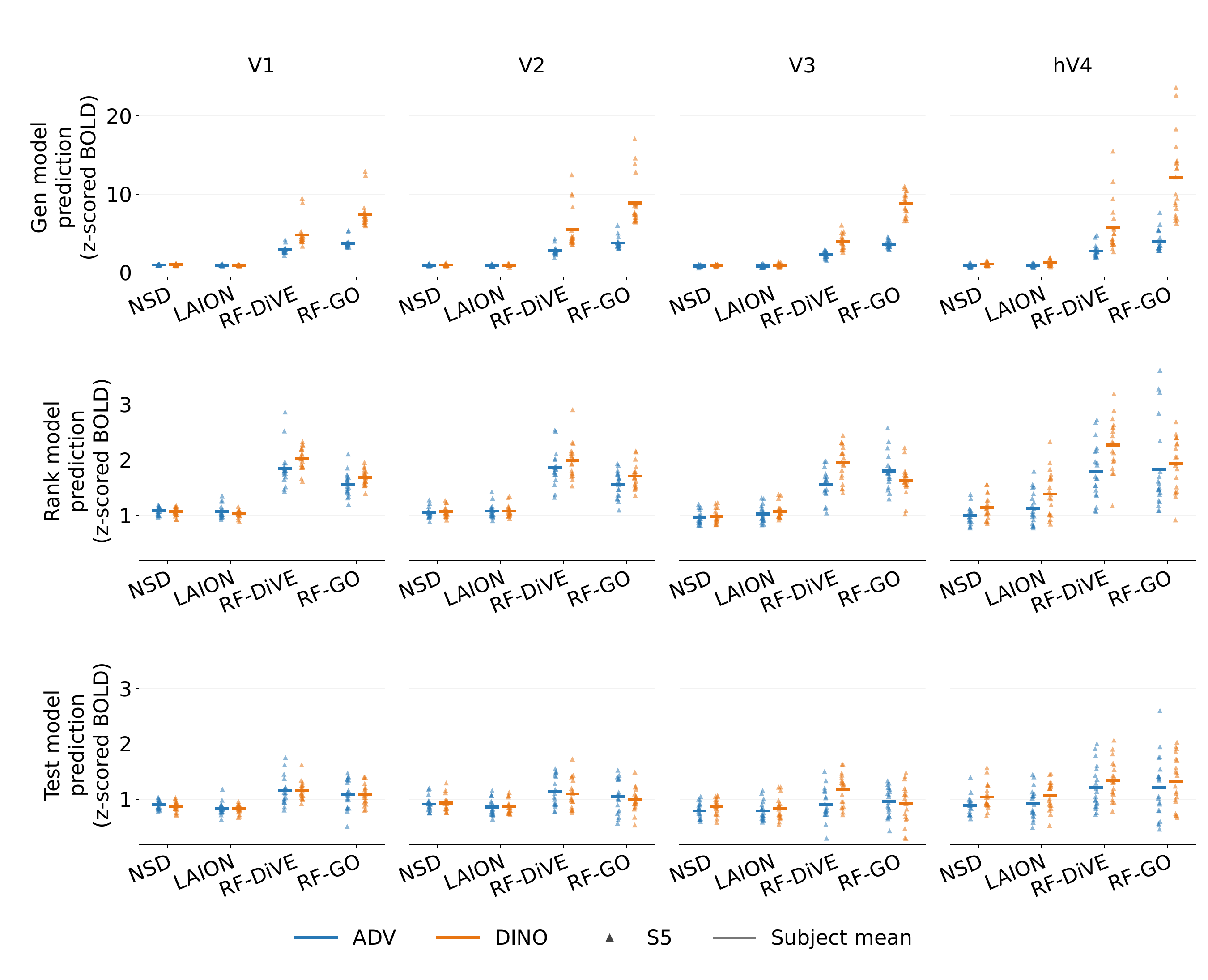}
  \caption{S5 responses: ranker-selected top-10 images; full-image constraints.}
  \label{fig:app_response_S5_rank_full}
  \vspace{6pt}
\includegraphics[width=0.94\linewidth,height=0.425\textheight,keepaspectratio]{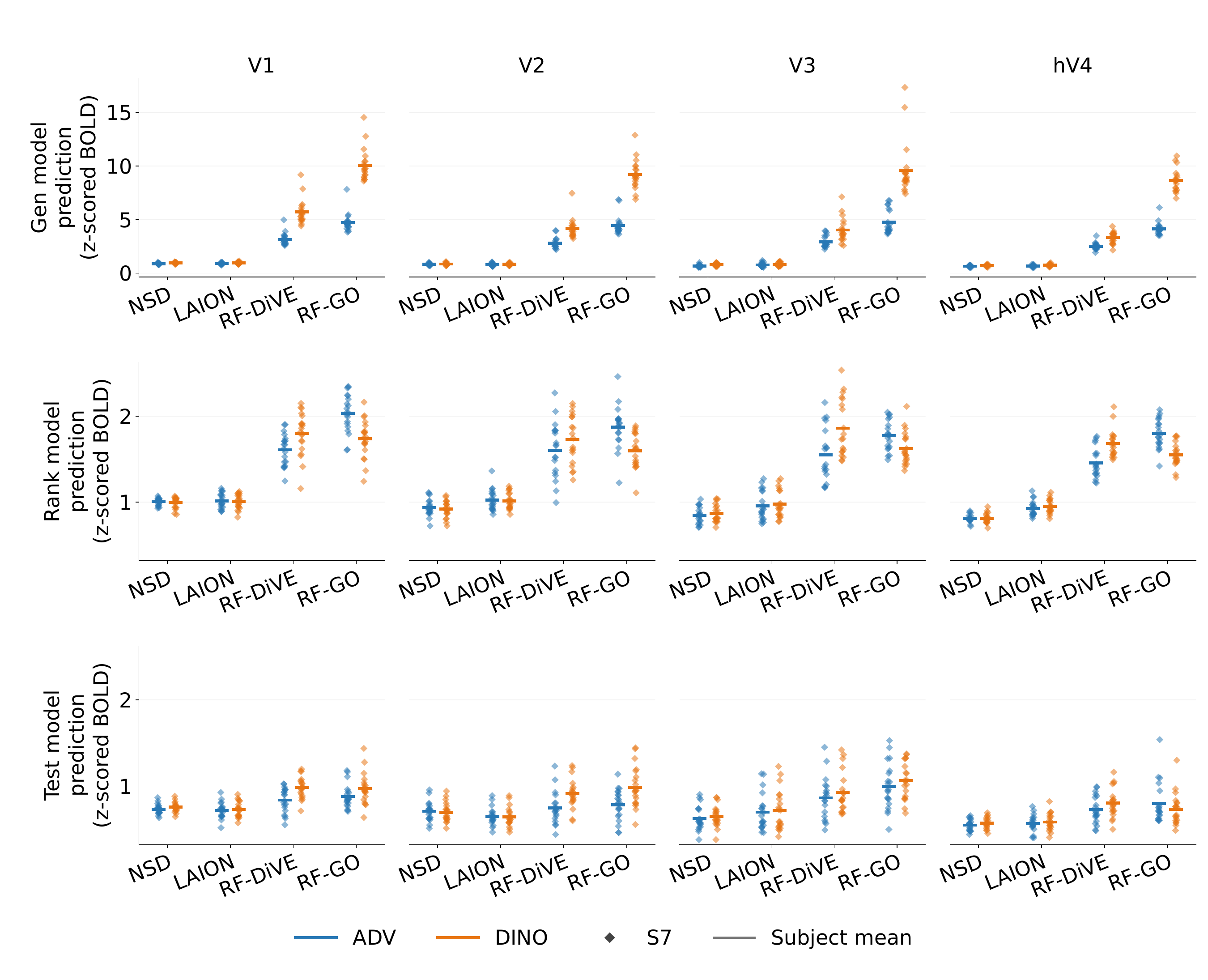}
  \caption{S7 responses: ranker-selected top-10 images; full-image constraints.}
  \label{fig:app_response_S7_rank_full}
\end{figure}

\begin{figure}[!htbp]
  \centering
  \small
  \setlength{\abovecaptionskip}{3pt}
  \setlength{\belowcaptionskip}{2pt}
\includegraphics[width=0.94\linewidth,height=0.425\textheight,keepaspectratio]{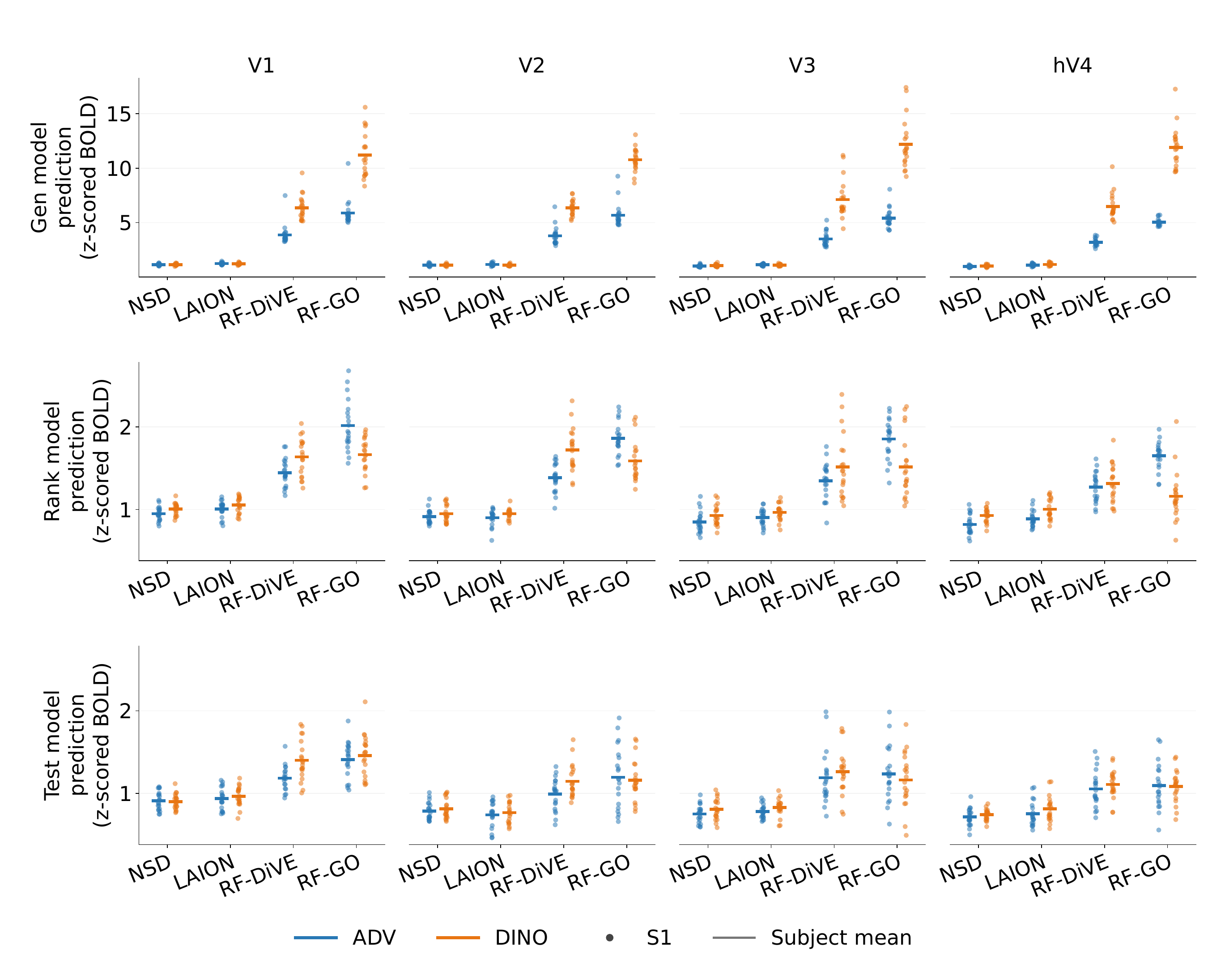}
  \caption{S1 responses: generator-selected top-10 images; full-image constraints.}
  \label{fig:app_response_S1_gen_full}
  \vspace{6pt}
\includegraphics[width=0.94\linewidth,height=0.425\textheight,keepaspectratio]{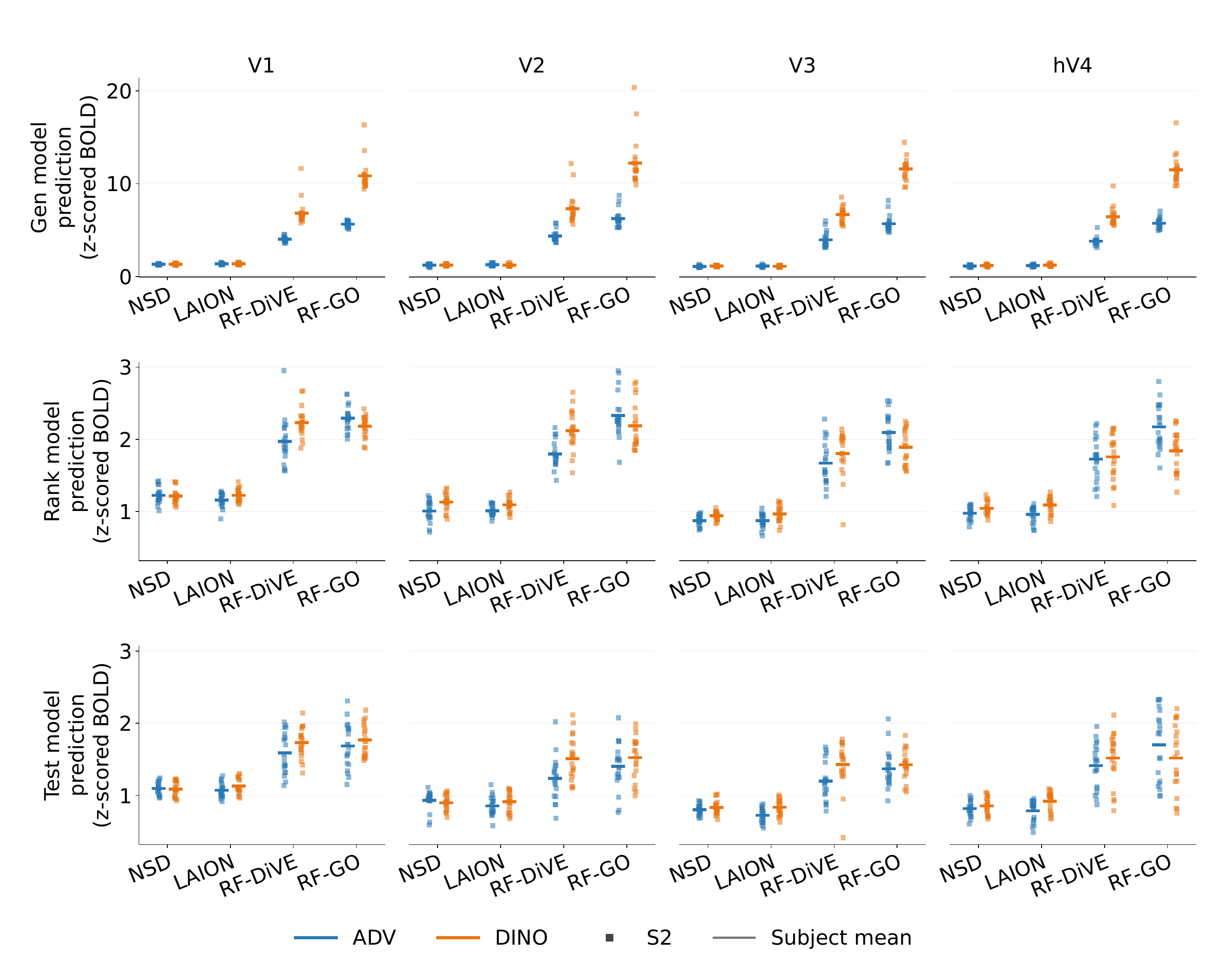}
  \caption{S2 responses: generator-selected top-10 images; full-image constraints.}
  \label{fig:app_response_S2_gen_full}
\end{figure}

\begin{figure}[!htbp]
  \centering
  \small
  \setlength{\abovecaptionskip}{3pt}
  \setlength{\belowcaptionskip}{2pt}
\includegraphics[width=0.94\linewidth,height=0.425\textheight,keepaspectratio]{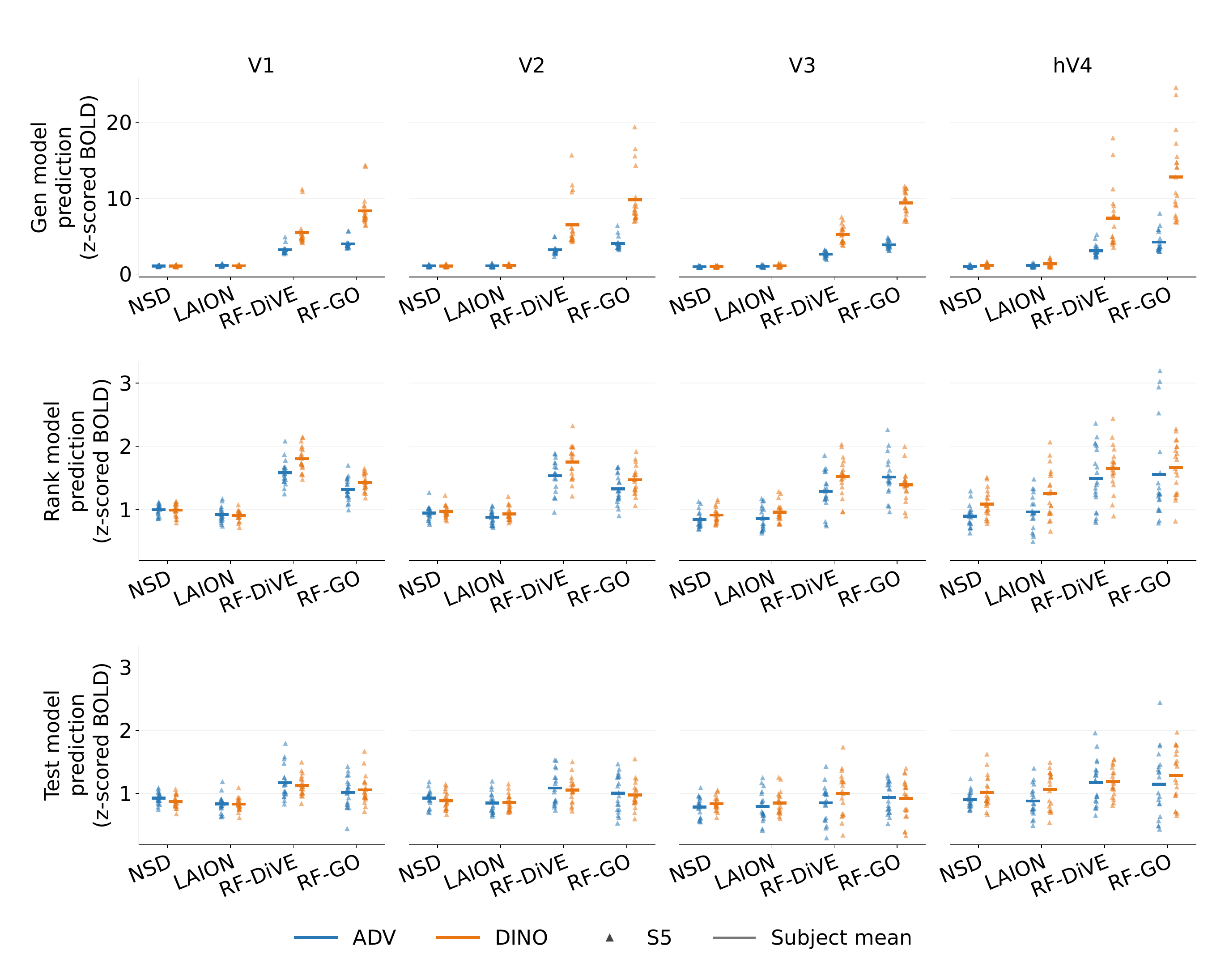}
  \caption{S5 responses: generator-selected top-10 images; full-image constraints.}
  \label{fig:app_response_S5_gen_full}
  \vspace{6pt}
\includegraphics[width=0.94\linewidth,height=0.425\textheight,keepaspectratio]{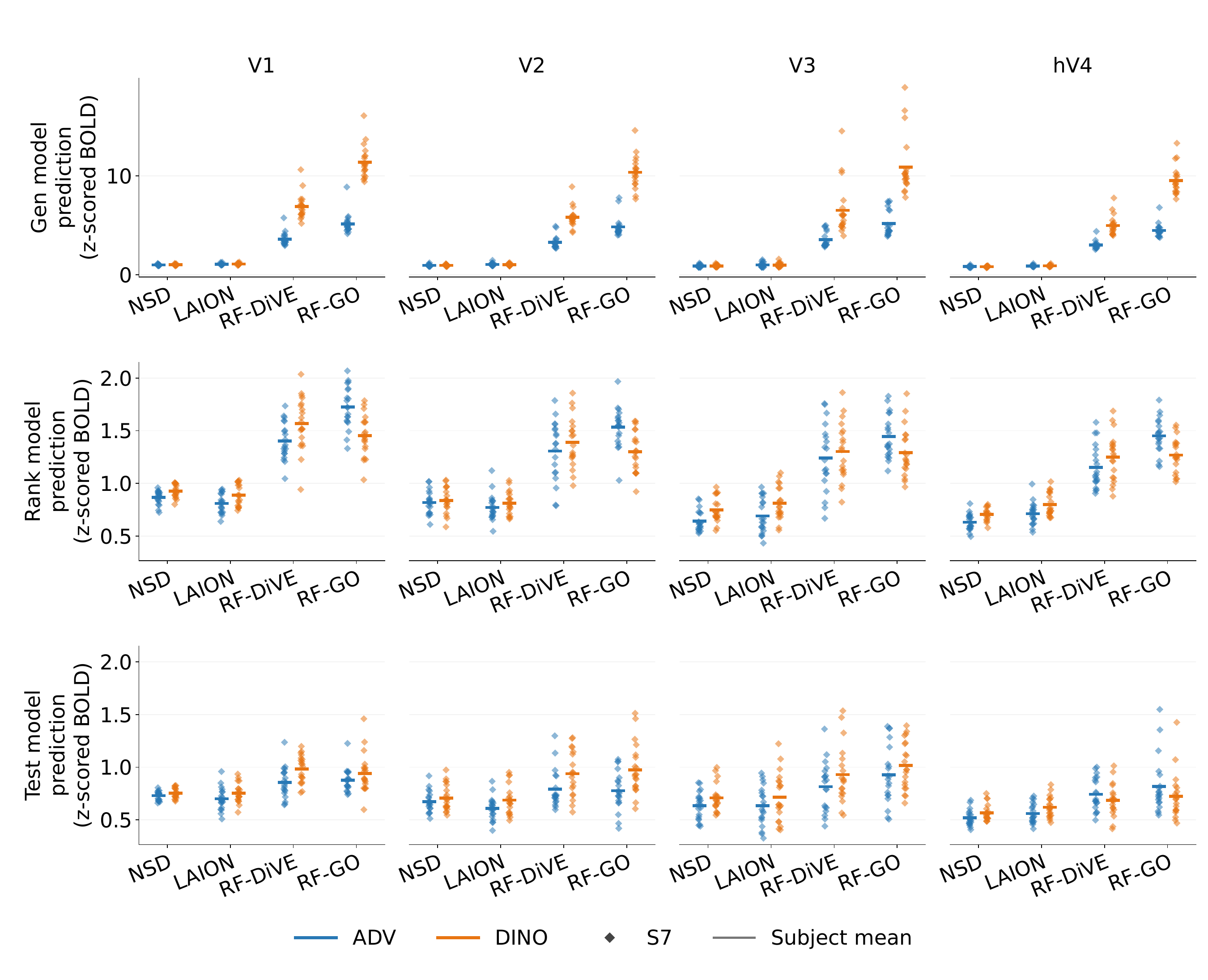}
  \caption{S7 responses: generator-selected top-10 images; full-image constraints.}
  \label{fig:app_response_S7_gen_full}
\end{figure}

\begin{figure}[!htbp]
  \centering
  \small
  \setlength{\abovecaptionskip}{3pt}
  \setlength{\belowcaptionskip}{2pt}
\includegraphics[width=0.94\linewidth,height=0.425\textheight,keepaspectratio]{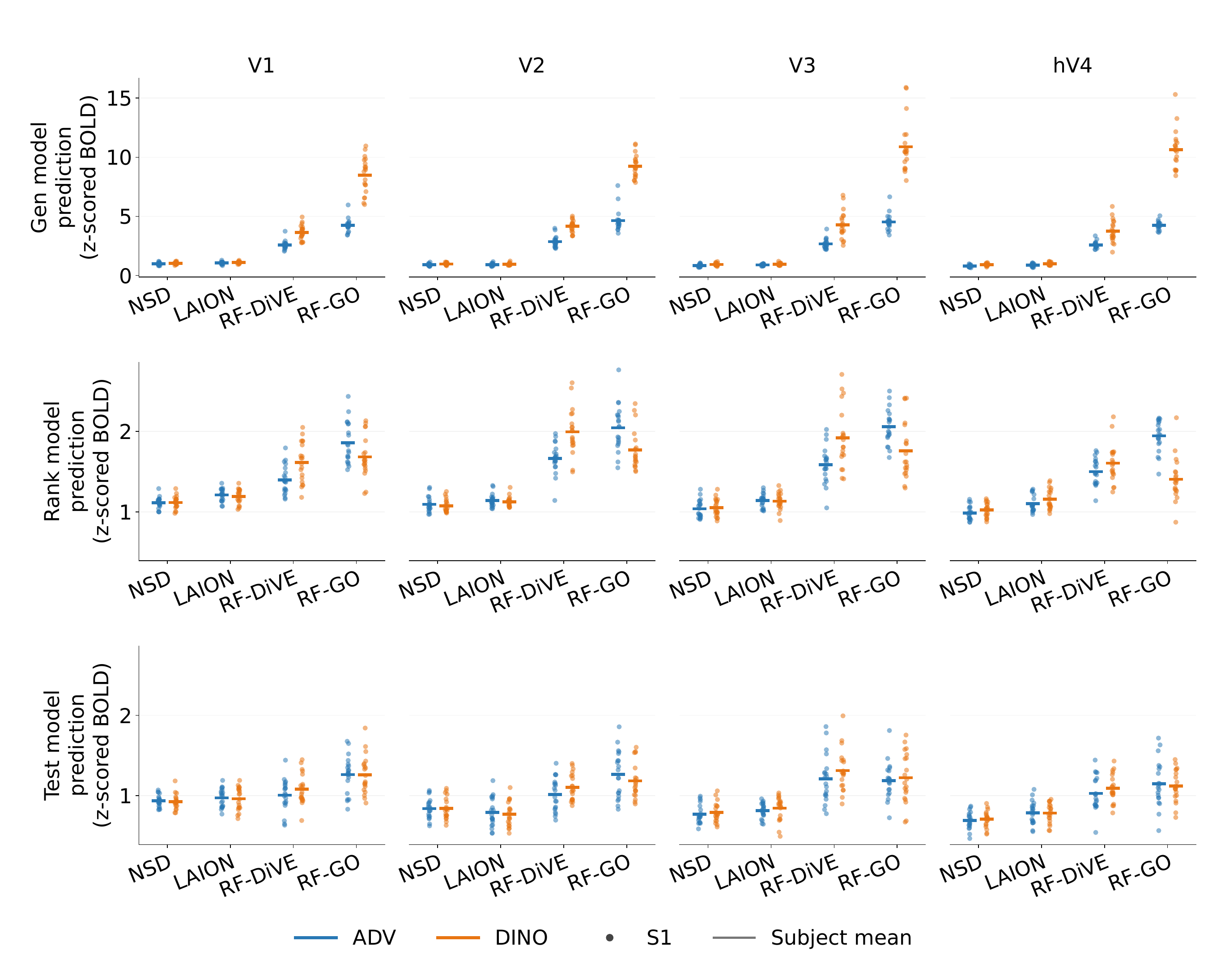}
  \caption{S1 responses: ranker-selected top-10 images; full-image + pRF constraints.}
  \label{fig:app_response_S1_rank_full_prf}
  \vspace{6pt}
\includegraphics[width=0.94\linewidth,height=0.425\textheight,keepaspectratio]{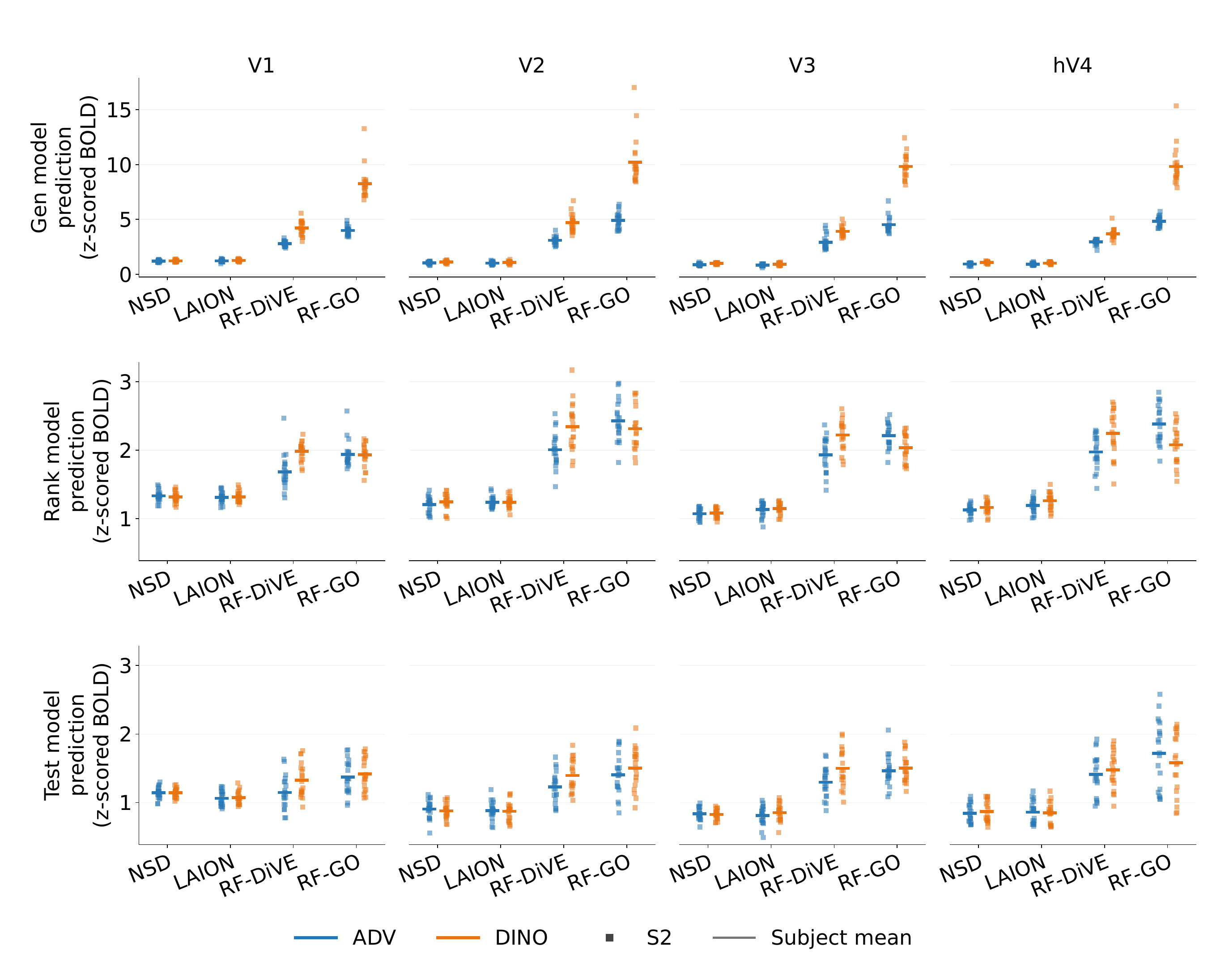}
  \caption{S2 responses: ranker-selected top-10 images; full-image + pRF constraints.}
  \label{fig:app_response_S2_rank_full_prf}
\end{figure}

\begin{figure}[!htbp]
  \centering
  \small
  \setlength{\abovecaptionskip}{3pt}
  \setlength{\belowcaptionskip}{2pt}
\includegraphics[width=0.94\linewidth,height=0.425\textheight,keepaspectratio]{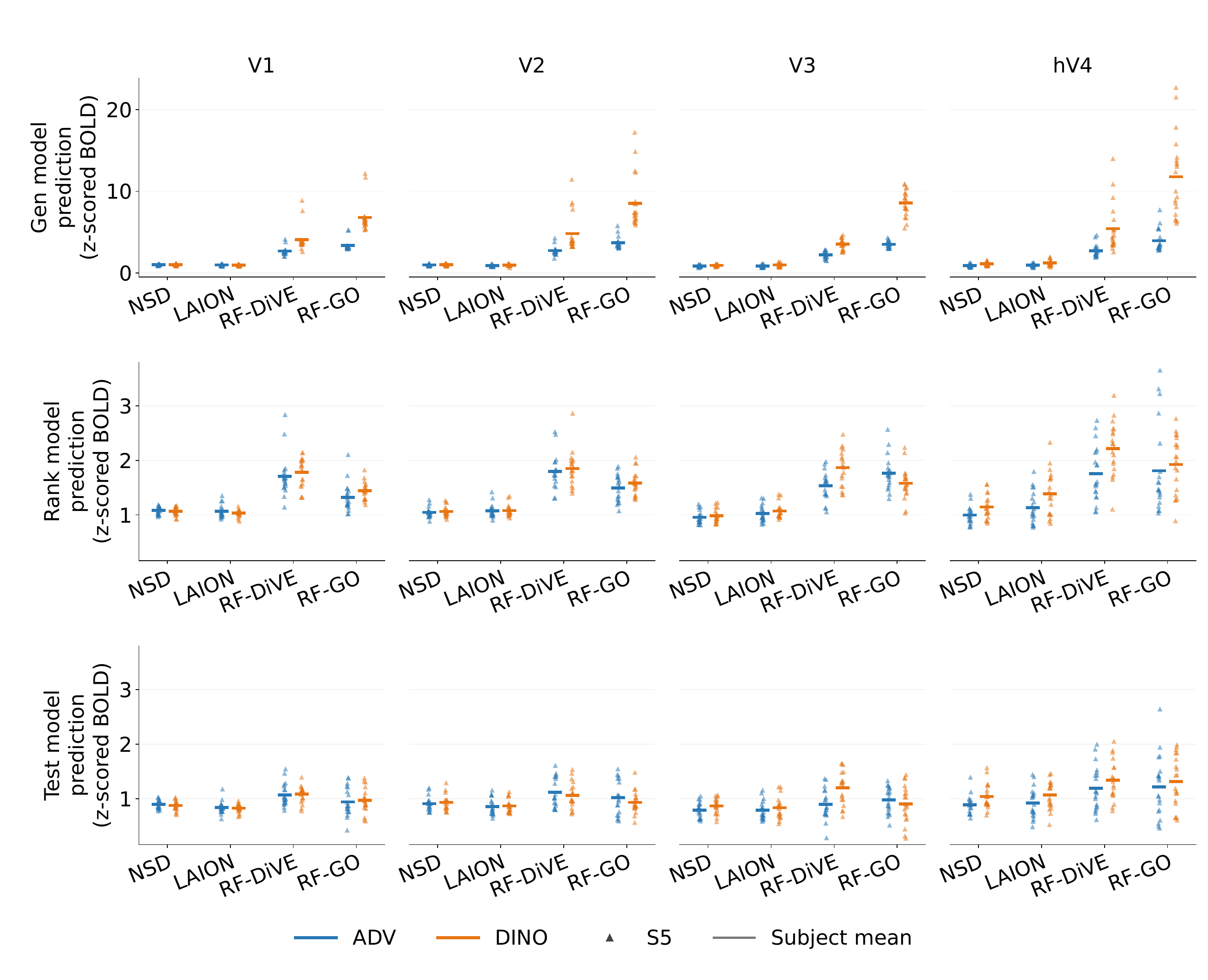}
  \caption{S5 responses: ranker-selected top-10 images; full-image + pRF constraints.}
  \label{fig:app_response_S5_rank_full_prf}
  \vspace{6pt}
\includegraphics[width=0.94\linewidth,height=0.425\textheight,keepaspectratio]{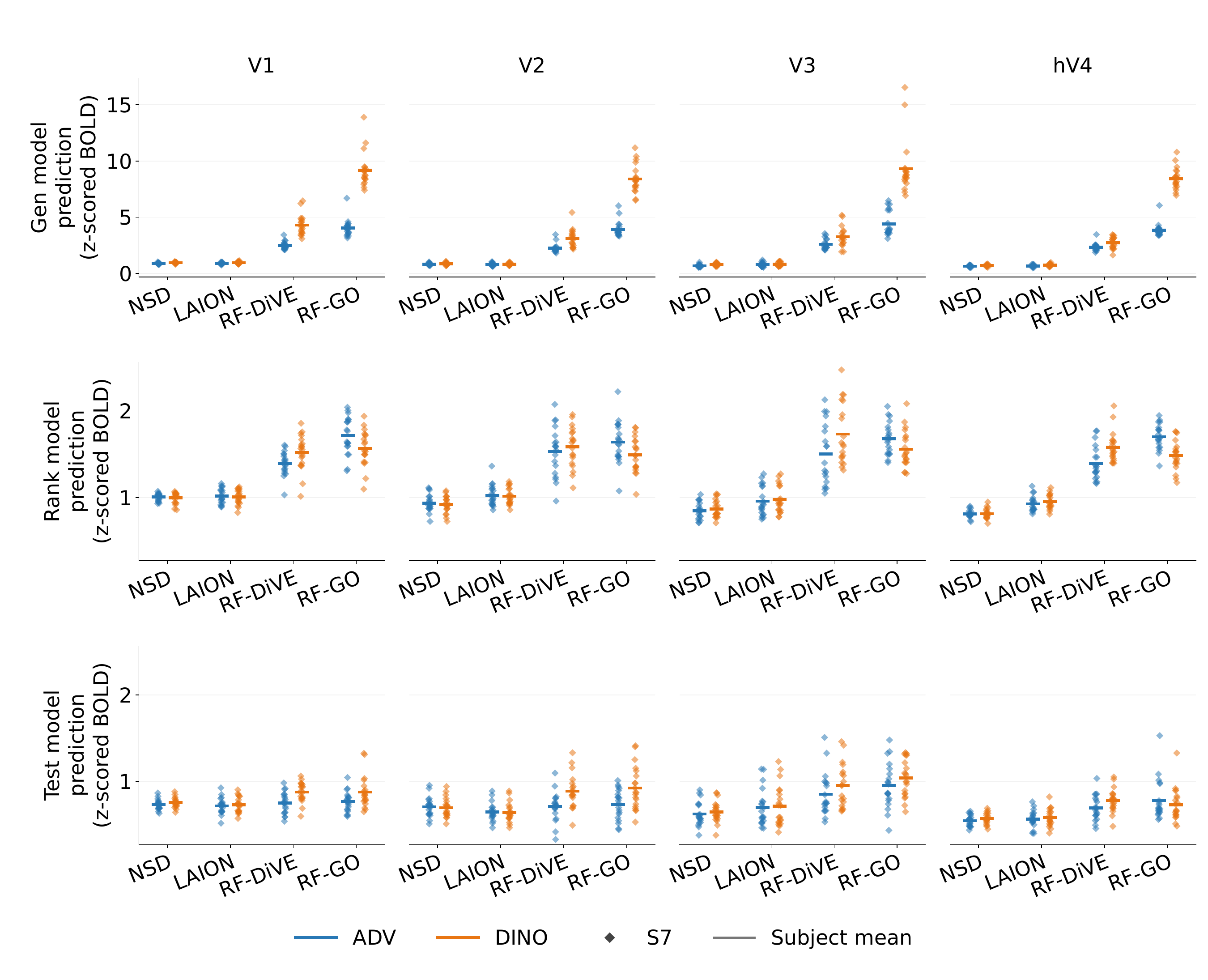}
  \caption{S7 responses: ranker-selected top-10 images; full-image + pRF constraints.}
  \label{fig:app_response_S7_rank_full_prf}
\end{figure}

\begin{figure}[!htbp]
  \centering
  \small
  \setlength{\abovecaptionskip}{3pt}
  \setlength{\belowcaptionskip}{2pt}
\includegraphics[width=0.94\linewidth,height=0.425\textheight,keepaspectratio]{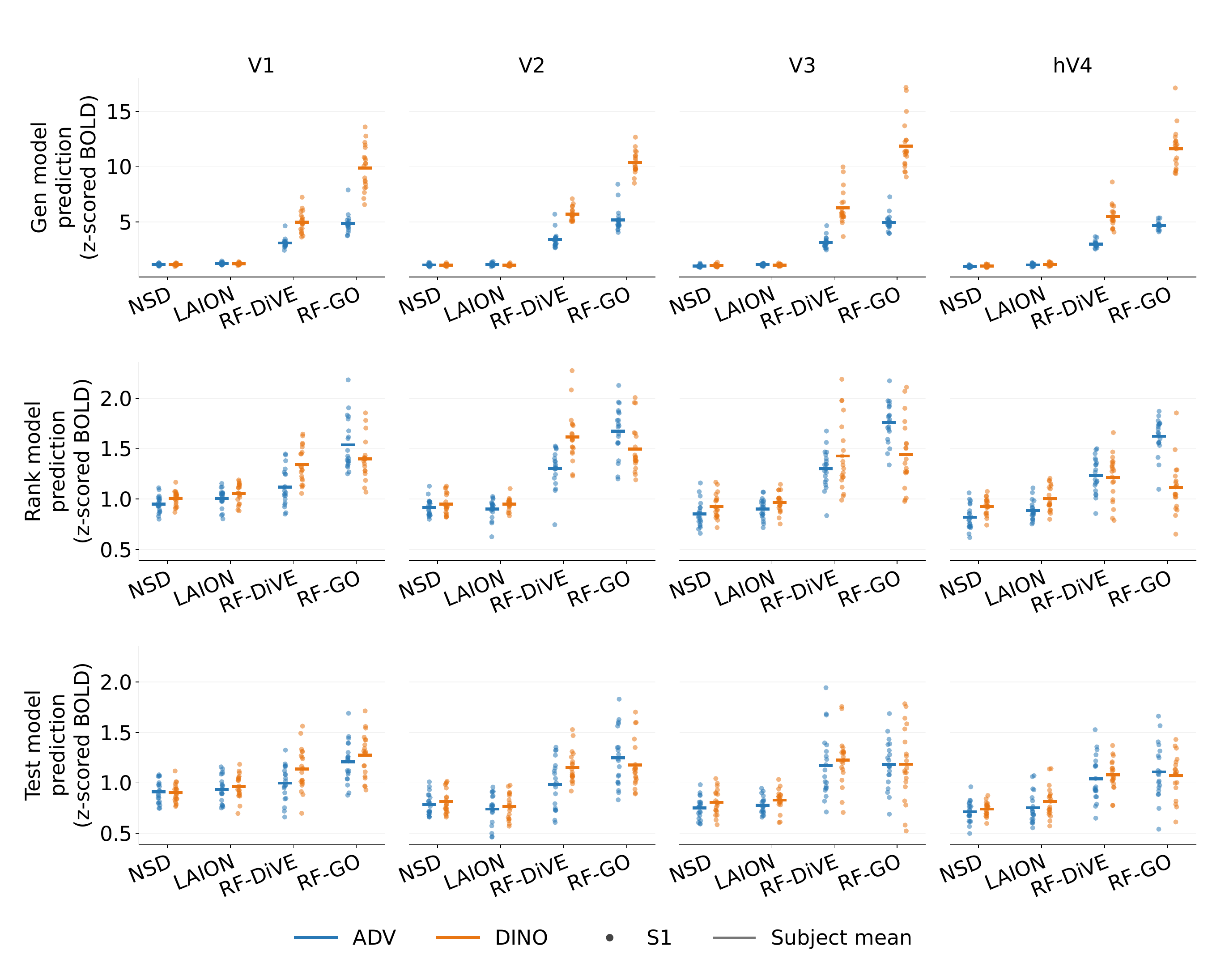}
  \caption{S1 responses: generator-selected top-10 images; full-image + pRF constraints.}
  \label{fig:app_response_S1_gen_full_prf}
  \vspace{6pt}
\includegraphics[width=0.94\linewidth,height=0.425\textheight,keepaspectratio]{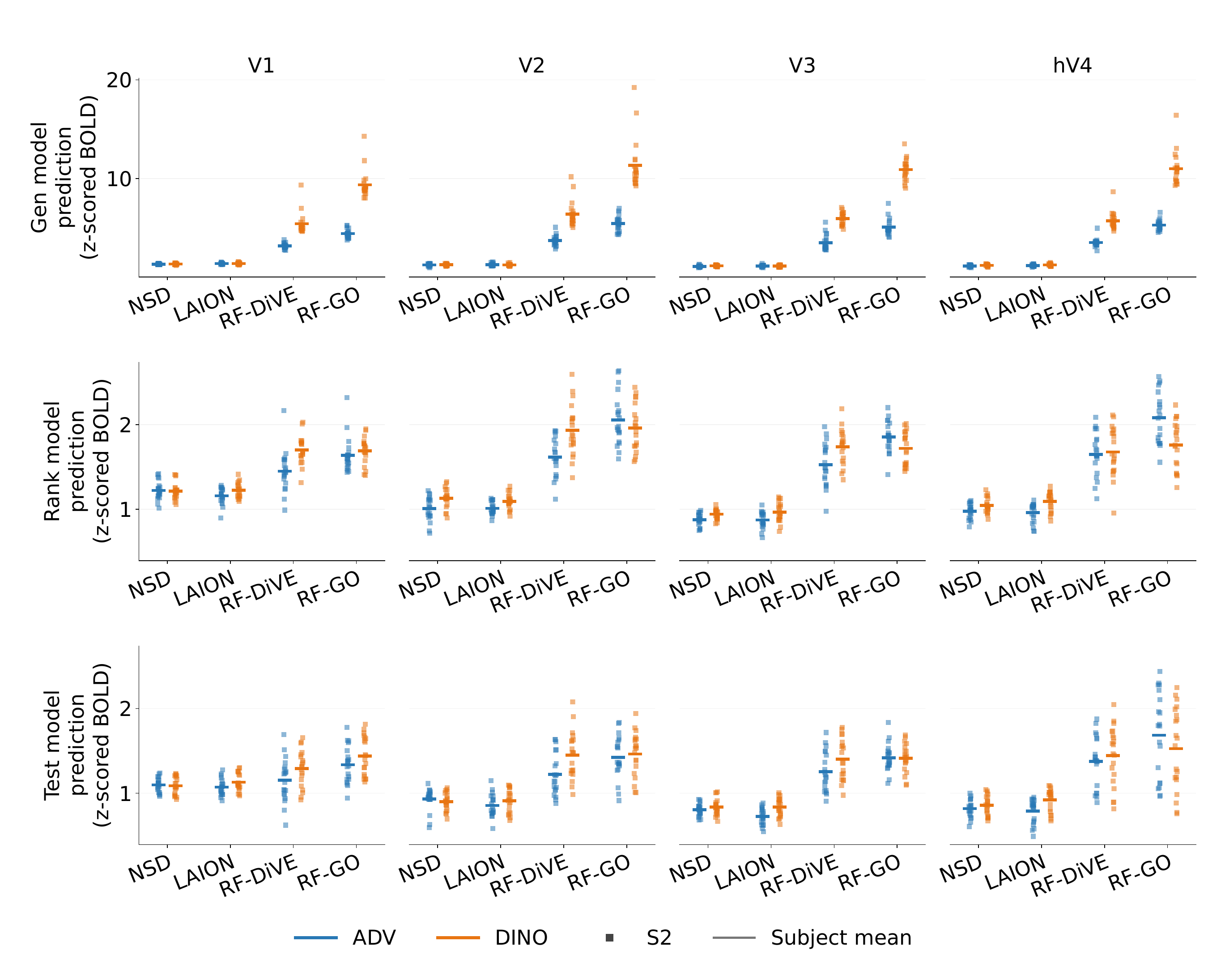}
  \caption{S2 responses: generator-selected top-10 images; full-image + pRF constraints.}
  \label{fig:app_response_S2_gen_full_prf}
\end{figure}

\begin{figure}[!htbp]
  \centering
  \small
  \setlength{\abovecaptionskip}{3pt}
  \setlength{\belowcaptionskip}{2pt}
\includegraphics[width=0.94\linewidth,height=0.425\textheight,keepaspectratio]{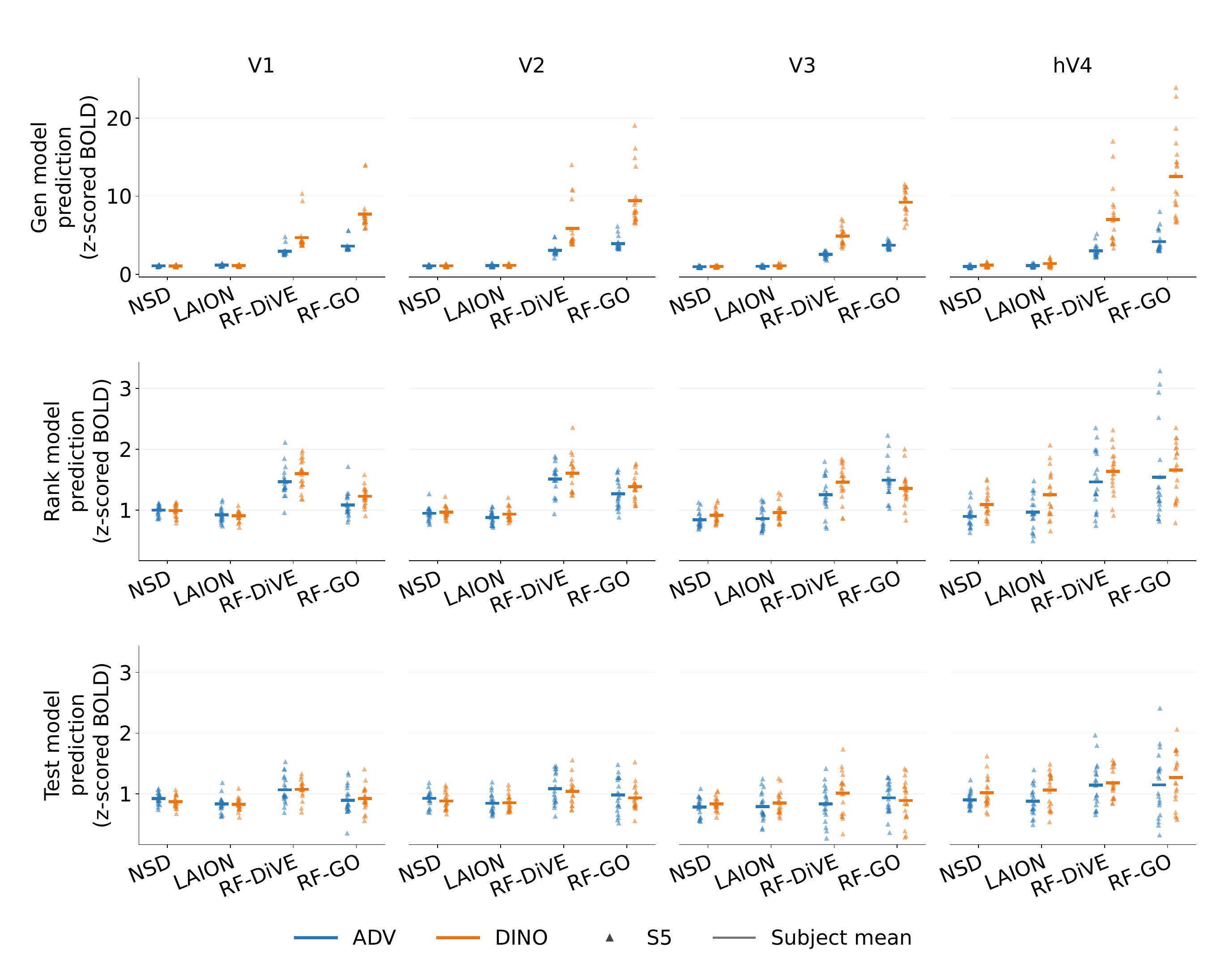}
  \caption{S5 responses: generator-selected top-10 images; full-image + pRF constraints.}
  \label{fig:app_response_S5_gen_full_prf}
  \vspace{6pt}
\includegraphics[width=0.94\linewidth,height=0.425\textheight,keepaspectratio]{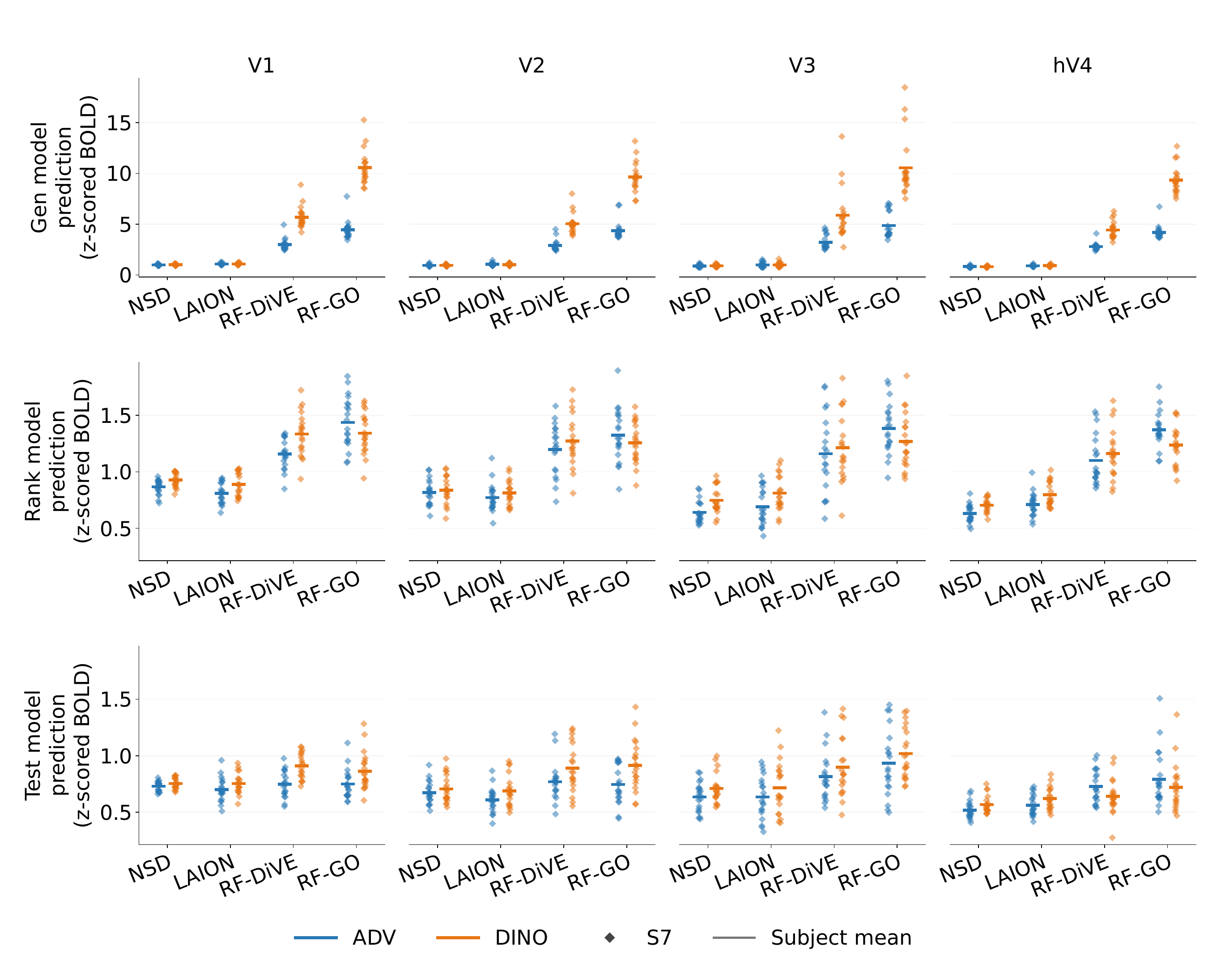}
  \caption{S7 responses: generator-selected top-10 images; full-image + pRF constraints.}
  \label{fig:app_response_S7_gen_full_prf}
\end{figure}

\clearpage
\subsection{Predicted-response advantages over LAION}
\label{app:laion_advantages}

For each voxel, we subtract the mean prediction for the selected LAION images from that for the selected MEIs, using ten images per source and the same evaluation encoder. Points show equal-weight means of these voxel-wise differences within each subject and ROI. Positive values favor MEIs. Color denotes backbone; circles, squares, triangles, and diamonds denote S1, S2, S5, and S7, respectively. Lines connect backbone summaries within a subject and method; the backbone-specific voxel selections need not be identical. The ranker-selected, full-image result appears in the main text; its three complementary conditions follow.

\begin{figure}[!htbp]
  \centering
  \small
  \setlength{\abovecaptionskip}{3pt}
  \setlength{\belowcaptionskip}{2pt}
\includegraphics[width=0.94\linewidth,height=0.425\textheight,keepaspectratio]{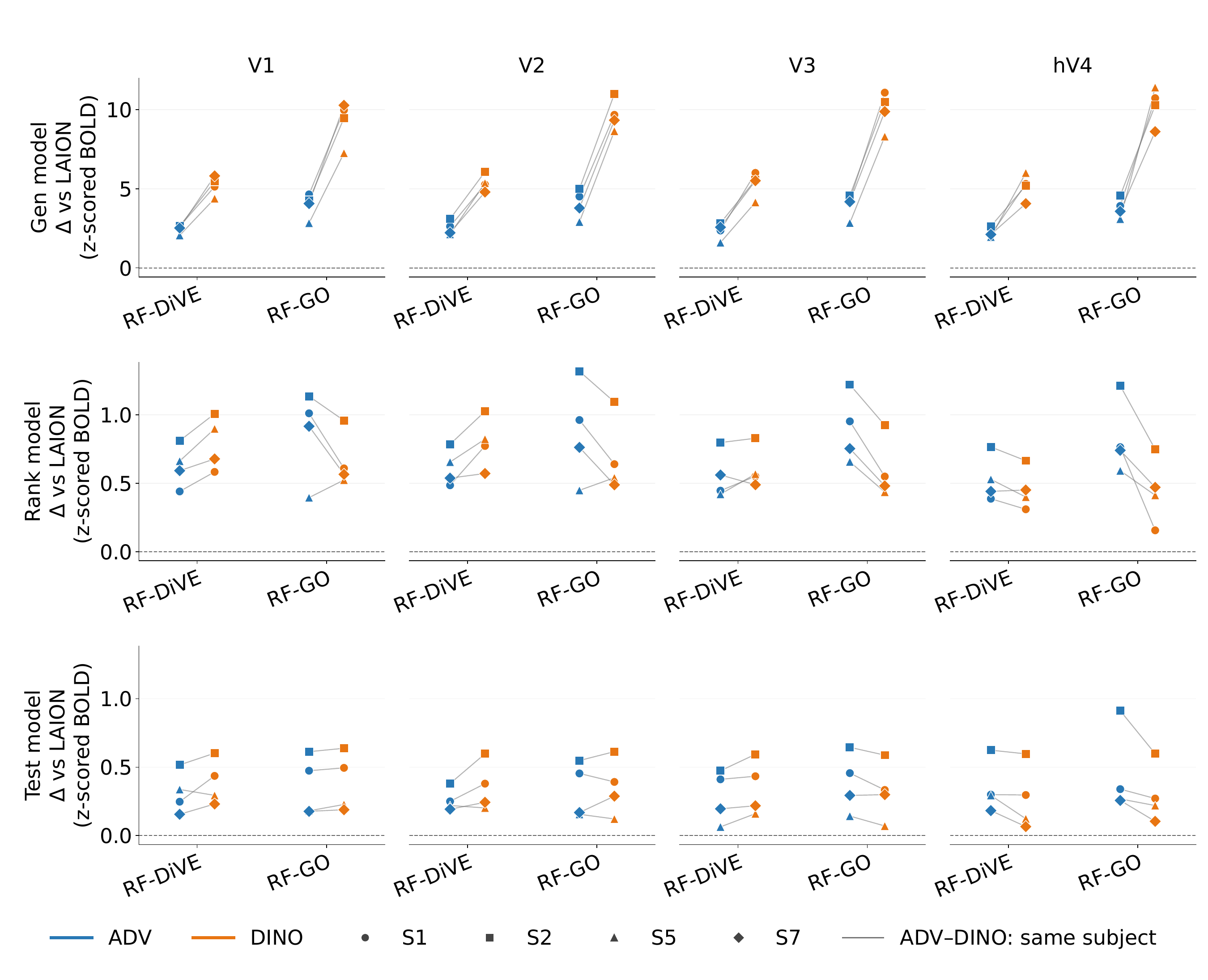}
  \caption{MEI--LAION response differences: generator-selected top-10 images; full-image constraints.}
  \label{fig:app_advantage_gen_full}
\end{figure}

\begin{figure}[!htbp]
  \centering
  \small
  \setlength{\abovecaptionskip}{3pt}
  \setlength{\belowcaptionskip}{2pt}
\includegraphics[width=0.94\linewidth,height=0.425\textheight,keepaspectratio]{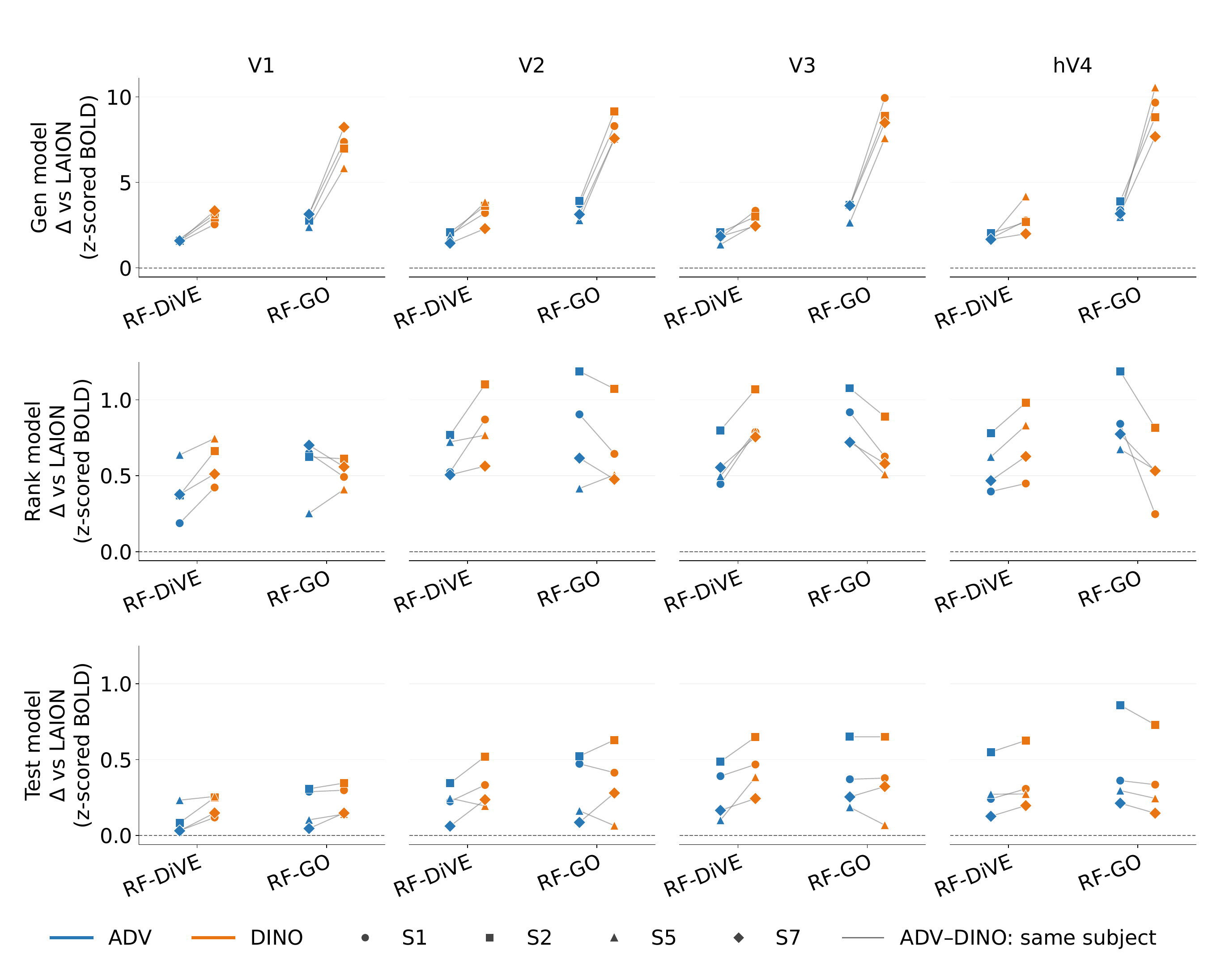}
  \caption{MEI--LAION response differences: ranker-selected top-10 images; full-image + pRF constraints.}
  \label{fig:app_advantage_rank_full_prf}
  \vspace{6pt}
\includegraphics[width=0.94\linewidth,height=0.425\textheight,keepaspectratio]{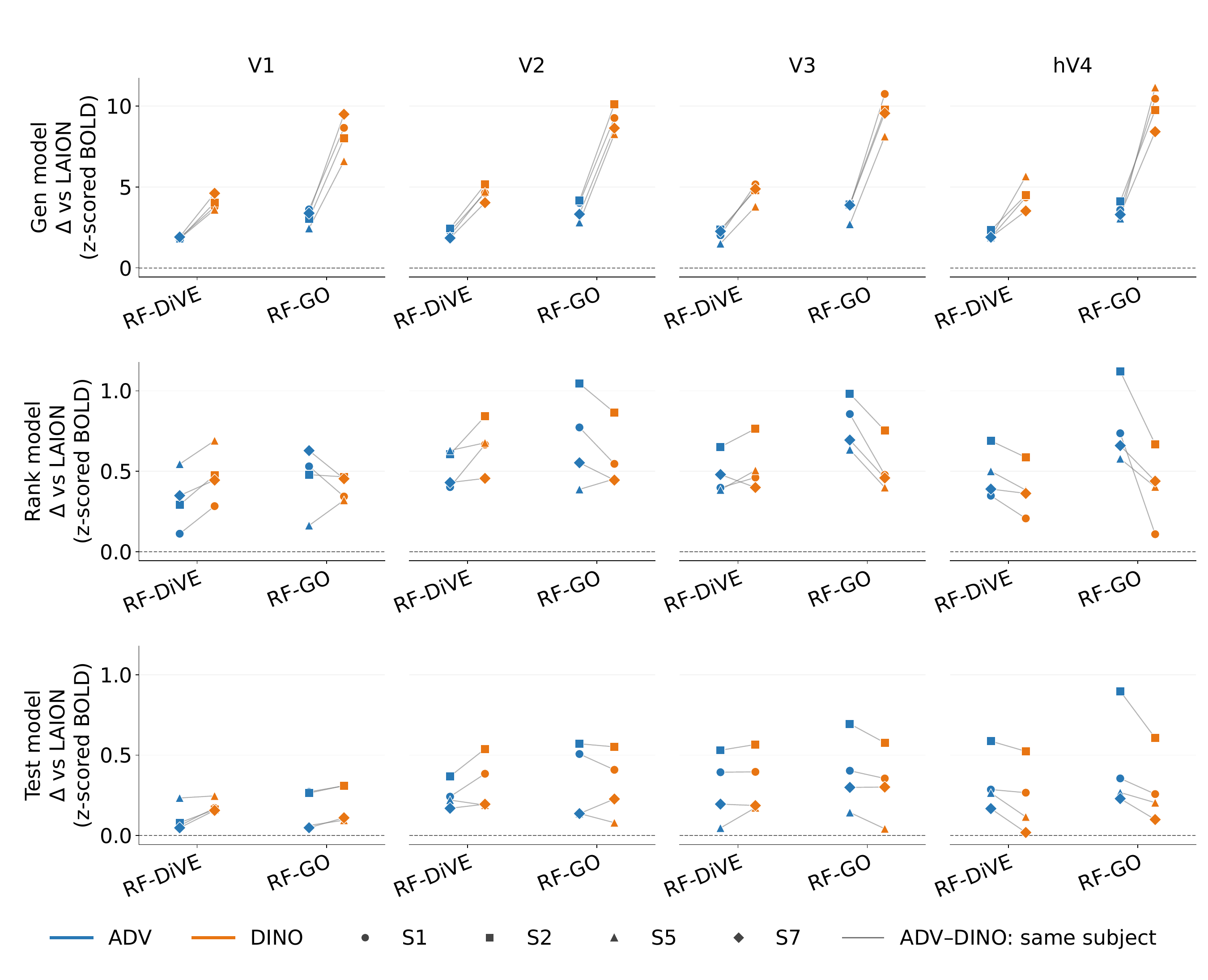}
  \caption{MEI--LAION response differences: generator-selected top-10 images; full-image + pRF constraints.}
  \label{fig:app_advantage_gen_full_prf}
\end{figure}

\clearpage

\subsection{Additional chromatic and orientation statistics}
\label{app:color_orientation}
We analyzed the top-10 ranker-selected images per voxel, using the full-image
final-constraint condition for generated images and unmodified natural images.
CIELAB $L^*,a^*,b^*$ (D65) and HSV saturation were computed per pixel and
averaged with normalized generator-pRF weights $w_p$.
Saturation was $S_p=(\max_c I_{pc}-\min_c I_{pc})/\max_c I_{pc}$, with $S_p=0$
for black pixels; thus the image statistic was $\sum_p w_p S_p$.
For orientation, unit-$L^2$-normalized Gabor filters were applied to the full
image before weighted pooling. For pooled magnitudes $A_i(f,\theta)$, each
image profile was $p_i(\theta)=\sum_f A_i(f,\theta)/\sum_{f,\theta}A_i(f,\theta)$.
Statistics were averaged over images within voxel, then available matched
voxels within subject (18--20 per subject, backbone, and ROI).
These are descriptive stimulus statistics, not direct neuronal tuning estimates.

\begin{figure}[htbp]
  \centering
  \includegraphics[width=\linewidth]{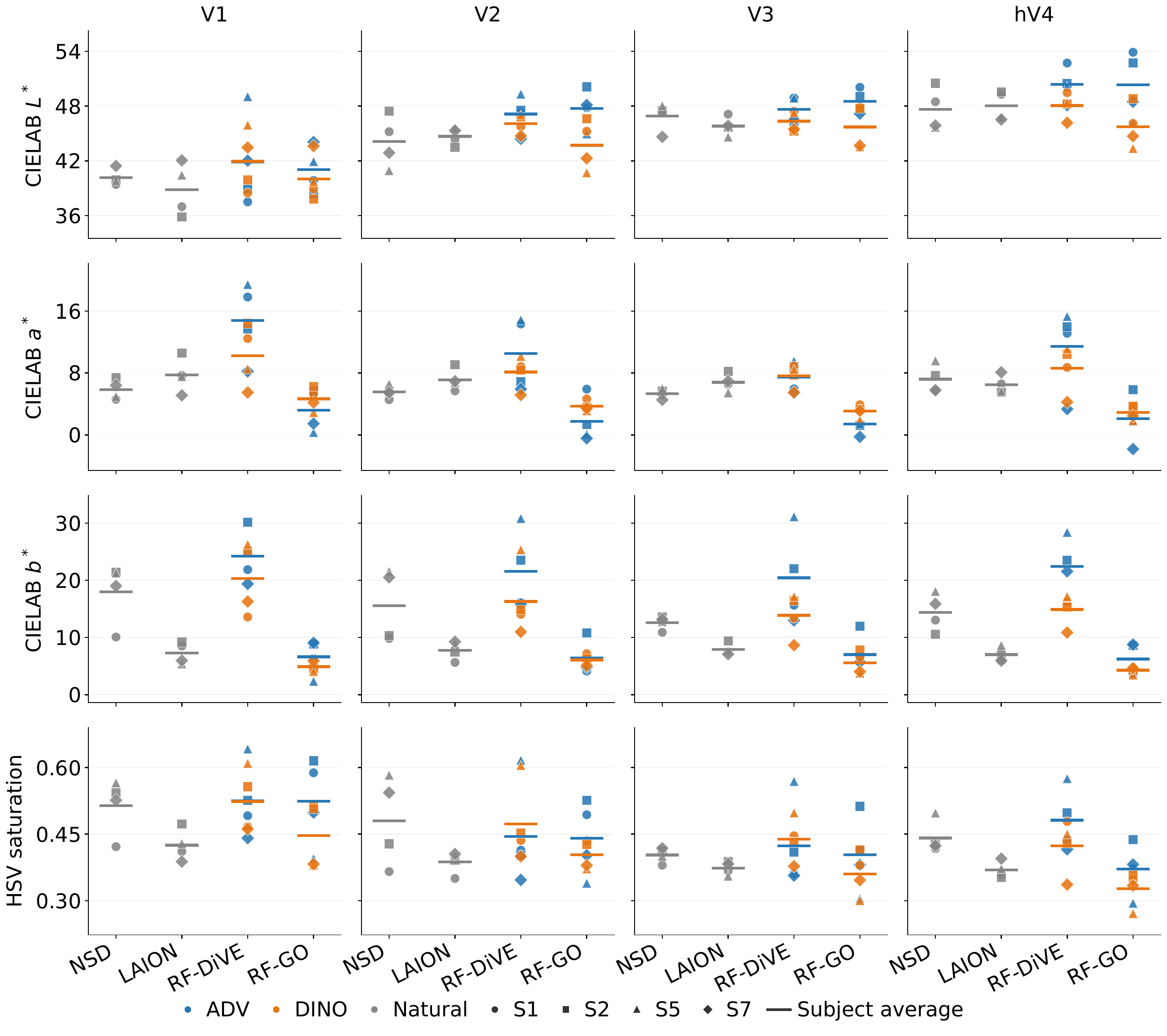}
  \caption{CIELAB components and HSV saturation across subjects and visual areas.
  Points denote subject means; short bars denote means across subjects.
  Natural references pool the two backbone-conditioned summaries.}
  \label{fig:app_color_all_rois}
\end{figure}

\begin{figure}[htbp]
  \centering
  \includegraphics[width=\linewidth]{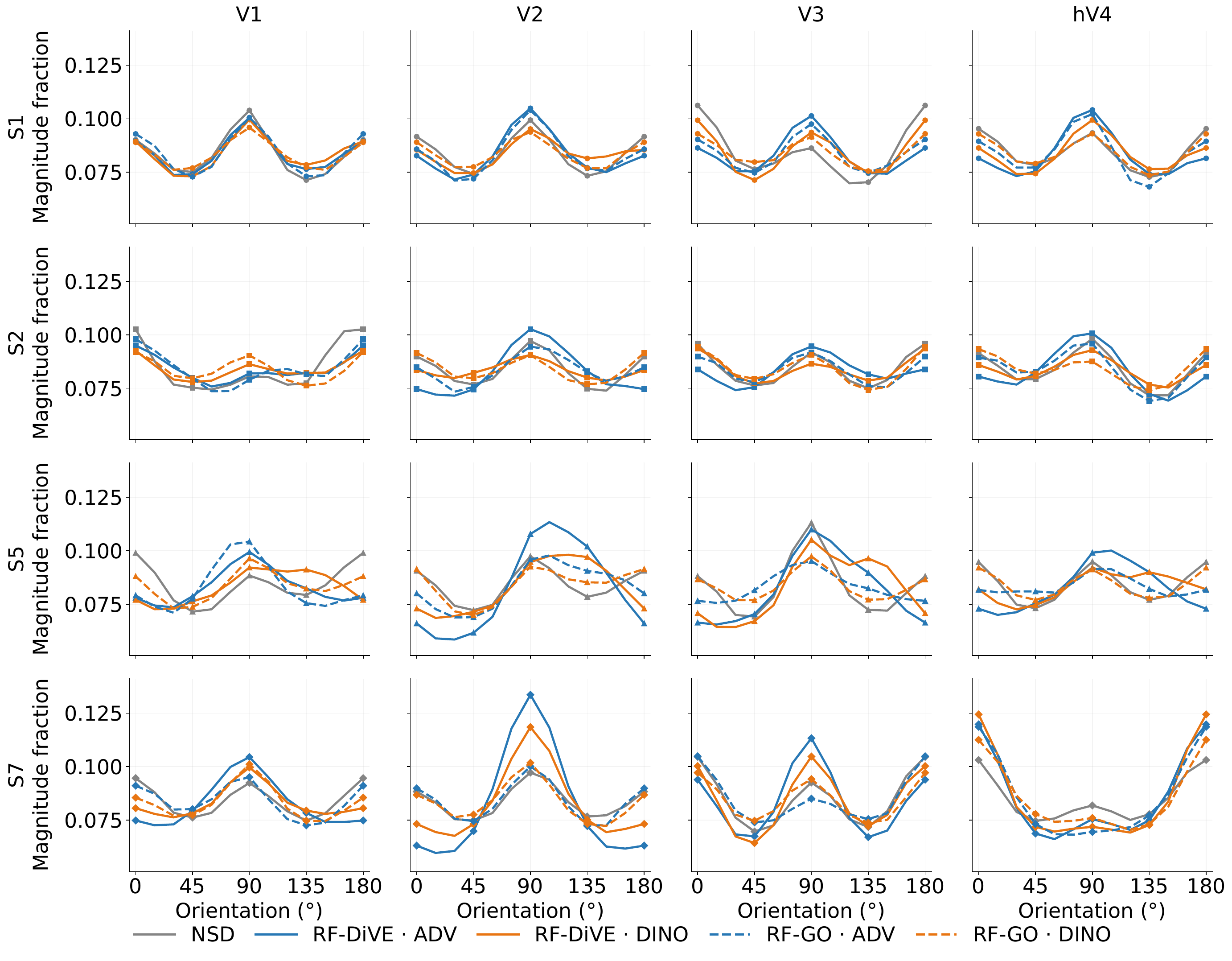}
  \caption{Normalized Gabor orientation profiles for each subject and visual area,
  using generator-pRF pooling. Solid and dashed colored lines denote RF-DiVE
  and RF-GO; gray denotes the pooled NSD reference.}
  \label{fig:app_orientation_all_rois}
\end{figure}

\begin{figure}[htbp]
  \centering
  \includegraphics[width=\linewidth]{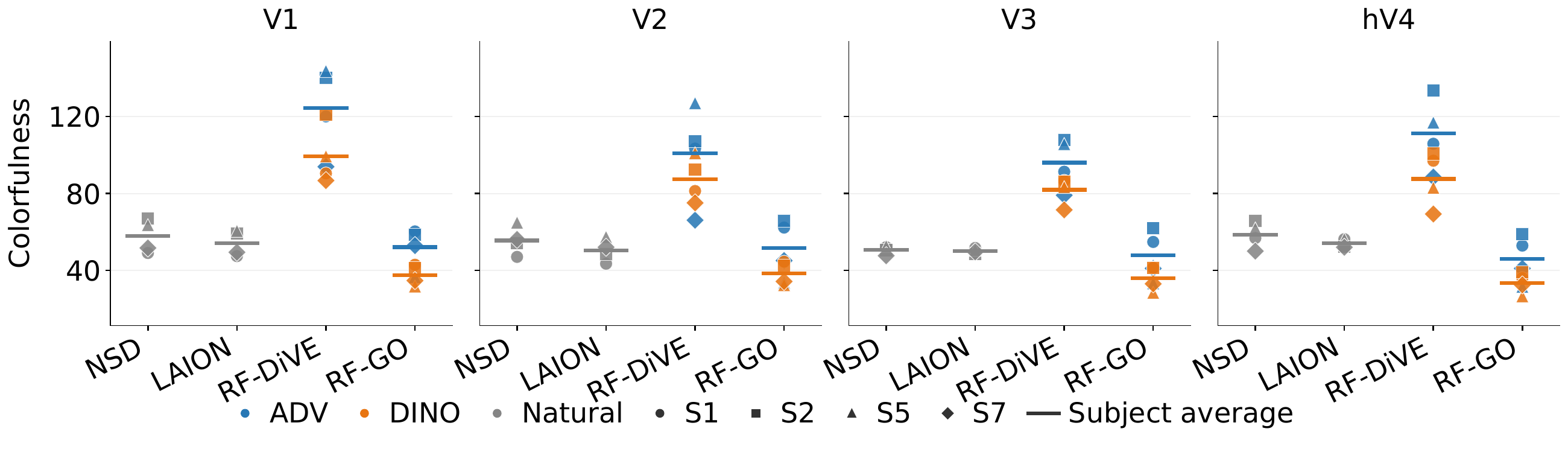}
  \caption{Colorfulness across subjects and visual areas, using generator-pRF weights. Points denote subject means; bars denote means across subjects.}
  \label{fig:app_colorfulness_all_rois_prf}
\end{figure}

\begin{figure}[htbp]
  \centering
  \includegraphics[width=\linewidth]{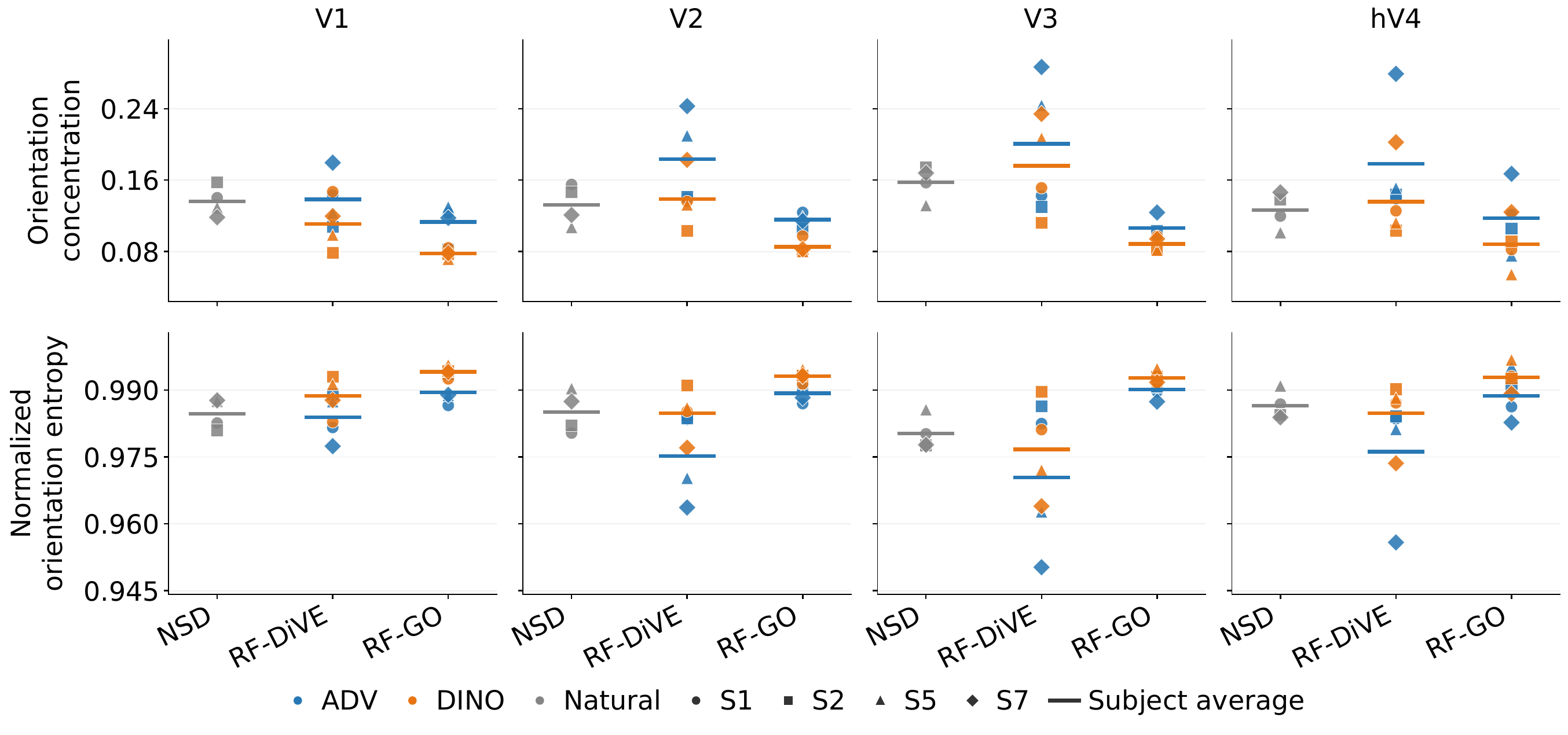}
  \caption{Per-image orientation concentration and normalized entropy, using generator-pRF weights. Points denote subject means; bars denote means across subjects.}
  \label{fig:app_orientation_summary_all_rois_prf}
\end{figure}

\begin{figure}[htbp]
  \centering
  \includegraphics[width=\linewidth]{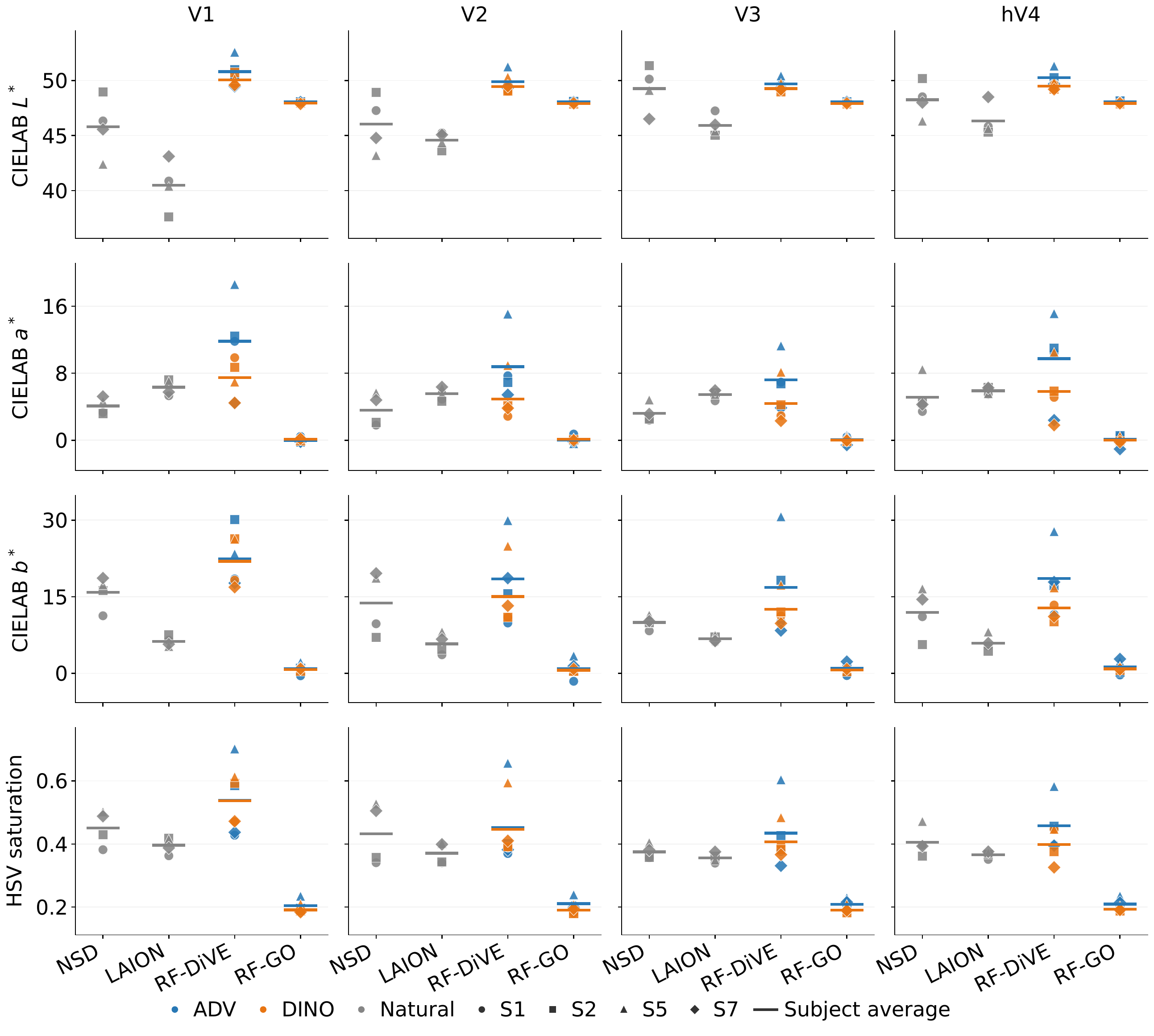}
  \caption{CIELAB components and HSV saturation with uniform full-field analysis weights. Points denote subject means; bars denote means across subjects.}
  \label{fig:app_cielab_saturation_all_rois_full_field}
\end{figure}

\begin{figure}[htbp]
  \centering
  \includegraphics[width=\linewidth]{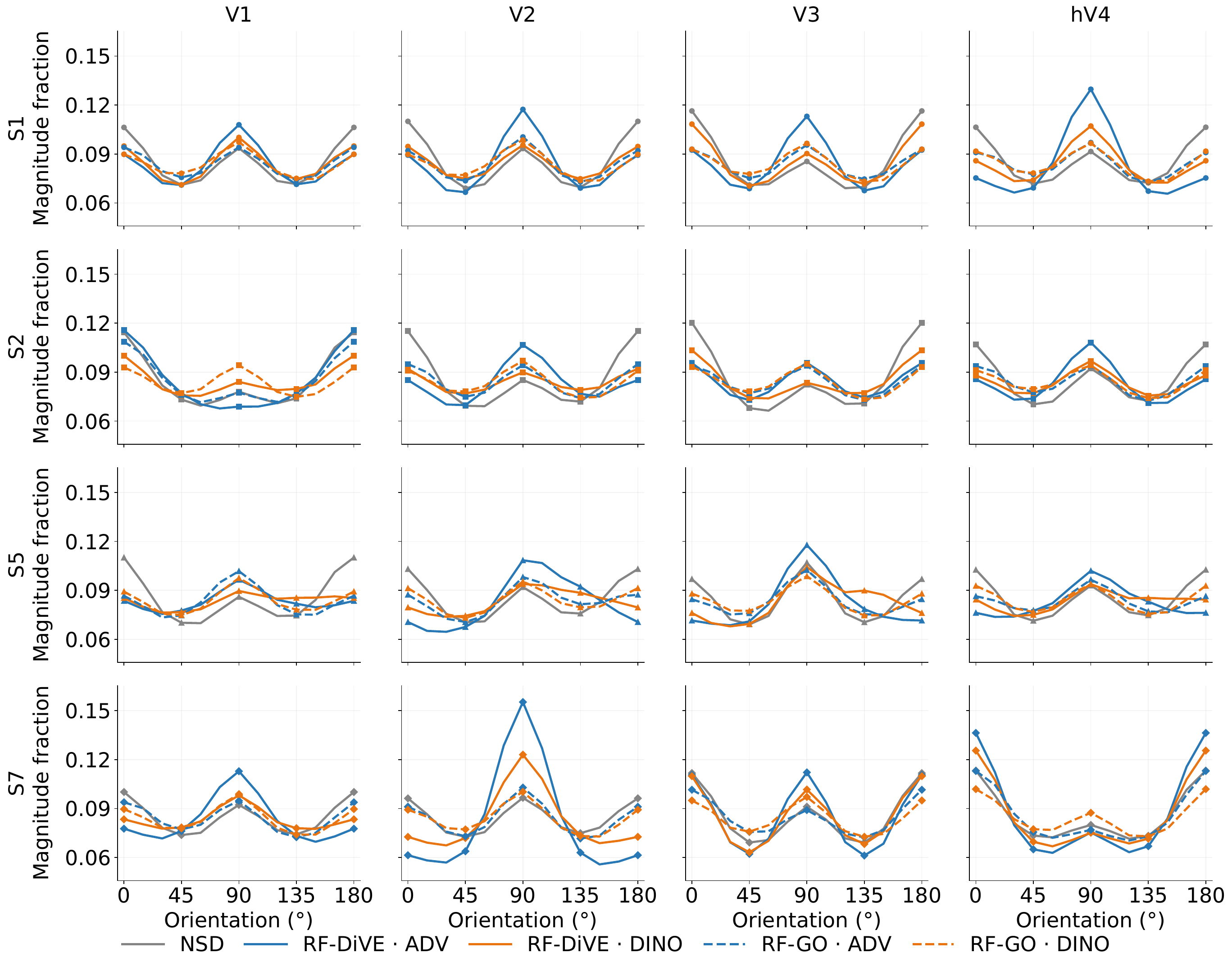}
  \caption{Normalized full-field Gabor orientation profiles for each subject and visual area. Solid and dashed colored lines denote RF-DiVE and RF-GO; gray denotes NSD.}
  \label{fig:app_orientation_profiles_all_subjects_rois_full_field}
\end{figure}

\begin{figure}[htbp]
  \centering
  \includegraphics[width=\linewidth]{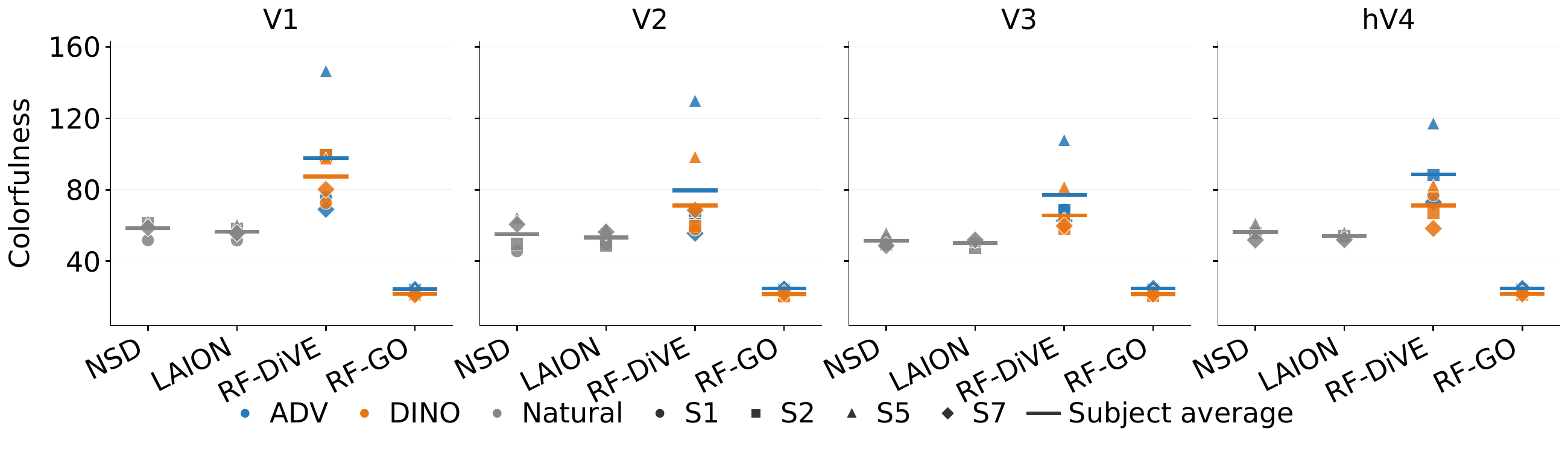}
  \caption{Colorfulness across subjects and visual areas, using uniform full-field weights. Points denote subject means; bars denote means across subjects.}
  \label{fig:app_colorfulness_all_rois_full_field}
\end{figure}

\begin{figure}[htbp]
  \centering
  \includegraphics[width=\linewidth]{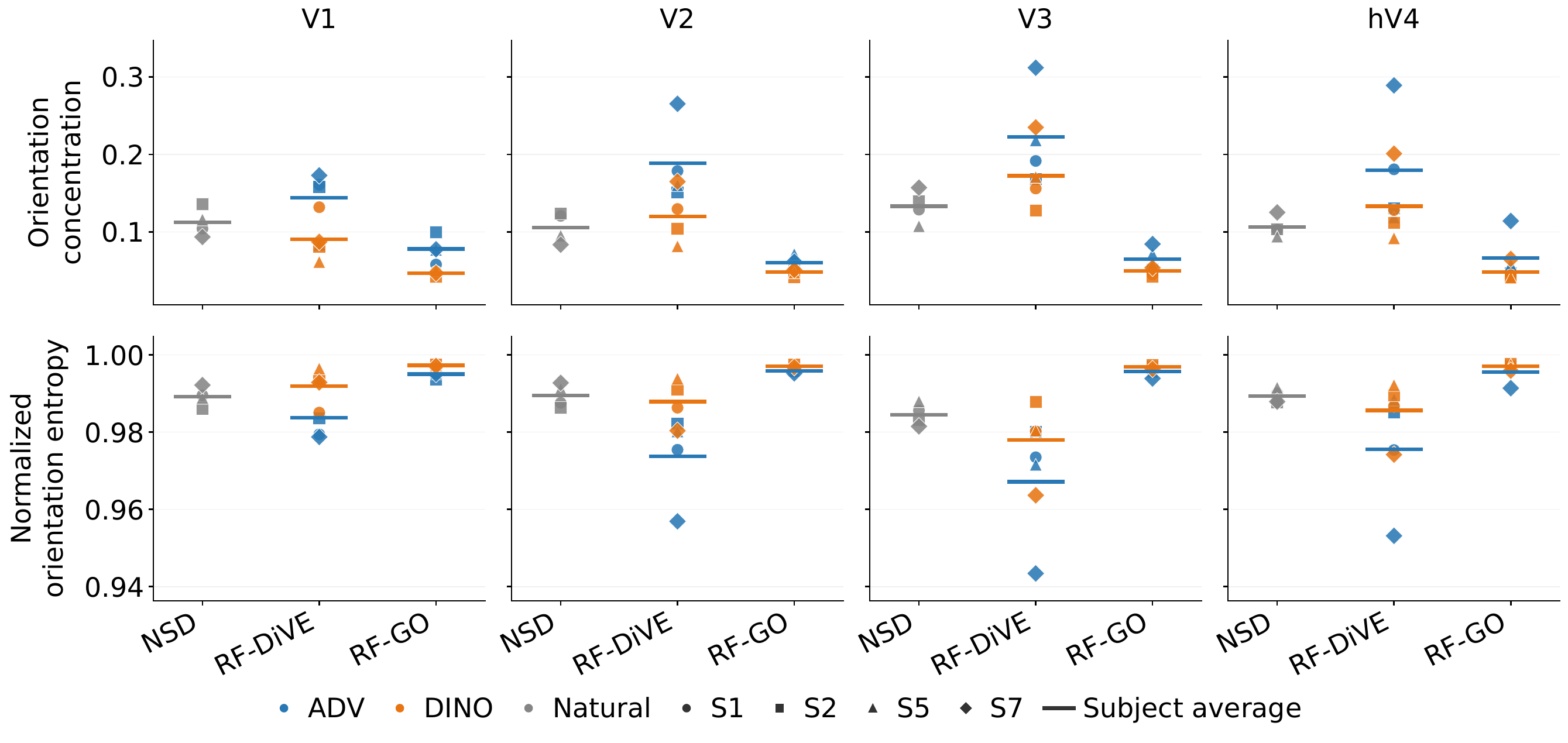}
  \caption{Per-image orientation concentration and normalized entropy with full-field weights. Points denote subject means; bars denote means across subjects.}
  \label{fig:app_orientation_summary_all_rois_full_field}
\end{figure}

\clearpage

\subsection{Additional structural image comparisons}
\label{supp:structure_comparisons}
We summarize ranker-selected top-10 images from the full-image final-constraint
condition, with unmodified NSD images as natural controls. These descriptive
comparisons supplement the main chromatic and spatial-frequency analyses.

\paragraph{Structural image statistics.}
For each voxel, the ranker-selected images form an array of shape
$[10,224,224,3]$ with RGB in $[0,1]$. Filters are applied to each full image before pooling
with normalized generator soft-pRF weights $w_p$ ($\sum_p w_p=1$).
Canny edge fraction is $\sum_p w_pE_p$, where $E_p$ is the binary edge map
(Rec.709 luma; smoothing $\sigma=1$ pixel; thresholds $0.1/0.2$).
GLCM joint entropy is $-\sum_{a,b}P_{ab}\log_2 P_{ab}$, averaged over offsets
$(0,1),(1,0),(1,1),(1,-1)$, using 32 luma levels and symmetric co-occurrence
probabilities weighted by $\sqrt{w_pw_{p+d}}$.
The bent-Gabor index is $\sum_p w_pM_pq_p/\sum_p w_pM_p$, where
$q_p\in\{0,\ldots,5\}$ is the winning bend-bin index among 144 filters and
$M_p$ selects pixels above the whole-image 90th-percentile Roberts edge
magnitude (BT.601 grayscale).
Effective edge support is $(\sum_p w_pM_p)^2/\sum_p(w_pM_p)^2$.
Metrics are averaged over selected images within voxel and then over voxels;
the four-subject summaries weight subjects equally. These are descriptive
image statistics, not measurements of neural tuning.

\begin{figure}[tbp]
\centering
\includegraphics[width=0.98\linewidth]{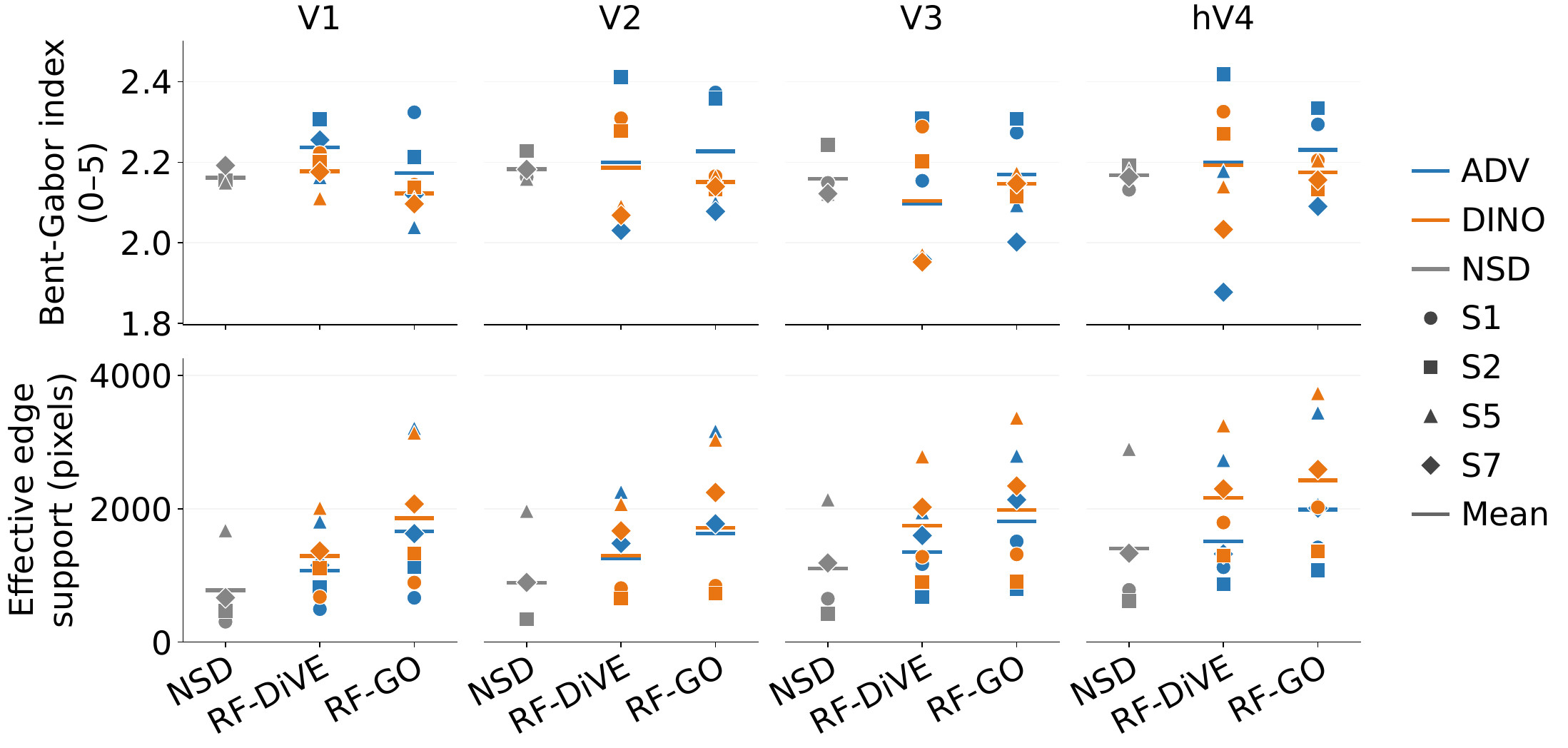}
\caption{Bent-Gabor curvature index and effective edge support across visual areas. Points denote subject means of voxel-wise top-10 image statistics; bars denote four-subject means. Statistics use generator soft-pRF weights.}
\label{fig:supp_curvature_prf}
\end{figure}
\begin{figure}[tbp]
\centering
\includegraphics[width=0.98\linewidth]{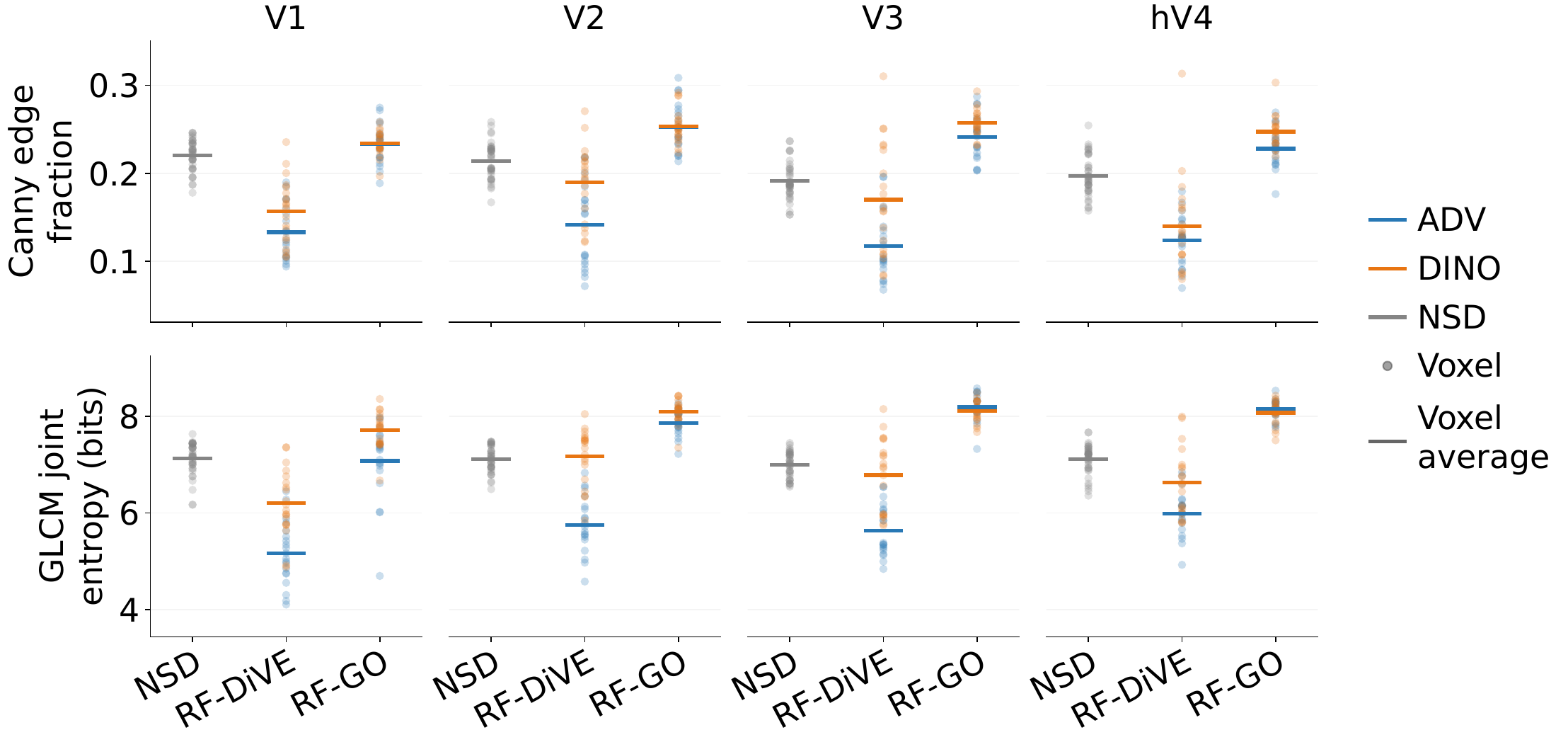}
\caption{Canny edge fraction and GLCM joint entropy in S1. Pale points denote voxel-wise top-10 image means; bars denote means across 20 voxels per backbone and ROI. Statistics use generator soft-pRF weights.}
\label{fig:supp_s1_edges_texture_prf}
\end{figure}

\clearpage

\subsection{Behavioral experiment to test differences between region MEIs}
\label{app:behavioral_experiment}

To evaluate which perceptual attributes differ between MEIs targeted to V1 and hV4, we collected human behavioral data using an online experiment conducted on Pavlovia, with participants recruited through Prolific (London, UK). We recruited 35 distinct participants for each of two task versions, and following exclusions due to incomplete data and quality control, we retained 32 participants in each task version. All participants provided informed consent, in accordance with a human subject protocol approved by the Institutional Review Board at the authors' institution.

In the experiment, participants performed trials of a two-alternative forced choice task comparing images that were targeted to V1 or hV4. On each trial, 10 images from a V1 voxel and 10 images from a hV4 voxel were shown, each presented in a 2x5 grid on the left and right sides of the screen. Below the images, a text question was presented, which asked: ``Which images have more three-dimensional form?''. Participants responded with keyboard buttons to indicate the left or right image group. The side (left/right) on which V1 and hV4 were shown was counterbalanced across trials. Participants were given a maximum of 10s to respond. On each trial, the images compared were always generated using the same method (RF-DiVE, RF-GO, or top LAION-fMRI images ranked by the ``ranking model'' encoder), and generated using the same DNN backbone, and were generated from voxels in the same NSD participant. Thus, the comparison isolated the effect of V1 and hV4 within an NSD participant and method. The pairing of which V1 voxel was compared to which hV4 voxel on each trial was randomly assigned using a different order within each participant. The two task versions (see above) referred to whether the participants viewed the full image or a cropped version that isolated the region within 3$\sigma$ of the voxel's Gaussian pRF.

\subsection{Additional MEIs}
\label{app:additional_meis}
\input{iclr2027_mei_top8_figures}

%% file: tables/paired_voxel_tests.tex

\begin{table*}[!htbp]
\centering
\footnotesize
\setlength{\tabcolsep}{3pt}
\renewcommand{\arraystretch}{1.08}
\caption{Paired-voxel tests of MEI responses against natural controls.
Positive $t$ favors MEIs; $n$ is the paired-voxel count. Bold adjusted
$p$ values indicate significantly higher MEI responses after BH correction
across all 256 comparisons. Results continue on the following pages.}
\label{tab:rf_paired_voxel_tests}
\begin{tabular}{lllrrrrr}
\hline
 & & & & \multicolumn{2}{c}{MEI $-$ NSD} & \multicolumn{2}{c}{MEI $-$ LAION}\\
\cline{5-6}\cline{7-8}
ROI & Backbone & Method & $n$ & $t$ & $p_{\mathrm{BH}}$ & $t$ & $p_{\mathrm{BH}}$\\
\hline
\multicolumn{8}{l}{\textbf{S1 --- Full image}}\\
V1 & ADV & RF-DiVE & 20 & $8.67$ & $\mathbf{1.7\!\times\!10^{-7}}$ & $6.76$ & $\mathbf{3.8\!\times\!10^{-6}}$\\
 &  & RF-GO & 20 & $11.08$ & $\mathbf{7.9\!\times\!10^{-9}}$ & $11.47$ & $\mathbf{6.1\!\times\!10^{-9}}$\\
 & DINO & RF-DiVE & 20 & $11.33$ & $\mathbf{6.7\!\times\!10^{-9}}$ & $8.38$ & $\mathbf{2.5\!\times\!10^{-7}}$\\
 &  & RF-GO & 20 & $11.25$ & $\mathbf{7.2\!\times\!10^{-9}}$ & $10.81$ & $\mathbf{9.8\!\times\!10^{-9}}$\\
V2 & ADV & RF-DiVE & 20 & $4.40$ & $\mathbf{4.3\!\times\!10^{-4}}$ & $6.51$ & $\mathbf{5.9\!\times\!10^{-6}}$\\
 &  & RF-GO & 20 & $4.73$ & $\mathbf{2.1\!\times\!10^{-4}}$ & $7.20$ & $\mathbf{1.7\!\times\!10^{-6}}$\\
 & DINO & RF-DiVE & 20 & $9.67$ & $\mathbf{4.0\!\times\!10^{-8}}$ & $11.18$ & $\mathbf{7.2\!\times\!10^{-9}}$\\
 &  & RF-GO & 20 & $8.73$ & $\mathbf{1.5\!\times\!10^{-7}}$ & $13.13$ & $\mathbf{1.4\!\times\!10^{-9}}$\\
V3 & ADV & RF-DiVE & 20 & $8.82$ & $\mathbf{1.4\!\times\!10^{-7}}$ & $6.68$ & $\mathbf{4.4\!\times\!10^{-6}}$\\
 &  & RF-GO & 20 & $8.75$ & $\mathbf{1.5\!\times\!10^{-7}}$ & $8.54$ & $\mathbf{2.0\!\times\!10^{-7}}$\\
 & DINO & RF-DiVE & 20 & $10.06$ & $\mathbf{2.3\!\times\!10^{-8}}$ & $11.44$ & $\mathbf{6.1\!\times\!10^{-9}}$\\
 &  & RF-GO & 20 & $8.04$ & $\mathbf{4.3\!\times\!10^{-7}}$ & $6.75$ & $\mathbf{3.8\!\times\!10^{-6}}$\\
hV4 & ADV & RF-DiVE & 20 & $7.71$ & $\mathbf{7.6\!\times\!10^{-7}}$ & $7.68$ & $\mathbf{7.7\!\times\!10^{-7}}$\\
 &  & RF-GO & 20 & $6.62$ & $\mathbf{4.9\!\times\!10^{-6}}$ & $8.21$ & $\mathbf{3.2\!\times\!10^{-7}}$\\
 & DINO & RF-DiVE & 20 & $10.53$ & $\mathbf{1.3\!\times\!10^{-8}}$ & $12.26$ & $\mathbf{3.0\!\times\!10^{-9}}$\\
 &  & RF-GO & 20 & $11.20$ & $\mathbf{7.2\!\times\!10^{-9}}$ & $10.44$ & $\mathbf{1.4\!\times\!10^{-8}}$\\
\hline
\multicolumn{8}{l}{\textbf{S1 --- Full image + pRF}}\\
V1 & ADV & RF-DiVE & 20 & $1.59$ & $0.136$ & $0.86$ & $0.415$\\
 &  & RF-GO & 20 & $6.24$ & $\mathbf{9.7\!\times\!10^{-6}}$ & $6.07$ & $\mathbf{1.4\!\times\!10^{-5}}$\\
 & DINO & RF-DiVE & 20 & $4.63$ & $\mathbf{2.6\!\times\!10^{-4}}$ & $3.07$ & $\mathbf{0.007}$\\
 &  & RF-GO & 20 & $7.08$ & $\mathbf{2.2\!\times\!10^{-6}}$ & $6.38$ & $\mathbf{7.5\!\times\!10^{-6}}$\\
V2 & ADV & RF-DiVE & 20 & $5.89$ & $\mathbf{1.9\!\times\!10^{-5}}$ & $6.73$ & $\mathbf{4.0\!\times\!10^{-6}}$\\
 &  & RF-GO & 20 & $7.42$ & $\mathbf{1.2\!\times\!10^{-6}}$ & $11.42$ & $\mathbf{6.1\!\times\!10^{-9}}$\\
 & DINO & RF-DiVE & 20 & $10.99$ & $\mathbf{8.5\!\times\!10^{-9}}$ & $10.60$ & $\mathbf{1.3\!\times\!10^{-8}}$\\
 &  & RF-GO & 20 & $11.57$ & $\mathbf{5.6\!\times\!10^{-9}}$ & $13.85$ & $\mathbf{8.8\!\times\!10^{-10}}$\\
V3 & ADV & RF-DiVE & 20 & $7.68$ & $\mathbf{7.7\!\times\!10^{-7}}$ & $6.09$ & $\mathbf{1.3\!\times\!10^{-5}}$\\
 &  & RF-GO & 20 & $7.57$ & $\mathbf{9.2\!\times\!10^{-7}}$ & $8.49$ & $\mathbf{2.1\!\times\!10^{-7}}$\\
 & DINO & RF-DiVE & 20 & $10.45$ & $\mathbf{1.4\!\times\!10^{-8}}$ & $10.91$ & $\mathbf{8.9\!\times\!10^{-9}}$\\
 &  & RF-GO & 20 & $7.79$ & $\mathbf{6.6\!\times\!10^{-7}}$ & $6.88$ & $\mathbf{3.1\!\times\!10^{-6}}$\\
hV4 & ADV & RF-DiVE & 20 & $7.55$ & $\mathbf{9.5\!\times\!10^{-7}}$ & $7.03$ & $\mathbf{2.3\!\times\!10^{-6}}$\\
 &  & RF-GO & 20 & $7.15$ & $\mathbf{1.9\!\times\!10^{-6}}$ & $9.28$ & $\mathbf{6.8\!\times\!10^{-8}}$\\
 & DINO & RF-DiVE & 20 & $10.22$ & $\mathbf{1.9\!\times\!10^{-8}}$ & $11.20$ & $\mathbf{7.2\!\times\!10^{-9}}$\\
 &  & RF-GO & 20 & $9.36$ & $\mathbf{6.1\!\times\!10^{-8}}$ & $8.35$ & $\mathbf{2.6\!\times\!10^{-7}}$\\
\hline
\end{tabular}
\end{table*}

\begin{table*}[!htbp]
\centering
\footnotesize
\setlength{\tabcolsep}{3pt}
\renewcommand{\arraystretch}{1.08}
{\small\tablename~\ref{tab:rf_paired_voxel_tests} (continued): S2.\par}
\smallskip
\begin{tabular}{lllrrrrr}
\hline
 & & & & \multicolumn{2}{c}{MEI $-$ NSD} & \multicolumn{2}{c}{MEI $-$ LAION}\\
\cline{5-6}\cline{7-8}
ROI & Backbone & Method & $n$ & $t$ & $p_{\mathrm{BH}}$ & $t$ & $p_{\mathrm{BH}}$\\
\hline
\multicolumn{8}{l}{\textbf{S2 --- Full image}}\\
V1 & ADV & RF-DiVE & 20 & $8.54$ & $\mathbf{2.0\!\times\!10^{-7}}$ & $10.00$ & $\mathbf{2.5\!\times\!10^{-8}}$\\
 &  & RF-GO & 20 & $8.63$ & $\mathbf{1.8\!\times\!10^{-7}}$ & $10.68$ & $\mathbf{1.2\!\times\!10^{-8}}$\\
 & DINO & RF-DiVE & 20 & $13.48$ & $\mathbf{1.1\!\times\!10^{-9}}$ & $20.69$ & $\mathbf{4.4\!\times\!10^{-12}}$\\
 &  & RF-GO & 20 & $14.30$ & $\mathbf{8.1\!\times\!10^{-10}}$ & $19.06$ & $\mathbf{9.8\!\times\!10^{-12}}$\\
V2 & ADV & RF-DiVE & 20 & $5.97$ & $\mathbf{1.6\!\times\!10^{-5}}$ & $5.54$ & $\mathbf{3.8\!\times\!10^{-5}}$\\
 &  & RF-GO & 20 & $8.50$ & $\mathbf{2.1\!\times\!10^{-7}}$ & $7.67$ & $\mathbf{7.8\!\times\!10^{-7}}$\\
 & DINO & RF-DiVE & 20 & $9.00$ & $\mathbf{1.0\!\times\!10^{-7}}$ & $9.34$ & $\mathbf{6.2\!\times\!10^{-8}}$\\
 &  & RF-GO & 20 & $8.80$ & $\mathbf{1.4\!\times\!10^{-7}}$ & $9.54$ & $\mathbf{4.8\!\times\!10^{-8}}$\\
V3 & ADV & RF-DiVE & 20 & $7.81$ & $\mathbf{6.4\!\times\!10^{-7}}$ & $6.24$ & $\mathbf{9.7\!\times\!10^{-6}}$\\
 &  & RF-GO & 20 & $10.58$ & $\mathbf{1.3\!\times\!10^{-8}}$ & $11.76$ & $\mathbf{4.4\!\times\!10^{-9}}$\\
 & DINO & RF-DiVE & 19 & $8.98$ & $\mathbf{1.5\!\times\!10^{-7}}$ & $10.59$ & $\mathbf{1.9\!\times\!10^{-8}}$\\
 &  & RF-GO & 20 & $11.08$ & $\mathbf{7.9\!\times\!10^{-9}}$ & $13.38$ & $\mathbf{1.2\!\times\!10^{-9}}$\\
hV4 & ADV & RF-DiVE & 20 & $12.27$ & $\mathbf{3.0\!\times\!10^{-9}}$ & $12.23$ & $\mathbf{3.0\!\times\!10^{-9}}$\\
 &  & RF-GO & 20 & $10.90$ & $\mathbf{8.9\!\times\!10^{-9}}$ & $10.88$ & $\mathbf{8.9\!\times\!10^{-9}}$\\
 & DINO & RF-DiVE & 20 & $11.04$ & $\mathbf{8.1\!\times\!10^{-9}}$ & $13.78$ & $\mathbf{8.8\!\times\!10^{-10}}$\\
 &  & RF-GO & 20 & $7.07$ & $\mathbf{2.2\!\times\!10^{-6}}$ & $8.38$ & $\mathbf{2.5\!\times\!10^{-7}}$\\
\hline
\multicolumn{8}{l}{\textbf{S2 --- Full image + pRF}}\\
V1 & ADV & RF-DiVE & 20 & $0.08$ & $0.939$ & $1.77$ & $0.100$\\
 &  & RF-GO & 20 & $4.23$ & $\mathbf{6.2\!\times\!10^{-4}}$ & $6.43$ & $\mathbf{6.9\!\times\!10^{-6}}$\\
 & DINO & RF-DiVE & 20 & $3.08$ & $\mathbf{0.007}$ & $5.69$ & $\mathbf{2.8\!\times\!10^{-5}}$\\
 &  & RF-GO & 20 & $4.99$ & $\mathbf{1.2\!\times\!10^{-4}}$ & $7.33$ & $\mathbf{1.4\!\times\!10^{-6}}$\\
V2 & ADV & RF-DiVE & 20 & $6.57$ & $\mathbf{5.4\!\times\!10^{-6}}$ & $5.84$ & $\mathbf{2.1\!\times\!10^{-5}}$\\
 &  & RF-GO & 20 & $8.51$ & $\mathbf{2.1\!\times\!10^{-7}}$ & $7.42$ & $\mathbf{1.2\!\times\!10^{-6}}$\\
 & DINO & RF-DiVE & 20 & $10.38$ & $\mathbf{1.5\!\times\!10^{-8}}$ & $9.40$ & $\mathbf{5.8\!\times\!10^{-8}}$\\
 &  & RF-GO & 20 & $9.05$ & $\mathbf{9.7\!\times\!10^{-8}}$ & $9.25$ & $\mathbf{7.0\!\times\!10^{-8}}$\\
V3 & ADV & RF-DiVE & 20 & $9.84$ & $\mathbf{3.2\!\times\!10^{-8}}$ & $7.82$ & $\mathbf{6.3\!\times\!10^{-7}}$\\
 &  & RF-GO & 20 & $11.83$ & $\mathbf{4.4\!\times\!10^{-9}}$ & $12.93$ & $\mathbf{1.6\!\times\!10^{-9}}$\\
 & DINO & RF-DiVE & 19 & $10.08$ & $\mathbf{3.6\!\times\!10^{-8}}$ & $12.26$ & $\mathbf{4.4\!\times\!10^{-9}}$\\
 &  & RF-GO & 20 & $12.97$ & $\mathbf{1.6\!\times\!10^{-9}}$ & $16.83$ & $\mathbf{6.1\!\times\!10^{-11}}$\\
hV4 & ADV & RF-DiVE & 20 & $12.18$ & $\mathbf{3.0\!\times\!10^{-9}}$ & $11.98$ & $\mathbf{3.8\!\times\!10^{-9}}$\\
 &  & RF-GO & 20 & $10.06$ & $\mathbf{2.3\!\times\!10^{-8}}$ & $10.19$ & $\mathbf{1.9\!\times\!10^{-8}}$\\
 & DINO & RF-DiVE & 20 & $12.47$ & $\mathbf{2.7\!\times\!10^{-9}}$ & $13.88$ & $\mathbf{8.8\!\times\!10^{-10}}$\\
 &  & RF-GO & 20 & $7.68$ & $\mathbf{7.7\!\times\!10^{-7}}$ & $9.22$ & $\mathbf{7.3\!\times\!10^{-8}}$\\
\hline
\end{tabular}
\end{table*}

\begin{table*}[!htbp]
\centering
\footnotesize
\setlength{\tabcolsep}{3pt}
\renewcommand{\arraystretch}{1.08}
{\small\tablename~\ref{tab:rf_paired_voxel_tests} (continued): S5.\par}
\smallskip
\begin{tabular}{lllrrrrr}
\hline
 & & & & \multicolumn{2}{c}{MEI $-$ NSD} & \multicolumn{2}{c}{MEI $-$ LAION}\\
\cline{5-6}\cline{7-8}
ROI & Backbone & Method & $n$ & $t$ & $p_{\mathrm{BH}}$ & $t$ & $p_{\mathrm{BH}}$\\
\hline
\multicolumn{8}{l}{\textbf{S5 --- Full image}}\\
V1 & ADV & RF-DiVE & 20 & $5.76$ & $\mathbf{2.5\!\times\!10^{-5}}$ & $8.27$ & $\mathbf{2.9\!\times\!10^{-7}}$\\
 &  & RF-GO & 20 & $3.67$ & $\mathbf{0.002}$ & $5.50$ & $\mathbf{4.1\!\times\!10^{-5}}$\\
 & DINO & RF-DiVE & 20 & $10.52$ & $\mathbf{1.3\!\times\!10^{-8}}$ & $10.90$ & $\mathbf{8.9\!\times\!10^{-9}}$\\
 &  & RF-GO & 20 & $5.87$ & $\mathbf{2.0\!\times\!10^{-5}}$ & $7.66$ & $\mathbf{7.8\!\times\!10^{-7}}$\\
V2 & ADV & RF-DiVE & 18 & $3.08$ & $\mathbf{0.008}$ & $3.48$ & $\mathbf{0.004}$\\
 &  & RF-GO & 20 & $1.85$ & $0.087$ & $2.67$ & $\mathbf{0.017}$\\
 & DINO & RF-DiVE & 19 & $2.79$ & $\mathbf{0.014}$ & $3.83$ & $\mathbf{0.002}$\\
 &  & RF-GO & 20 & $0.94$ & $0.375$ & $2.10$ & $0.055$\\
V3 & ADV & RF-DiVE & 19 & $2.14$ & $0.052$ & $1.92$ & $0.077$\\
 &  & RF-GO & 20 & $3.87$ & $\mathbf{0.001}$ & $3.06$ & $\mathbf{0.008}$\\
 & DINO & RF-DiVE & 19 & $6.64$ & $\mathbf{5.9\!\times\!10^{-6}}$ & $5.49$ & $\mathbf{5.1\!\times\!10^{-5}}$\\
 &  & RF-GO & 20 & $0.69$ & $0.510$ & $0.97$ & $0.363$\\
hV4 & ADV & RF-DiVE & 20 & $4.09$ & $\mathbf{8.4\!\times\!10^{-4}}$ & $5.59$ & $\mathbf{3.5\!\times\!10^{-5}}$\\
 &  & RF-GO & 20 & $2.57$ & $\mathbf{0.022}$ & $3.26$ & $\mathbf{0.005}$\\
 & DINO & RF-DiVE & 20 & $3.58$ & $\mathbf{0.003}$ & $6.23$ & $\mathbf{9.8\!\times\!10^{-6}}$\\
 &  & RF-GO & 20 & $3.49$ & $\mathbf{0.003}$ & $4.01$ & $\mathbf{9.9\!\times\!10^{-4}}$\\
\hline
\multicolumn{8}{l}{\textbf{S5 --- Full image + pRF}}\\
V1 & ADV & RF-DiVE & 20 & $4.19$ & $\mathbf{6.8\!\times\!10^{-4}}$ & $8.19$ & $\mathbf{3.4\!\times\!10^{-7}}$\\
 &  & RF-GO & 20 & $0.78$ & $0.457$ & $2.23$ & $\mathbf{0.043}$\\
 & DINO & RF-DiVE & 20 & $7.54$ & $\mathbf{9.5\!\times\!10^{-7}}$ & $10.50$ & $\mathbf{1.4\!\times\!10^{-8}}$\\
 &  & RF-GO & 20 & $1.97$ & $0.070$ & $3.55$ & $\mathbf{0.003}$\\
V2 & ADV & RF-DiVE & 18 & $2.71$ & $\mathbf{0.017}$ & $3.14$ & $\mathbf{0.007}$\\
 &  & RF-GO & 20 & $1.50$ & $0.160$ & $2.31$ & $\mathbf{0.037}$\\
 & DINO & RF-DiVE & 19 & $2.40$ & $\mathbf{0.031}$ & $3.69$ & $\mathbf{0.002}$\\
 &  & RF-GO & 20 & $0.00$ & $1.000$ & $1.29$ & $0.224$\\
V3 & ADV & RF-DiVE & 19 & $2.12$ & $0.054$ & $1.86$ & $0.085$\\
 &  & RF-GO & 20 & $4.54$ & $\mathbf{3.2\!\times\!10^{-4}}$ & $3.57$ & $\mathbf{0.003}$\\
 & DINO & RF-DiVE & 19 & $6.64$ & $\mathbf{5.9\!\times\!10^{-6}}$ & $6.26$ & $\mathbf{1.2\!\times\!10^{-5}}$\\
 &  & RF-GO & 20 & $0.52$ & $0.616$ & $0.84$ & $0.423$\\
hV4 & ADV & RF-DiVE & 20 & $3.68$ & $\mathbf{0.002}$ & $5.15$ & $\mathbf{8.5\!\times\!10^{-5}}$\\
 &  & RF-GO & 20 & $2.56$ & $\mathbf{0.022}$ & $3.26$ & $\mathbf{0.005}$\\
 & DINO & RF-DiVE & 20 & $3.38$ & $\mathbf{0.004}$ & $5.38$ & $\mathbf{5.3\!\times\!10^{-5}}$\\
 &  & RF-GO & 20 & $3.20$ & $\mathbf{0.006}$ & $3.73$ & $\mathbf{0.002}$\\
\hline
\end{tabular}
\end{table*}

\begin{table*}[!htbp]
\centering
\footnotesize
\setlength{\tabcolsep}{3pt}
\renewcommand{\arraystretch}{1.08}
{\small\tablename~\ref{tab:rf_paired_voxel_tests} (continued): S7.\par}
\smallskip
\begin{tabular}{lllrrrrr}
\hline
 & & & & \multicolumn{2}{c}{MEI $-$ NSD} & \multicolumn{2}{c}{MEI $-$ LAION}\\
\cline{5-6}\cline{7-8}
ROI & Backbone & Method & $n$ & $t$ & $p_{\mathrm{BH}}$ & $t$ & $p_{\mathrm{BH}}$\\
\hline
\multicolumn{8}{l}{\textbf{S7 --- Full image}}\\
V1 & ADV & RF-DiVE & 20 & $3.67$ & $\mathbf{0.002}$ & $3.96$ & $\mathbf{0.001}$\\
 &  & RF-GO & 20 & $4.55$ & $\mathbf{3.1\!\times\!10^{-4}}$ & $6.35$ & $\mathbf{8.0\!\times\!10^{-6}}$\\
 & DINO & RF-DiVE & 20 & $9.57$ & $\mathbf{4.6\!\times\!10^{-8}}$ & $9.46$ & $\mathbf{5.4\!\times\!10^{-8}}$\\
 &  & RF-GO & 20 & $6.43$ & $\mathbf{6.8\!\times\!10^{-6}}$ & $6.74$ & $\mathbf{3.9\!\times\!10^{-6}}$\\
V2 & ADV & RF-DiVE & 19 & $0.92$ & $0.382$ & $3.40$ & $\mathbf{0.004}$\\
 &  & RF-GO & 20 & $2.37$ & $\mathbf{0.032}$ & $4.03$ & $\mathbf{9.6\!\times\!10^{-4}}$\\
 & DINO & RF-DiVE & 19 & $7.67$ & $\mathbf{1.1\!\times\!10^{-6}}$ & $9.18$ & $\mathbf{1.2\!\times\!10^{-7}}$\\
 &  & RF-GO & 20 & $8.17$ & $\mathbf{3.4\!\times\!10^{-7}}$ & $7.37$ & $\mathbf{1.3\!\times\!10^{-6}}$\\
V3 & ADV & RF-DiVE & 19 & $5.41$ & $\mathbf{5.9\!\times\!10^{-5}}$ & $5.41$ & $\mathbf{5.9\!\times\!10^{-5}}$\\
 &  & RF-GO & 20 & $5.86$ & $\mathbf{2.0\!\times\!10^{-5}}$ & $4.56$ & $\mathbf{3.1\!\times\!10^{-4}}$\\
 & DINO & RF-DiVE & 19 & $6.73$ & $\mathbf{5.2\!\times\!10^{-6}}$ & $8.16$ & $\mathbf{5.1\!\times\!10^{-7}}$\\
 &  & RF-GO & 20 & $8.45$ & $\mathbf{2.3\!\times\!10^{-7}}$ & $6.91$ & $\mathbf{3.0\!\times\!10^{-6}}$\\
hV4 & ADV & RF-DiVE & 20 & $5.97$ & $\mathbf{1.6\!\times\!10^{-5}}$ & $5.50$ & $\mathbf{4.2\!\times\!10^{-5}}$\\
 &  & RF-GO & 20 & $4.87$ & $\mathbf{1.6\!\times\!10^{-4}}$ & $5.29$ & $\mathbf{6.4\!\times\!10^{-5}}$\\
 & DINO & RF-DiVE & 20 & $6.85$ & $\mathbf{3.2\!\times\!10^{-6}}$ & $7.68$ & $\mathbf{7.7\!\times\!10^{-7}}$\\
 &  & RF-GO & 20 & $4.15$ & $\mathbf{7.4\!\times\!10^{-4}}$ & $4.54$ & $\mathbf{3.2\!\times\!10^{-4}}$\\
\hline
\multicolumn{8}{l}{\textbf{S7 --- Full image + pRF}}\\
V1 & ADV & RF-DiVE & 20 & $0.68$ & $0.510$ & $1.12$ & $0.293$\\
 &  & RF-GO & 20 & $1.21$ & $0.255$ & $1.94$ & $0.074$\\
 & DINO & RF-DiVE & 20 & $6.06$ & $\mathbf{1.4\!\times\!10^{-5}}$ & $6.48$ & $\mathbf{6.2\!\times\!10^{-6}}$\\
 &  & RF-GO & 20 & $3.59$ & $\mathbf{0.002}$ & $4.08$ & $\mathbf{8.5\!\times\!10^{-4}}$\\
V2 & ADV & RF-DiVE & 19 & $-0.18$ & $0.862$ & $2.01$ & $0.066$\\
 &  & RF-GO & 20 & $0.93$ & $0.379$ & $2.61$ & $\mathbf{0.020}$\\
 & DINO & RF-DiVE & 19 & $5.84$ & $\mathbf{2.5\!\times\!10^{-5}}$ & $6.94$ & $\mathbf{3.6\!\times\!10^{-6}}$\\
 &  & RF-GO & 20 & $6.02$ & $\mathbf{1.5\!\times\!10^{-5}}$ & $5.73$ & $\mathbf{2.6\!\times\!10^{-5}}$\\
V3 & ADV & RF-DiVE & 19 & $5.10$ & $\mathbf{1.1\!\times\!10^{-4}}$ & $5.33$ & $\mathbf{6.9\!\times\!10^{-5}}$\\
 &  & RF-GO & 20 & $5.27$ & $\mathbf{6.6\!\times\!10^{-5}}$ & $4.16$ & $\mathbf{7.2\!\times\!10^{-4}}$\\
 & DINO & RF-DiVE & 19 & $7.06$ & $\mathbf{3.0\!\times\!10^{-6}}$ & $9.90$ & $\mathbf{4.6\!\times\!10^{-8}}$\\
 &  & RF-GO & 20 & $8.44$ & $\mathbf{2.3\!\times\!10^{-7}}$ & $6.25$ & $\mathbf{9.7\!\times\!10^{-6}}$\\
hV4 & ADV & RF-DiVE & 20 & $5.78$ & $\mathbf{2.4\!\times\!10^{-5}}$ & $4.75$ & $\mathbf{2.0\!\times\!10^{-4}}$\\
 &  & RF-GO & 20 & $4.62$ & $\mathbf{2.7\!\times\!10^{-4}}$ & $5.08$ & $\mathbf{9.8\!\times\!10^{-5}}$\\
 & DINO & RF-DiVE & 20 & $7.43$ & $\mathbf{1.2\!\times\!10^{-6}}$ & $7.82$ & $\mathbf{6.3\!\times\!10^{-7}}$\\
 &  & RF-GO & 20 & $4.03$ & $\mathbf{9.5\!\times\!10^{-4}}$ & $4.21$ & $\mathbf{6.5\!\times\!10^{-4}}$\\
\hline
\end{tabular}
\end{table*}

\clearpage

%% file: iclr2027_mei_top8_figures.tex

\newcommand{\MEIPath}{appendix_figures/MEIs/ICLR2026_mei_top8_figures}

\newcommand{\SingleMEIFigure}[5]{%
  \clearpage
  \begin{figure}[p]
    \centering
    \includegraphics[
      height=\textheight,
      width=\linewidth,
      keepaspectratio
    ]{\MEIPath/sub-#1__#2__#3__top8.pdf}
    \caption{MEIs of top-8 voxels for Subject #1, visual area #2, using the
      #4 brain-encoder backbone.}
    \label{fig:mei-#5}
  \end{figure}
  \clearpage
}

\SingleMEIFigure{01}{V1}{ADV_RN50}{ADV-RN50}{s01-v1-adv}
\SingleMEIFigure{01}{V2}{ADV_RN50}{ADV-RN50}{s01-v2-adv}
\SingleMEIFigure{01}{V3}{ADV_RN50}{ADV-RN50}{s01-v3-adv}
\SingleMEIFigure{01}{hV4}{ADV_RN50}{ADV-RN50}{s01-hv4-adv}

\SingleMEIFigure{01}{V1}{DINO_RN50}{DINO-RN50}{s01-v1-dino}
\SingleMEIFigure{01}{V2}{DINO_RN50}{DINO-RN50}{s01-v2-dino}
\SingleMEIFigure{01}{V3}{DINO_RN50}{DINO-RN50}{s01-v3-dino}
\SingleMEIFigure{01}{hV4}{DINO_RN50}{DINO-RN50}{s01-hv4-dino}

\SingleMEIFigure{02}{V1}{ADV_RN50}{ADV-RN50}{s02-v1-adv}
\SingleMEIFigure{02}{V2}{ADV_RN50}{ADV-RN50}{s02-v2-adv}
\SingleMEIFigure{02}{V3}{ADV_RN50}{ADV-RN50}{s02-v3-adv}
\SingleMEIFigure{02}{hV4}{ADV_RN50}{ADV-RN50}{s02-hv4-adv}

\SingleMEIFigure{02}{V1}{DINO_RN50}{DINO-RN50}{s02-v1-dino}
\SingleMEIFigure{02}{V2}{DINO_RN50}{DINO-RN50}{s02-v2-dino}
\SingleMEIFigure{02}{V3}{DINO_RN50}{DINO-RN50}{s02-v3-dino}
\SingleMEIFigure{02}{hV4}{DINO_RN50}{DINO-RN50}{s02-hv4-dino}

\SingleMEIFigure{05}{V1}{ADV_RN50}{ADV-RN50}{s05-v1-adv}
\SingleMEIFigure{05}{V2}{ADV_RN50}{ADV-RN50}{s05-v2-adv}
\SingleMEIFigure{05}{V3}{ADV_RN50}{ADV-RN50}{s05-v3-adv}
\SingleMEIFigure{05}{hV4}{ADV_RN50}{ADV-RN50}{s05-hv4-adv}

\SingleMEIFigure{05}{V1}{DINO_RN50}{DINO-RN50}{s05-v1-dino}
\SingleMEIFigure{05}{V2}{DINO_RN50}{DINO-RN50}{s05-v2-dino}
\SingleMEIFigure{05}{V3}{DINO_RN50}{DINO-RN50}{s05-v3-dino}
\SingleMEIFigure{05}{hV4}{DINO_RN50}{DINO-RN50}{s05-hv4-dino}

\SingleMEIFigure{07}{V1}{ADV_RN50}{ADV-RN50}{s07-v1-adv}
\SingleMEIFigure{07}{V2}{ADV_RN50}{ADV-RN50}{s07-v2-adv}
\SingleMEIFigure{07}{V3}{ADV_RN50}{ADV-RN50}{s07-v3-adv}
\SingleMEIFigure{07}{hV4}{ADV_RN50}{ADV-RN50}{s07-hv4-adv}

\SingleMEIFigure{07}{V1}{DINO_RN50}{DINO-RN50}{s07-v1-dino}
\SingleMEIFigure{07}{V2}{DINO_RN50}{DINO-RN50}{s07-v2-dino}
\SingleMEIFigure{07}{V3}{DINO_RN50}{DINO-RN50}{s07-v3-dino}
\SingleMEIFigure{07}{hV4}{DINO_RN50}{DINO-RN50}{s07-hv4-dino}